\documentclass{aa}

\usepackage[varg]{txfonts}
\usepackage{graphicx}
\usepackage{amsmath}
\usepackage{multirow}
\usepackage[colorlinks,linkcolor=blue,citecolor=blue]{hyperref}

\bibpunct{(}{)}{;}{a}{}{,}	% defines punctuation in citations according to A&A style

\usepackage[flushleft]{threeparttable}
\usepackage{color}

\newcommand{\changeslan}[1]{#1}

\begin{document}

\title{Realistic collisional water transport during terrestrial planet formation}
\subtitle{Self-consistent modeling by an N-body--SPH hybrid code}

\author{C. Burger\inst{1,2} \and \'A. Bazs\'o\inst{1} \and C. M. Sch{\"a}fer\inst{2}}
\institute{Department of Astrophysics, University of Vienna, T{\"u}rkenschanzstra{\ss}e 17, 1180 Vienna, Austria \and Institute of Astronomy and Astrophysics, University of T{\"u}bingen, Auf der Morgenstelle 10, 72076 T{\"u}bingen, Germany, \email{christoph.burger@uni-tuebingen.de}}

\date{Received \dots / Accepted \dots}

\abstract
%context
{According to the latest theoretical and isotopic evidence, \changeslan{Earth's water content} originates mainly from today's asteroid belt region, or at least from the same precursor material. This suggests that water was transported inwards to Earth, and to similar planets in their habitable zone, via (giant) collisions of planetary embryos and planetesimals during the chaotic final phase of planet formation.}
% aims
{In current dynamical simulations water delivery to terrestrial planets is still studied almost exclusively by assuming oversimplified perfect merging, even though water and other volatiles are particularly prone to collisional transfer and loss.
To close this gap we have developed a computational framework to model collisional water transport by direct combination of long-term N-body computations with dedicated 3D smooth particle hydrodynamics (SPH) collision simulations of differentiated, self-gravitating bodies for each event.}
% methods
{Post-collision water inventories are traced self-consistently in the further dynamical evolution, in accretionary or erosive as well as hit-and-run encounters with two large surviving bodies, where besides collisional losses, water transfer between the encountering bodies has to be considered. This hybrid approach enables us for the first time to trace the full dynamical and collisional evolution of a system of approximately 200 bodies throughout the whole late-stage accretion phase (several hundred Myr). As a first application we choose a Solar System-like architecture with already formed giant planets on either circular or eccentric orbits and a debris disk spanning the whole terrestrial planet region (0.5\,-\,4\,au).}
% results
{Including realistic collision treatment leads to considerably different results than simple perfect merging, with lower mass planets and water inventories reduced regularly by a factor of two or more.
Due to a combination of collisional losses and a considerably lengthened accretion phase, \changeslan{final water content, especially with giant planets on circular orbits,} is strongly reduced to more Earth-like values, and closer to results with eccentric giant planets.
Water delivery to potentially habitable planets is dominated by very few decisive collisions, mostly with embryo-sized or larger objects and only rarely with smaller bodies, at least if embryos have formed throughout the whole disk initially. \changeslan{The high frequency of hit-and-run collisions and the differences to predominantly accretionary encounters, such as generally low water (and mass) transfer efficiencies, are a crucial part of water delivery, and of system-wide evolution in general.}}
%conclusions
{}


\maketitle

\section{Introduction and motivation}
\label{sect:introduction}
The origin of water on Earth, but also on terrestrial planets in general, is a long-standing problem in planetary science, especially for planets with (current) orbital distances where protoplanetary disk material is believed to have been practically dry. All four terrestrial planets in the Solar System are examples thereof.
Against \changeslan{those odds}, Earth managed to acquire (and retain) a substantial amount of water during its formation, at least that stored in some form on the surface and in the crust, referred to as an Earth-ocean\footnote{The sum of surface and crustal water inventory amounts to $\sim$\,$1.7\times10^{21}$\,kg, which is $\sim$\,$2.8\times10^{-4}$ of Earth's mass \citep{Lecuyer1998_hydrogen_isotope_composition_seawater_and_global_water_cycle}.}, and there is evidence for considerable water inventories on early Venus and Mars as well.
In the case of Earth, even the current total water content is highly uncertain, due to suspected large reservoirs in the mantle, estimated to contain up to ten Earth-oceans \citep[e.g.,][]{Marty2012_Origins_of_volatiles_on_Earth}.
Alternative theories, where growing planets inside the ice line (snow line) acquired their water content from local disk material via direct incorporation into growing planetesimals, suffer various difficulties \citep[see, e.g.,][]{Izidoro2013_Compound_model_origin_water, OBrien2018_The_Delivery_of_Water_During_Terrestrial_Planet_Formation, DAngelo2019_Monte_Carlo_sims_of_forsterite_hydration}.
The same is true for the hypothesis of delivery by comets, due to nonmatching isotopic fingerprints (D/H ratios), but more severely because the amount of water that can be dynamically delivered is significantly too low \citep{Morbidelli2000_Source_regions_for_water}.
Therefore, dynamical delivery from an external source at the current location of the (outer) asteroid belt (or at least from the same source material) appears to be the most likely explanation at present.
From the point of view of habitability, liquid water is believed to be one of the few absolutely necessary conditions for life.
Excessive water content however can also have adverse effects for habitability \citep[e.g.,][]{Noack2016_Water-rich_planets_how_habitable}, can impede sea-floor volcanism, and the formation of high-pressure ices at the bottom of global oceans may cut off connection to the underlying silicates.
Therefore, an intermediate water content (like Earth) appears to be the most favorable for habitability.
The increasing number of known low-mass exoplanets has unraveled a broad compositional diversity, ranging from probably iron-dominated, Mercury-like bodies to sub-Neptunes with extended envelopes.
Proper characterization of their physical structure requires knowledge of their bulk composition, which is shaped by collisions on all scales, up to (and particularly) giant impacts between planet-sized bodies \citep[e.g.,][]{Marcus2009_Collisional_stripping_of_Super-Earths, Marcus2010_Icy_Super_Earths_max_water_content, Inamdar2016_Stealing_the_gas_diversity_exoplanet_densities, Bonomo2019_GI_likely_origin_Kepler-107_twins}.
For all these reasons it is important to understand the origin and evolution of water and other volatiles, which is intimately linked to their fate in collision events.

Several theories for the growth from dust to planetary embryos have been put forward, from the classical scenario of planetesimal- followed by oligarchic growth \citep[e.g.,][]{Kokubo2002_Formation_of_protoplanet_systems_oligarchic_growth, McNeil2005_Effects_of_typeI_migration_on_terrestrial_planet_formation}, to more recent alternatives like direct accretion from pebbles \citep[see][for an overview]{Johansen2017_Forming_planets_via_pebble_accretion}.
However, all these theories cumulate in a common final phase of accretion, where embryos and remaining planetesimals gradually grow into planets. This stage begins roughly when the gas disk disappears and the remaining smaller bodies cannot provide enough dynamical friction to keep embryos on separated orbits. It is characterized by chaotic interaction and giant collisions between embryos and/or protoplanets, which shape the final characteristics of planets.
Indirect evidence for such events has been found in young extrasolar systems \citep[e.g.,][]{Wyatt2016_Insights_into_Planet_Formation_from_debris_disks_II_GI_exosystems}, mainly as observable infrared excess of warm dust attributed to debris of recent collisions \citep[e.g.,][]{Meng2014_Large_impacts_around_solar-analog_star, Morlok2014_Dust_from_collisions_probe_composition_exoplanets, Leinhardt2015_Numerically_predicted_signatures_of_TPF, Bonomo2019_GI_likely_origin_Kepler-107_twins}.
A direct consequence of late-stage accretion is radial mixing, which allows the inwards transport of water-rich material from beyond the ice line.
For the Solar System, dynamical as well as isotopic evidence indicates that most of Earth's water originates from the asteroid belt region, or at least from the same precursor material. This basic means of water delivery to growing terrestrial planets is also the underlying working hypothesis in this study.
A potentially crucial influence comes from giant planets, which form quickly (within a few Myr), and can excite the orbits of planetary building blocks even if they remain on static orbits themselves and do not run into dynamical instabilities or migrate larger distances \citep[as proposed in the framework of Grand Tack models for example;][]{Walsh2011_Grand_Tack, OBrien2014_Water_delivery_and_giant_impacts_in_Grand_Tack}.
It stands to reason that similar mechanisms of water transport are at work during the formation of many extrasolar systems despite large variations in individual histories.

Late-stage accretion is commonly studied by means of N-body simulations, where mostly the sole gravitational interaction of up to a few thousand bodies is followed for typically a few hundred Myr \citep[e.g.,][as some of the more recent work]{Raymond2006_High_res_sims_planet_formation_I, Kokubo2006_Formation_of_terrestrial_planets_from_protoplanets_I, Morishima2010_N-body_sims_including_gas_and_giant_planets, Chambers2013_Late-stage_accretion_hit-and-run_fragmentation, Fischer2014_Dynamics_terrestrial_planets_large_number_N-body_sims, OBrien2006_Terrestrial_planet_formation_with_strong_dynamical_friction, OBrien2014_Water_delivery_and_giant_impacts_in_Grand_Tack, Quintana2014_Effect_of_planets_beyond_ice_line_on_volatile_accretion, Quintana2016_Frequency_giant_impacts_Earth-like_worlds}.
Planets grow in principle by collisions, and at least for the fate of water, \changeslan{simple perfect merging} has been and is still assumed almost exclusively \citep[e.g.,][]{Haghighipour2007_Hab_planet_formation_binaries, Raymond2004_Dynamical_sims_water_delivery, Raymond2007_High_res_sims_planet_formation_II, Izidoro2013_Compound_model_origin_water, Fischer2014_Dynamics_terrestrial_planets_large_number_N-body_sims, OBrien2014_Water_delivery_and_giant_impacts_in_Grand_Tack, Quintana2014_Effect_of_planets_beyond_ice_line_on_volatile_accretion}, even though volatiles can be expected to be particularly affected by collisional processes.
More sophisticated collision outcome models based on semi-analytical scaling laws have been developed, first distinguishing only perfect accretion and hit-and-run \citep{Kokubo2010_Formation_under_realistic_accretion, Genda2012_Merging_criteria_for_giant_impacts}, and later including a more extended range of outcomes \citep{Leinhardt2012_Collisions_between_gravity-dominated_bodies_I, Stewart2012_Collisions_between_gravity-dominated_bodies_II, Leinhardt2015_Numerically_predicted_signatures_of_TPF}, and also an approximate treatment of compositional changes following \citet{Marcus2009_Collisional_stripping_of_Super-Earths, Marcus2010_Icy_Super_Earths_max_water_content}.
These models have been applied in various studies on planetary accretion \citep[e.g.,][]{Chambers2013_Late-stage_accretion_hit-and-run_fragmentation, Bonsor2015_Collisional_origin_to_Earths_non-chondritic_composition, Carter2015_Compositional_evolution_during_accretion, Carter2018_Collisional_stripping_of_crusts, Quintana2016_Frequency_giant_impacts_Earth-like_worlds}.
\changeslan{However, they seem to be currently not suited to also account for minor constituents like surface volatile layers}, or the more complicated behavior of hit-and-run with individual (water) losses and transfer between the colliding bodies, as tested by \citet{Burger2018_Hit-and-run}.
A very recent approach is the use of machine learning to predict collision outcomes, but this has not yet progressed beyond basic parameters like the mass accretion efficiency \citep{Cambioni2019_Machine_learning_applied_to_GI}.
A combination of dynamical and impact simulations to study the flux of water-rich material to habitable zone planets has been applied by \citet{Bancelin2017_Orbital_resonances_water_transport_in_binaries}, albeit only at the end point of the dynamical evolution of inward-scattered objects.
Direct coupling of N-body computations and individual collision simulations -- the approach we use in this work -- has \changeslan{rarely been applied} so far \citep{Genda2011_Direct_coupling_N-body_SPH, Genda2017_Hybrid_code_Ejection_of_iron-bearing_fragments}, and only to follow nonvolatile material (iron and silicates).

While assuming complete accretion (perfect merging) of the colliding bodies is typically a good approximation for large-enough mass ratios, it is often not justified for more similar-sized bodies, with a diverse range of possible outcomes between (partial) accretion, hit-and-run, and (partial) erosion \citep[see, e.g.,][]{Leinhardt2012_Collisions_between_gravity-dominated_bodies_I}.
There have only been a few studies on water transfer and loss in collisions of large, gravity-dominated bodies.
In most cases, only suites of individual, separate encounters were investigated, from planetesimal- to embryo-sized \citep{Maindl2014_Fragmentation_of_planetesimals_water, MaindlSchaeferHaghighipourBurger2017_water_transport, DvorakMaindlBurger2015_Nonlinear_phenomena_complex_systems_article}, up to planet-sized \citep{Canup2006_Water_planet-scale_collisions, Burger2017_hydrodynamic_scaling, Burger2018_Hit-and-run, Burger2019_IAU_proceedings_water_delivery_to_dry_protoplanets_hit_and_run} and even in the super-Earth regime \citep{Marcus2010_Icy_Super_Earths_max_water_content}.
These studies showed that water losses in a single event can reach tens of percent, but often considered only combined losses and did not distinguish the fate of individual survivors.
\changeslan{The relevant collision parameter space is large, consisting at least of impact velocity, impact angle, the mass ratio, and the combined mass, but also further ones, such as the composition of the colliding bodies.
For this reason, such studies are naturally forced to fix one or more parameters.}
Unlike the other collision regimes, in hit-and-run there are two large post-collision bodies, where not only water losses but also transfer (such as from a water-rich impactor to a water-poor target) are important \citep{Burger2018_Hit-and-run, Burger2019_IAU_proceedings_water_delivery_to_dry_protoplanets_hit_and_run}.
Hit-and-run collisions can also be very transformative and strip especially the smaller body of its outer layers, such as atmosphere, ocean, or parts of the mantle \citep{Asphaug2010_Similar_sized_collisions_diversity_of_planets}, or result in large-scale degassing of the interior due to pressure unloading \citep{Asphaug2006_hit-and-run}. Since the bodies continue on separated trajectories after hit-and-run, it is imperative to track them both independently \citep{Emsenhuber2019_Fate_of_Runner_in_har}, which rules out approaches where collisions are evaluated only as a post-processing step (e.g., after perfect merging runs).

Even though we focus on the collisional aspects of water transport, the physical state of mantle, surface, and atmosphere can all influence water inventories and their evolution.
Large collisions, in combination with heating by short-lived radioisotopes and differentiation, can lead to local or even global magma oceans.
Existing water is either vaporized into a steam atmosphere, or gradually outgassed during magma ocean solidification, after which it is susceptible to atmospheric loss processes and impact erosion \citep[][and references therein]{Schlichting2015_Atmospheric_mass_loss_impacts, Schlichting2018_Atmosphere_Impact_Losses}.
Current models for coupled magma ocean and atmosphere evolution on approximately Earth-sized planets \citep{Hamano2013_Oceans_after_solidification_of_magma_ocean, Lebrun2013_Magma_ocean_with_atmosphere_water_condensation, Salvador2017_Influence_of_H20_CO2_on_water_evolution_of_rocky_planets} suggest that beyond a critical distance ($\sim$\,0.7\,au in the Solar System) rapid magma ocean solidification within a few Myr and subsequent water ocean formation are possible.
Smaller bodies can experience considerable volatile losses also farther from the star before re-condensation \citep{Odert2018_Escape_and_fractionation_of_volatiles_from_growing_protoplanets}.
However, if most water is gradually transported inwards from larger orbital distances, the likelihood of retention also by smaller protoplanets is increased.
With typical time-spans between giant collisions of several Myr (Sect.~\ref{sect:timing_accretion_water_delivery_pot_hab_planets}), repeated surface solidification and water ocean formation appear plausible, at least for already sufficiently large bodies in the habitable zone and beyond.

In the following we first describe the applied physical model including the initial conditions in Sect.~\ref{sect:physical_model}, while aspects of its numerical realization are discussed separately in Sect.~\ref{sect:numerical_model}.
An overview of the computed scenarios is provided in Sect.~\ref{sect:scenarios}.
The presentation of our results is split into two logical parts, where collisional water transport from a system-wide point of view is covered in Sect.~\ref{sect:results_I}, followed by a detailed analysis focused on potentially habitable planets in Sect.~\ref{sect:results_II}.
A combined discussion in Sect.~\ref{sect:discussion_and_conclusions} and a summary in Sect.~\ref{sect:summary} conclude the paper.

\section{The physical model}
\label{sect:physical_model}
Our simulations focus on the gravitational and collisional dynamics of the late-stage accretion (giant impact) phase of terrestrial planet formation, and are set in an already gas-free environment. The initial conditions represent the transition from oligarchic growth to late-stage accretion, with a bimodal system of a relatively small number of approximately Mars-sized oligarchs (planetary embryos) embedded in a large number of smaller bodies \citep[see, e.g.,][]{Kokubo2002_Formation_of_protoplanet_systems_oligarchic_growth}. Embryos are initially still on dynamically cold orbits due to dynamical friction with small bodies (in the following referred to as `planetesimals') and possibly previous dampening by the gas disk, but become quickly excited by the influence of giant planets and by self-stirring.
\changeslan{In general, all bodies interact gravitationally, except for planetesimals with each other.} Planetesimal--planetesimal interactions can be either switched off entirely (also collisions), or they interact only in close encounters (gravitationally, and they can collide), which is further described in Sect.~\ref{sect:N-body_numerics}.

We include a 1\,$M_\odot$ star, and already formed nonmigrating Jupiter- and Saturn-like gas giants at 5.2 and 9.6\,au.
These start on either circular orbits \citep[similar to the Nice model,][but with their current semi-major axes]{Tsiganis2005_Nice_model}, or eccentric orbits with today's eccentricities of those planets ($\sim$\,0.05), and are referred to as `circular Jupiter and Saturn' (cJS) and `eccentric Jupiter and Saturn' (eJS), respectively. Similar configurations have been frequently used in the literature to study different giant planet architectures \citep[e.g.,][]{OBrien2006_Terrestrial_planet_formation_with_strong_dynamical_friction, Fischer2014_Dynamics_terrestrial_planets_large_number_N-body_sims, Quintana2014_Effect_of_planets_beyond_ice_line_on_volatile_accretion}.
In these works, eccentric gas giants have generally proved detrimental for water delivery (by dynamically removing water-rich material), at least in the absence of remaining gas \citep{Morishima2010_N-body_sims_including_gas_and_giant_planets}.
We note that it is not our principal aim to study (or reproduce) the formation of the Solar System in detail, but rather to investigate the implications of realistic treatment of collisional water transport during late-stage accretion in general.

\subsection{Initial conditions}
\label{sect:initial_conditions}
Our main scenarios include an initial debris disk of approximately Mars-sized planetary embryos and approximately Moon-sized `planetesimals', extending from 0.5 to 4\,au. Both of these populations follow the same disk profile, with the surface density in solids
\begin{equation} \label{eq:surfdens}
    \Sigma_\mathrm{solid} (r) = \Sigma_0 \left( \frac{r}{1\text{au}} \right)^{-\alpha}.
\end{equation}
We use $\Sigma_0 =$ 10~g/cm$^2$ ($1.125 \times 10^{-6} \, M_{\odot}$/au$^2$) and $\alpha = 1.5$, consistent with the Minimum Mass Solar Nebula, in all scenarios, and do not consider discontinuities across the ice line.
These parameters result in a total disk mass in solids of $\sim$\,6\,$M_\oplus$.

The generation of planetary embryos follows the concept of the isolation mass $M_\mathrm{iso} = 2\pi a b \Sigma_\mathrm{solid}(a)$, which is simply the total mass available in some disk annulus with width $b$ at semi-major axis $a$.
The initial mass fraction in embryos is independent of the location in the disk and they are placed at separations of several times their mutual Hill radii
\begin{equation} \label{eq:mutualhillradius}
    R_\mathrm{H,m} = \frac{a_1 + a_2}{2} \left( \frac{m_1 + m_2}{3 M_0} \right)^{1/3},
\end{equation}
with semi-major axes $a_1$ and $a_2$, masses $m_1$ and $m_2$, and the mass of the star $M_0$.
The precise parameter setups for all our scenarios are described in Sect.~\ref{sect:scenarios}.
The above introduced disk structure (Eq.\,\ref{eq:surfdens}) corrected for the actual mass fraction in embryos, along with the isolation mass and their mutual spacings, constrains the distribution of embryos to a single unique solution consisting typically of a few tens of bodies.
For the chosen disk architecture the masses and separations of embryos both increase with semi-major axis.
The remaining fraction of disk-solids is assumed to be still in equal-mass planetesimals that have not yet been accreted.
The supposition that planetesimals follow the same surface density profile as embryos (Eq.\,\ref{eq:surfdens}) results also in increasing separations with semi-major axis.
Initial eccentricities and inclinations are drawn from Rayleigh distributions, with scale factors $\sigma = 0.005$ for embryos, and $\sigma = 0.05$ for planetesimals, meaning almost co-planar, low-eccentricity orbits.
The three remaining orbital elements (argument of perihelion, longitude of ascending node, mean anomaly) are distributed uniformly over $[0,2\pi]$.

\begin{figure}
	\centering{\includegraphics[width=7.5cm]{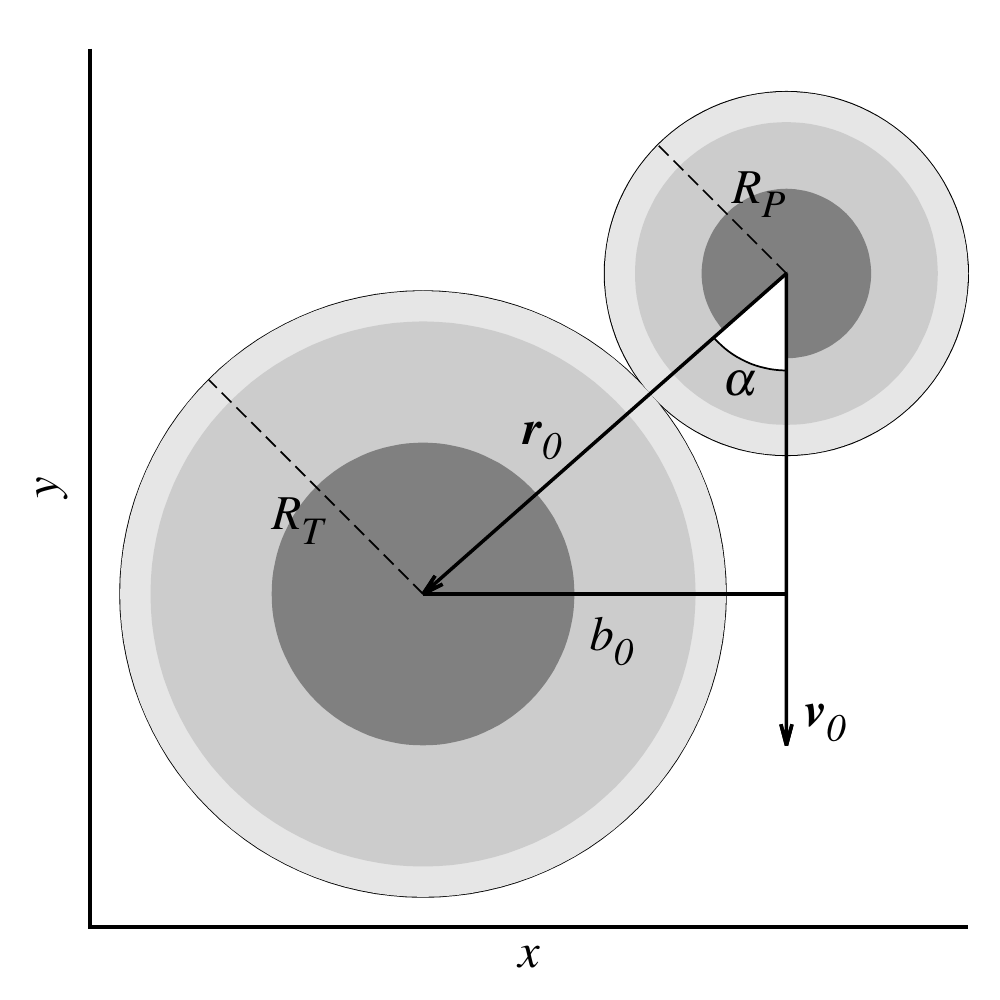}}
	\caption{Collision geometry of two differentiated bodies in a frame centered on the target. The impact angle $\alpha$ ranges from $0^\circ$ for head-on to $90^\circ$ for just grazing. The impact velocity is $|\mathbf{v}_0|$. The common convention is to label the larger body as the \emph{target}, but in some situations the smaller object is the one followed (e.g., on its growth path towards a planet) and is thus considered (and referred to as) the target.}
	\label{fig:collision_geometry_differentiated}
	\end{figure}

The embryos and planetesimals themselves generally have a differentiated three-layer structure with an iron core, a silicate mantle, and a water shell (cf.\ Fig.~\ref{fig:collision_geometry_differentiated}).
Initially, all bodies are assigned an equal core mass fraction of 0.25, but water mass fractions (WMF) vary with distance to the star, following a step function:
\begin{equation}
\mathrm{WMF} = \left\{ \begin{array}{lc} 10^{-5} & \qquad a < 2\,\mathrm{au}\\ 10^{-3} & \qquad 2\,\mathrm{au} < a < 2.5\,\mathrm{au}\\ 0.05 & \qquad a > 2.5\,\mathrm{au} \end{array} \right.
\end{equation}
This particular distribution is based on meteorite data and has been \changeslan{regularly used in previous studies} \citep[e.g.,][]{Raymond2004_Dynamical_sims_water_delivery,
Raymond2006_High_res_sims_planet_formation_I,
Raymond2007_High_res_sims_planet_formation_II, Haghighipour2007_Hab_planet_formation_binaries,
Fischer2014_Dynamics_terrestrial_planets_large_number_N-body_sims,
Quintana2014_Effect_of_planets_beyond_ice_line_on_volatile_accretion}.

\subsection{Collision model}
\label{sect:collision_model}
Our collision model is based on dedicated hydrodynamical simulations among differentiated embryos and planetesimals.
The bodies do not possess atmospheric components, and therefore the water inventory is represented by a surface liquid or solid layer, where all material is modeled as a fluid without material strength. We have chosen such properties not because of their simplicity (we could have included solid-body behavior quite easily and this usually is also not much more computationally expensive), but rather due to the expected broad diversity of material properties in the simulated environment.
It is not only the current physical state (mass, composition, thermal state, and distance to the star), but especially also its history (thermal and orbital evolution, previous collisions) that would have to be modeled in order to reliably assess the current state of a given body.
We believe that a fluid description is the best compromise, as long as those factors are not accounted for additionally.

In general, all embryos and planetesimals can collide once they come closer than the sum of their (uncompressed) radii, hence we consider only real (physical) collisions.
In addition there is also the possibility to switch off interactions between planetesimals, meaning that they cannot collide or interact otherwise (see Sect.~\ref{sect:N-body_numerics}).
Each collision results in either 0, 1, or 2 post-collision bodies, which allows us to capture all main collision outcome regimes (total disruption, accretion, erosion, hit-and-run), while still assuring that the number of bodies does not increase.
The outcome is determined by computing the largest (most massive) and second-largest gravitationally bound aggregates of collision fragments (spatially connected material). These each remain in the simulation if their mass is above a lower threshold ($0.5-1.5\times 10^{21}\,\mathrm{kg} \cong 1-3\times 10^{-4}$\,$M_\oplus$, depending on the scenario), and are discarded when below.
Gravitational aggregates above $1.5\times 10^{23}\,\mathrm{kg} \cong 2.5\times 10^{-2}$\,$M_\oplus$ are treated as fully interacting embryos or protoplanets, and those below are included as planetesimals\footnote{\changeslan{We generally use the term `embryo' for planetary embryos from the initial configuration, and the term `protoplanet' for already larger bodies but not yet fully formed planets. However, there is no distinction from the perspective of the numerical model. The term `planetesimal' is used for the population of small bodies.}}.
The simulation then continues with the aggregates' positions, velocities, masses, and compositions.
Collisionally induced rotation is currently not tracked.
It occasionally happens that two bodies evolve into a (mutually) bound configuration (if the apocenter distance of their relative orbit is smaller than their mutual Hill radius) but do not physically collide. In those situations the bodies are still assumed to accrete (computed via perfect merging) in order to avoid computational slow down by a long-term bound configuration.
In collisions of protoplanets, embryos, or planetesimals with the star or one of the giant planets, the smaller body is completely accreted by the larger one. This happens in approaches closer than $\sim$\,3\,$\times\,R_\mathrm{G}$ for the giant planets, and closer than $\sim$\,5\,-\,10\,$\times\,R_\odot$ (depending on the scenario) for the star, with radii of the giant planet $R_\mathrm{G}$ and the Sun $R_\odot$.
Due to numerical reasons, bodies are also merged into the star if their semi-major axis falls below a threshold of $\sim$\,0.23\,au.

In addition to this model we also performed simulations with simple perfect merging in all collisions for comparison. The physical model and the initial conditions are the same, except for the treatment (or the lack thereof) of collisions.

\subsection{Ejections}
In the dynamical environment of our scenarios a considerable number of bodies are ejected from the system, or at least evolve onto orbits with large perihelia where they very likely would no longer participate in terrestrial planet formation.
We remove bodies from the simulation if they have distances to the star above 50\,au and are on unbound (hyperbolic) orbits, and also those that are still bound but have perihelia above $\sim$\,11\,au.

\section{The numerical hybrid model}
\label{sect:numerical_model}
Our simulations are based on three numerical pillars: long-term N-body integrations, smooth particle hydrodynamics (SPH) collision simulations, and the interface between them.

\subsection{N-body numerics}
\label{sect:N-body_numerics}
For the N-body integrations we use the \texttt{REBOUND} package \citep{Rein2012_REBOUND_introduction}, which offers several different integrators and a highly modularized design, which helped to incorporate it into our hybrid framework.
The bulk of integrations are carried out with the symplectic Wisdom-Holman integrator \texttt{WHFast} \citep{Rein2015_WHFAST_integrator}, using modified democratic heliocentric (\texttt{WHDS}) coordinates \citep{Hernandez2017_Symplectic_integrators_for_planetary_systems_error_analysis_and_comparison}, and a timestep of always less than 1\,day.
The embryo population is fully self-interacting, while two different models for the behavior of planetesimals are used:
\begin{itemize}
\item `pp0' -- planetesimals never interact with each other (neither gravitationally nor collisionally), only with the other types of bodies;
\item `pp1' -- planetesimals interact gravitationally with each other in close encounters, but not elsewhere, and they are allowed to collide with each other.
\end{itemize}
Model pp1 is intended to allow small bodies some self-interaction (e.g., mutual dynamical friction), while it simultaneously provides gravitational focusing for more realistic planetesimal--planetesimal collisions.
In order to accurately resolve close encounters, the integration switches to the \texttt{IAS15} integrator \citep{Rein2015_IAS15_integrator}, a \changeslan{fifteenth order} Gau{\ss}-Radau scheme, whenever two bodies come close.
The changeover distance is chosen as $\sim$\,2\,-\,3 $\times$ $R_\mathrm{H,m}$ for all encounters except those between planetesimals, where a somewhat lower value of $\sim$\,2 $\times$ $R_\mathrm{H,m}$ is used, with the mutual Hill radius $R_\mathrm{H,m}$ (Eq.~\ref{eq:mutualhillradius}).
Whether protoplanets, embryos, and planetesimals actually (physically) collide is determined only once they approach each other to within approximately 10 to 20 times the sum of their radii, which has proven suitable such that their (theoretical) two-body orbit is no longer significantly disturbed.
The pericenter distance of the (analytically computed) relative two-body orbit is then compared to the sum of the bodies' uncompressed radii, which are also used to include potential `tidal collisions' \citep[see, e.g.,][]{Asphaug2010_Similar_sized_collisions_diversity_of_planets}, even though they may still also collide physically due to tidal deformation.
If on a colliding trajectory then the simulation is continued until they have approached to the desired starting distance of the SPH simulation, set to 2.5 times the sum of their radii, which still allows the approximate buildup of tidal deformation before contact.
At this point the SPH simulation is started (see Sect.~\ref{sect:SPH_numerics}), while the whole N-body system is additionally integrated more accurately with \texttt{IAS15} for its duration (with the colliding bodies replaced by a merged placeholder), before SPH results are reinserted and the N-body integration continues.
Using this procedure allows us to precisely control the timing of the N-body integration and to have a well-defined starting distance for SPH.

We took great care to constantly monitor and check all conservation quantities and the consistency of results in all simulations. This is particularly important for the long-term N-body integration, including encounter handling, as well as the handover to and from the SPH simulations.
As a good indicator for overall accuracy, the relative error in total energy is on the order of $10^{-5}$ for the whole evolution over several hundred Myr (naturally excluding collisions and ejections or removals, where for example mass is removed from the system).
%... within 0.01\% over 200 Myr \citep{Genda2017_Hybrid_code_Ejection_of_iron-bearing_fragments}
%... energy better than 1e-4 and angular momentum better than 1e-11 \citep{Raymond2004_Dynamical_sims_water_delivery}
%... energy better than 1e-3 \citep{Raymond2006_High_res_sims_planet_formation_I}
%... 1e-5 to 1e-3 (Morishima et al., 2010), 

\subsection{Numerics of SPH collisions}
\label{sect:SPH_numerics}
Our hybrid approach requires a large number of dedicated hydrodynamical simulations performed with our SPH code \texttt{miluphcuda}.
It allows for fast and efficient computation by utilizing highly parallel graphics processing unit (GPU) hardware and has already been applied in various studies of collision and impact physics, including nonviscous fluids \citep[e.g.,][]{Burger2018_Hit-and-run, Malamud2018_Moonfalls}, solid bodies \citep[e.g.,][]{Maindl2014_Fragmentation_of_planetesimals_water, Burger2017_hydrodynamic_scaling}, granular media \citep[e.g.,][]{Schaefer2017_Regolith_sampling}, and porous continua \citep[e.g.,][]{Haghighipour2018_Main-belt_comets_porosity}.
A comprehensive description of the code is given in \citet{Schaefer2016_miluphcuda}, and we provide only a brief summary of the most relevant features here.

The colliding bodies themselves are fully self-gravitating and in addition the star and all other bodies in the underlying N-body simulation are included as point masses to ensure a consistent environment.
However, the kinematic state of the other bodies at the end of the SPH run is taken from the (usually more accurate) N-body integration which computes the whole system in parallel (see Sect.~\ref{sect:N-body_numerics}).
SPH runs start once two bodies are found to be on collision trajectories and have approached to the desired initial distance (2.5 times the sum of their radii).
The total simulation time is at least eight times the time until impact (roughly the collision timescale), which is intended to allow all main post-collision fragments to form and clearly separate.
The thermodynamic response of simulated materials is modeled by the nonlinear \citet{Tillotson1962_Metallic_EOS} equation of state (EoS), which is applicable over a wide range of conditions, but is still relatively simple and efficient \citep[see][for a summary]{Melosh1989_Impact_cratering}.
The reason for not using a more sophisticated EoS is similar to the rationale for a general fluid description for all materials (see Sect.~\ref{sect:collision_model}) -- the rather large uncertainties in the actual physical state of bodies before the collision.
In addition, it has been shown that especially for lower-density regions very high particle numbers are required for accurate results, where outcomes often depend on resolution rather than precise modeling by the EoS \citep{Reinhardt2017_Numerical_aspects_of_giant_impact}.
We use Tillotson parameters for iron from \citet{Melosh1989_Impact_cratering} and for basalt and water(-ice) from \citet{Benz1999_Catastrophic_disruptions_revisited}.

The differentiated bodies (iron core, silicate mantle, and water shell) are set up in an already relaxed state by using self-consistent hydrostatic radial profiles.
Those can be computed very rapidly and eliminate the need for lengthy numerical relaxation. We refer to \citet{Burger2018_Hit-and-run} for details on this procedure.
A numerical issue that requires special care is the spatial extension (thickness) of water layers.
Water mass fractions in the initial conditions range from $10^{-5}$ to 0.05 (Sect.~\ref{sect:initial_conditions}) and can and do naturally decrease further in collisions.
We apply a lower WMF limit of $10^{-9}$, below which objects are considered practically dry and their WMF is set to zero.
However, actual water layers are regularly thin, and would be made up of only a few SPH particles.
In order to maintain a reasonable extent in terms of SPH numerics and resolution, we artificially increase simulated water layers to a thickness of at least approximately three SPH particles where necessary (by replacing mantle mass by water).
This increase is corrected for after the simulation based on the assumption that relative water loss and transfer (e.g., in hit-and-run) is approximately independent of the actual thickness of the water layer.
\changeslan{Therefore, putting $\alpha$ times as much water mass on a body means that about $\alpha$ times as much (as would actually have been) is lost or transferred.}
This approximate scale-invariance has been tested and confirmed by \citet{Burger2018_Hit-and-run}.
The underlying physical reasoning is that if some part of a body's water shell is affected (lost, transferred, etc.) then the fate of this part is relatively independent of its thickness, where the error naturally decreases towards thinner surface layers.
The respective correction factor (1/$\alpha$) can be directly applied to the WMF of post-collision bodies, because the WMF is directly proportional to the (absolute) water content (for a fixed total mass).
Since the colliding bodies can have vastly different WMFs (either prior to the artificial increase, after it, or both), it is necessary to additionally distinguish between projectile and target water and to keep track of the origin of SPH particles. However, for these individual reservoirs the above considerations hold again \citep{Burger2018_Hit-and-run}.
This procedure leads to typical WMFs in SPH runs of around 0.05\,-\,0.2.

The post-collision analysis of identifying gravitationally bound objects and their dynamics and composition comprises two main steps.
First, a friends-of-friends algorithm identifies all directly (spatially) connected clumps of SPH particles (we refer to them as `fragments'), before gravitationally bound `aggregates' of those fragments are identified in the second step.
Starting from a seed aggregate consisting of the most massive fragment and all those mutually gravitationally bound to it, an iterative algorithm adds\,/\,removes fragments to\,/\,from this seed aggregate (its barycenter) until convergence.
After identification of the largest aggregate, this algorithm is run once again to find the second-largest one, starting from the most massive fragment not yet incorporated into the largest aggregate.
We refer to \citet{Burger2018_Hit-and-run} for more details on this procedure.
For hit-and-run, the largest aggregate usually originates in the target (the more massive pre-collision body) and the second-largest one in the projectile.
Each of the two aggregates is then either included as embryo (protoplanet), as planetesimal, or is otherwise discarded (see Sect.~\ref{sect:collision_model}).
If kept, the N-body run continues with the aggregate's barycenter position and velocity.

All SPH simulations use between 25\,000 and 75\,000 SPH particles, depending on the mass ratio of the colliding bodies, with 25k particles for mass ratios down to 1:25, 50k between 1:25 and 1:100, and 75k below 1:100.
This ensures a reasonable resolution even for scenarios involving large targets and comparatively small projectiles, while saving computing time in more similar-sized collisions.
The runtime depends mainly on particle number and impact velocity (which determines the overall simulated time), where a single SPH simulation, including all pre- and post-processing, takes on the order of an hour to complete.

\section{The scenarios}
\label{sect:scenarios}

\begin{table}
	\centering
	\small
	\caption{Summary of all scenarios.}
	\label{tab:scenarios}
	\begin{tabular}[h!]{p{2.67cm} p{1.2cm} p{0.4cm} cp{0.5cm} cp{0.5cm} rp{0.55cm}}
	\hline
	\hline
	Scenario name		&	$N_\mathrm{emb}$\,/\,$N_\mathrm{plan}$	&	$f_\mathrm{emb}$	&	Model	&	Arch.	&	Pla-Pla	\\
	\hline
	SPH[01-05]-cJS-pp1	&	35\,/\,150	&	0.7	&	SPH	&	cJS	&	yes	\\
	SPH[06-10]-cJS-pp0	&	35\,/\,150	&	0.7	&	SPH	&	cJS	&	no	\\
	SPH[11-15]-eJS-pp1	&	35\,/\,150	&	0.7	&	SPH	&	eJS	&	yes	\\
	SPH[16-20]-eJS-pp0	&	35\,/\,150	&	0.7	&	SPH	&	eJS	&	no	\\
	\hline
	PM[01-03]-cJS-pp1	&	35\,/\,150	&	0.7	&	PM	&	cJS	&	yes	\\
	PM[04-06]-cJS-pp0	&	35\,/\,150	&	0.7	&	PM	&	cJS	&	no	\\
	PM[07-09]-eJS-pp1	&	35\,/\,150	&	0.7	&	PM	&	eJS	&	yes	\\
	PM[10-12]-eJS-pp0	&	35\,/\,150	&	0.7	&	PM	&	eJS	&	no	\\
	\hline
	\hline
	EO-SPH[21-25]-cJS	&	29\,/\,0		&	1.0	&	SPH	&	cJS	&	yes	\\
	EO-SPH[26-30]-eJS	&	29\,/\,0		&	1.0	&	SPH	&	eJS	&	yes	\\
	\hline
	EO-PM[13-15]-cJS	&	29\,/\,0		&	1.0	&	PM	&	cJS	&	n/a	\\
	EO-PM[16-18]-eJS	&	29\,/\,0		&	1.0	&	PM	&	eJS	&	n/a	\\
	\hline
	\end{tabular}
	\tablefoot{$N_\mathrm{emb}$ and $N_\mathrm{plan}$ are the initial numbers of embryos and planetesimals. $f_\mathrm{emb}$ gives the mass fraction of the disk in embryos. `SPH' indicates our collision model, while `PM' indicates that all collisions are resolved with perfect merging. `Pla-Pla' indicates whether planetesimal--planetesimal interactions are included (pp1 vs.\ pp0). The embryo-only runs in the lower part of the table are marked by an `EO' prefix.}
	\end{table}

Our scenarios, all summarized in Table~\ref{tab:scenarios}, consist in general of a disk of embryos and planetesimals in the terrestrial planet region (0.5\,-\,4\,au), embedded in a Solar System-like architecture, with a 1\,$M_\odot$ star and two giant planets resembling Jupiter and Saturn. The latter are either on the current orbits of those planets with today's eccentricities (eJS) or initialized on circular orbits (cJS).
The surface density profile, total mass, and extent of the embryo\,+\,planetesimal disk are in principle the same in all simulations (see Sect.~\ref{sect:initial_conditions}).

For our regular scenarios, the mass fraction of the disk initially in embryos is set to 0.7, leaving 0.3 in planetesimals. With an embryo spacing of 10 mutual Hill radii, this gives a system of 35 embryos (between $\sim$\,0.06 and 0.25\,$M_\oplus$), and 150 equal-mass ($\sim$\,0.012\,$M_\oplus$\,$\sim$\,$M_\mathrm{Moon}$) planetesimals with approximately 1/10 of the average embryo mass.
These parameters provide a significant fraction of the total mass initially still in planetesimals combined with planetesimals that are considerably smaller than embryos in order to have two distinct populations, while also maintaining acceptable runtimes.
In addition to the two giant planet architectures (cJS and eJS), the two different models for planetesimal--planetesimal interactions (pp1 and pp0, Sect.~\ref{sect:N-body_numerics}) were applied.
Due to the extremely stochastic nature of late-stage accretion, we performed five simulations with the SPH collision model for each of these parameter combinations, and three additional ones with perfect merging (PM), for comparison. This gives a total of 32 simulations for the regular scenarios (upper part of Table~\ref{tab:scenarios}).

In addition, we also consider scenarios without planetesimals, that is, starting with a pure embryo disk, to study the principal influence of the planetesimal population on the dynamical as well as the collisional evolution. With the same disk parameters this results in 29 embryos, with masses between $\sim$\,0.1 and 0.4\,$M_\oplus$.
While for PM runs, planetesimals cannot form at all now, this is still possible with our collision model, but since usually a few planetesimals are around at most we do not expect a significant difference between pp0 and pp1. Therefore, we allow such interactions (pp1) in all embryo-only scenarios.
As for the regular runs we performed five simulations with the SPH collision model and three with PM, for both the cJS and eJS architectures, which gives a total of 16 scenarios here (lower part of Table~\ref{tab:scenarios}).
We note that the embryo-only initial conditions are similar to those in the few existing and comparable direct hybrid simulations \citep{Genda2011_Direct_coupling_N-body_SPH, Genda2017_Hybrid_code_Ejection_of_iron-bearing_fragments}, where our disk spans a wider range in semi-major axis and thus includes more embryos than in these studies.

The total disk mass in the regular runs amounts to 6.1\,$M_\oplus$, including a total of 0.1\,$M_\oplus$ of water, equal to 345 Earth-oceans\footnote{We define 1\,$O_\oplus$ as the sum of Earth's current inventory of surface and crustal water $\sim$\,$1.7\times10^{21}$\,kg $\sim$\,$2.8\times10^{-4}$\,$M_\oplus$ \citep{Lecuyer1998_hydrogen_isotope_composition_seawater_and_global_water_cycle}.
The term `Earth-ocean' is used somehow ambiguously in the literature, with varying definitions (e.g., often as $1.4\times10^{21}$\,kg), possibly accounting only for literal ocean water.}~($O_\oplus$).
Except for $\sim$\,0.1\,$O_\oplus$ inside 2\,au, all remaining water is located in the outer disk initially.
These figures are in general the same for the embryo-only scenarios but can differ somewhat due to the initial configurations' setup algorithm (see Sect.~\ref{sect:initial_conditions}), with a slightly smaller disk mass (5.9\,$M_\oplus$), and a somewhat higher initial water inventory (375\,$O_\oplus$). The latter difference is found almost only in the outer disk beyond 2\,au.
Each scenario was computed for at least 200 and at most 700~Myr -- until a system of final planets on approximately stable orbits had formed.
The runtime per simulation with regular (embryos + planetesimals) initial conditions was on the order of 2\,-\,4~months, where those with the SPH collision model took at least twice as long as with PM.
The main reason for that is not directly the extra computing time of the SPH runs (on the order of a week of combined GPU time), but the considerably slower decrease in the number of bodies in the N-body simulation (see Fig.~\ref{fig:body_numbers_over_t}).
In each of the main scenarios about 150\,-\,200 collisions were treated with SPH, in addition to approximately 50 PM collisions, mainly with the star and the inner giant planet.
For the embryo-only scenarios these figures are about 40\,-\,50 SPH-treated collisions, in addition to about 10 with PM.
Therefore, altogether we have simulated around 4000 collision scenarios with dedicated SPH simulations.

\section{Results I: System-wide evolution and water transport}
\label{sect:results_I}
	\begin{figure*}
	\centering{\includegraphics[width=0.95\hsize]{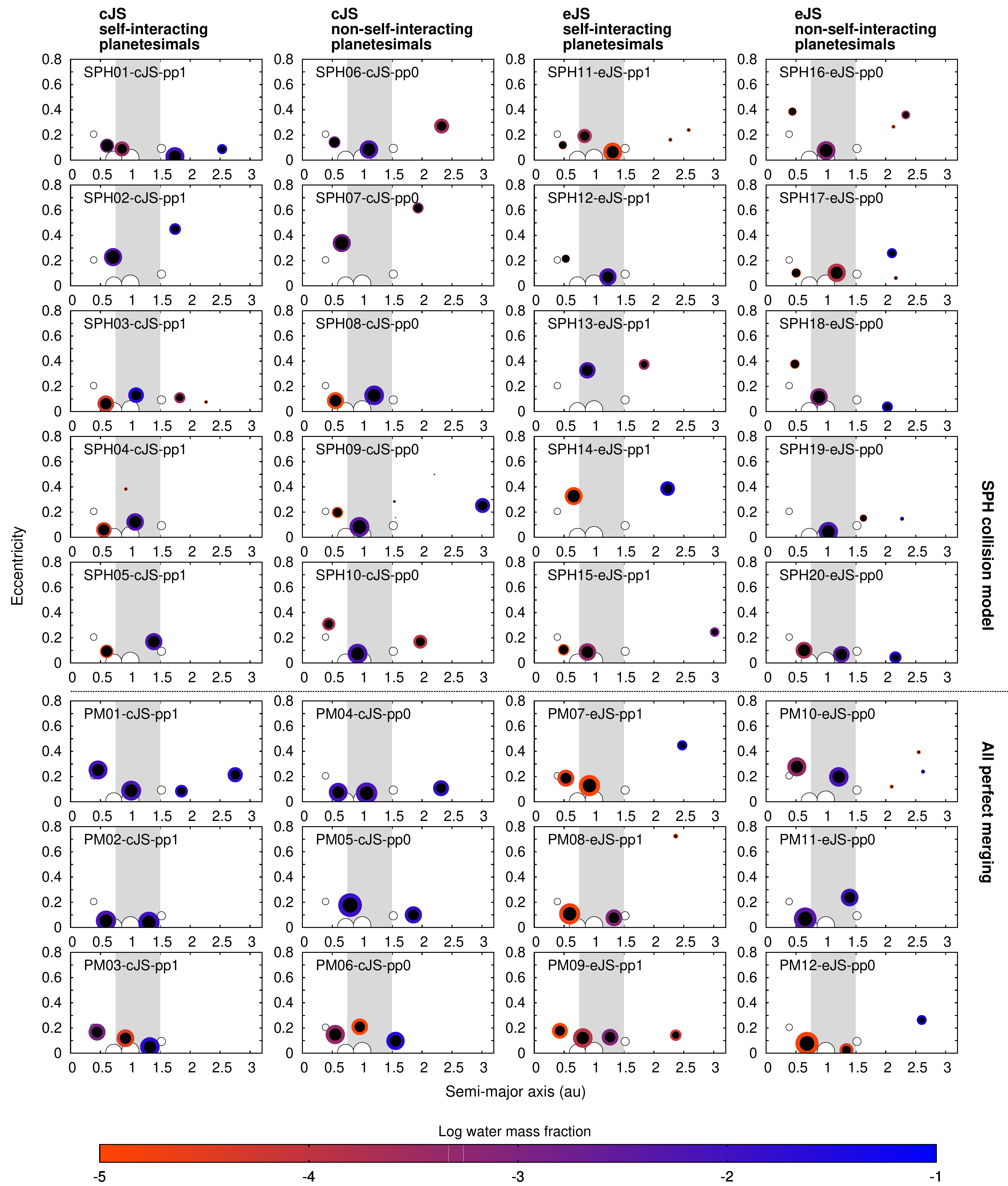}}
	\caption{Final systems for all regular scenarios (cf.\ Table~\ref{tab:scenarios}). Each column represents a particular giant planet architecture (cJS or eJS) and mode of planetesimal--planetesimal interaction (pp1 or pp0). The upper five rows are results based on the SPH collision model, and the lower three are from PM comparison runs. The sizes of planets and their iron cores (black) are proportional to mass$^{1/3}$. The region corresponding to our definition of potentially habitable planets (see Sect.~\ref{sect:potentially_habitable_planets}) is indicated as a gray bar, and the Solar System planets are plotted as open circles as a reference.}
	\label{fig:final_planets_medN}
	\end{figure*}
	
	\begin{figure}[h!]
	\centering{\includegraphics[width=8.7cm]{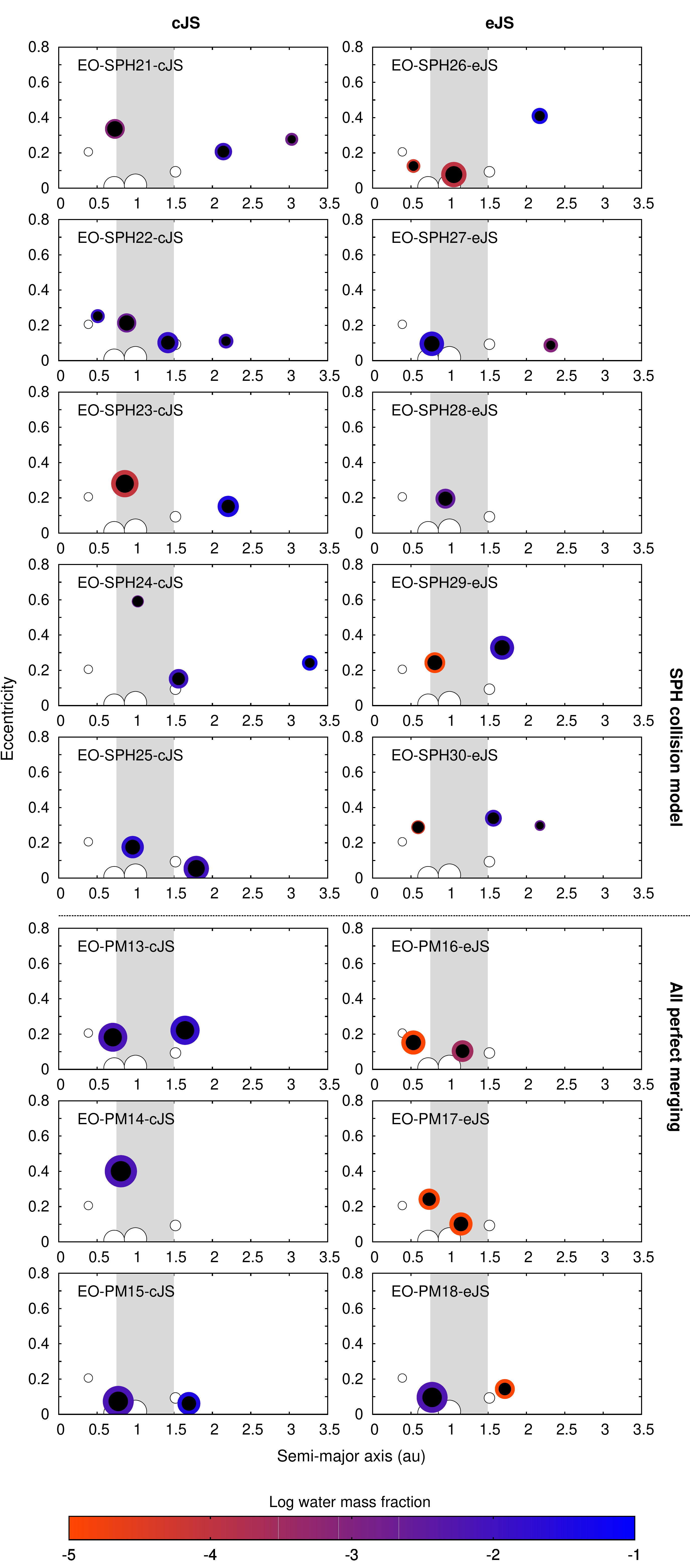}}
	\caption{Final systems for all embryo-only scenarios (cf.\ Table~\ref{tab:scenarios}). The columns represent cJS and eJS architectures, while planetesimal--planetesimal interactions are always included (pp1). The upper five rows are results based on the SPH collision model, and the lower three are from PM comparison runs. Colors, symbols, and their sizes have the same meanings as in Fig.~\ref{fig:final_planets_medN}.}
	\label{fig:final_planets_lowN}
	\end{figure}

In this first part we primarily present results from a system-wide point of view, while the accretion histories and water delivery to potentially habitable planets are presented in more detail in Sect.~\ref{sect:results_II}.
An overview of all scenarios and their proper names used throughout this paper can be found in Table~\ref{tab:scenarios}.
Since this is the first time a hybrid code with dedicated collision simulations is applied to debris disks with embryos and an additional planetesimal population, we ran scenarios including planetesimal--planetesimal interactions (pp1) and also without them (pp0) in order to compare this novel aspect.
However, for the most part, the outcomes of pp0 and pp1 are fairly similar, particularly compared to variations among other parameters (further discussed in Sect.~\ref{sect:discussion_collision_model}).
Therefore, we often combine pp0 and pp1 results in the following, unless stating otherwise.
A similar policy is applied to the embryo-only simulations, which are only referred to when relevant to highlight effects of omitting the initial planetesimal population, while we focus on the (probably more realistic) regular scenarios otherwise (but see Sect.~\ref{sect:discussion_initial_conditions} for discussion).
Results for embryo-only scenarios are included in Tables~\ref{tab:systems}, \ref{tab:pot_hab_planets}, and \ref{tab:pot_hab_planets_collisions}, and in Fig.~\ref{fig:final_planets_lowN}, but not in the other figures.

For quantifying water contents we mostly use absolute values (in easy-to-interpret units of $O_\oplus$) instead of WMFs, with the exceptions of Figs.~\ref{fig:final_planets_medN}, \ref{fig:final_planets_lowN}, and \ref{fig:N_body_snapshots} because using WMFs is common in this kind of plot in the literature.
Absolute amounts, for example of water delivered to potentially habitable planets, are useful in relation to the known initially available supply, but also because the thickness of (surface) water layers is usually better indicated by the absolute inventory\footnote{\label{footnote:wmf}This is in principle because surface area grows slower than volume, and therefore approximate mass (cube-square law), where for constant density $A(M)\propto M^{2/3}$, and even more so with increased hydrostatic compression of more massive bodies. Literally speaking this would mean ever deeper oceans for more massive planets with equal WMF for example. We note however that further factors like geological activity and increased surface gravity (and thus seafloor pressure) can become important for determining ocean depth and the general topography \citep[see, e.g.,][]{Cowan2014_Water_cycling_ocean_mantle_Super_Earths_need_not_be_waterworlds}.}.
In addition, the limitations of our collision model suggest that some neglected refractory debris is likely re-accreted at some point, which adds mass and could therefore change the WMF but not directly the absolute water mass.

To enhance clarity we apply a uniform color coding and symbols for the different parameter combinations in all plots throughout the paper.
Results from the SPH collision model are always shown as crosses or solid lines, and those from PM runs are indicated by circles or dashed lines.
cJS scenarios are color-coded in black (pp1) and dark-green (pp0), and eJS simulations are shown in red (pp1) and orange (pp0).

\subsection{Final terrestrial planet systems}
\label{sect:final_terrestrial_planet_systems}
\begin{figure}[h!]
	\centering{\includegraphics[width=8.3cm]{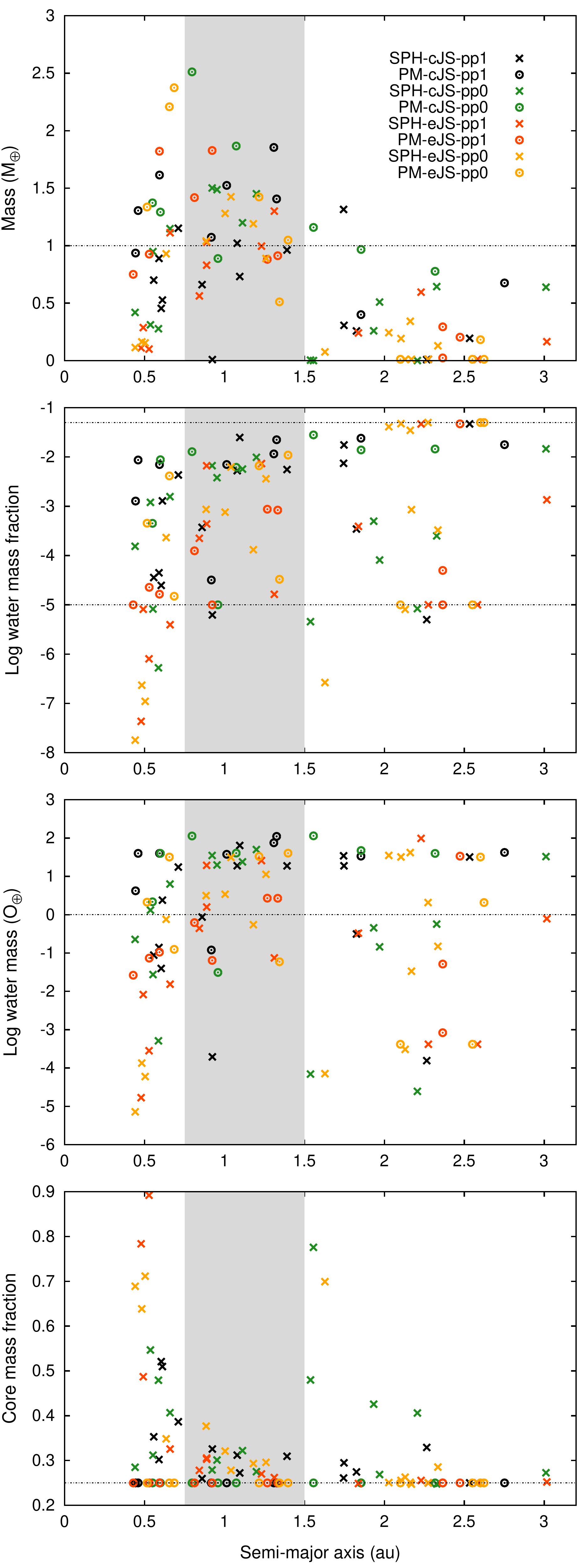}}
	\caption{Various properties of final planets in the regular scenarios, as a function of semi-major axis. Mass and core mass fraction (first and last panel) are plotted linearly, while water mass fraction and total water mass (middle panels) are log-plots. The region of potentially habitable planets (see Sect.~\ref{sect:potentially_habitable_planets}) is indicated as a gray bar.}
	\label{fig:final_planets_III}
	\end{figure}

The final planetary systems are illustrated in the a\,--\,e plane in Figs.~\ref{fig:final_planets_medN} and \ref{fig:final_planets_lowN} for the regular and embryo-only scenarios, respectively. The upper five rows show results from our SPH collision model (mostly referred to simply as `our model' in the following), with the perfect merging (PM) comparison runs below.
The main properties of formed planets as a function of semi-major axis are plotted in Fig.~\ref{fig:final_planets_III}, and basic data and statistics on all scenarios are listed in Table~\ref{tab:systems}.
The final planets' orbital parameters are always averaged values over the last few ten Myr to smooth out naturally occurring fluctuations (oscillations) caused by mutual perturbations.
One of the first things that comes to mind when looking at Figs.~\ref{fig:final_planets_medN} and \ref{fig:final_planets_lowN} is the well-known highly stochastic nature of late-stage accretion, irrespective of whether our model or simple PM is used. The sample size of 5\,+\,3 scenarios (our model + PM) per parameter set (imposed by computational limits) is naturally insufficient to derive robust statistics, but general trends, and an assessment of the degree of scatter can nevertheless be obtained.
Even though this is not our main intention, it is noteworthy that the formed systems rarely resemble the inner Solar System as a whole, but individual planets do, a well-known characteristic of the stochastic formation process.

	\begin{table*}
	\centering
	\small
%	\scriptsize
	\caption{Summarized data on all formed planet systems.}
	\label{tab:systems}
	\begin{tabular}[b]{p{2.1cm} p{1.0cm} p{1.6cm} p{1.7cm} p{1.7cm} p{1.4cm} p{1.7cm} p{1.6cm} p{1.4cm}}
	\hline
	\hline
Scenario name       &  $N_\mathrm{planets}$  &  $M_\mathrm{planets}$  &  $M_\mathrm{water}$  &   $M_\mathrm{col-star}$ &   $M_\mathrm{outward-lost}$  &   $M_\mathrm{col-losses}$  &  $M_\mathrm{water,col-losses}$  &  $t_\mathrm{last-emb-col}$ \\
	    	       &  \scriptsize{(pot-hab)} &  \scriptsize{(pot-hab)}  &  \scriptsize{(pot-hab)}  &   \scriptsize{(water)} &   \scriptsize{(water)} &   \scriptsize{(iron-mass-frac)}  &    &  ($t_\mathrm{last-col}$) \\
\hline                                                                                                                                                                                         
SPH01-cJS-pp1  &  4 (1)        &  2.70 (0.66)   &  70   (0.87)   &   0.11 (0.063)      &   2.17                  &   1.12 (0.09)              &  45              &  158 (235)             \\
SPH02-cJS-pp1  &  2 (0)        &  1.46 (0.0)    &  36   (0.0)    &   0.41 (7.3)        &   2.48                  &   1.73 (0.09)              &  104             &  311 (311)             \\
SPH03-cJS-pp1  &  3 (1)        &  1.89 (0.73)   &  64.6 (64.1)   &   0.19 (2.1)        &   2.69                  &   1.32 (0.13)              &  61              &  116 (217)             \\
SPH04-cJS-pp1  &  2 (1)        &  1.73 (1.03)   &  19   (18.9)   &   0.1  (4.2)        &   2.79                  &   1.47 (0.11)              &  48              &  258 (258)             \\
SPH05-cJS-pp1  &  2 (1)        &  1.42 (0.96)   &  18.8 (18.8)   &   0.24 (6.3)        &   2.67                  &   1.76 (0.12)              &  38              &  254 (254)             \\
               &               &  \bf{1.84 (0.68)}   &  \bf{42   (20.5)}   &   \bf{0.21 (4.0)}        &   \bf{2.56 (240)}            &   \bf{1.48}              &  \bf{59}              &  \bf{219 (255)}             \\
                                                                                                                                                                                         
SPH06-cJS-pp0  &  3 (1)        &  2.15 (1.20)   &  25.8 (23.9)   &   0.28 (2.5)        &   2.23                  &   1.42 (0.08)              &  53              &  144 (306)             \\
SPH07-cJS-pp0  &  2 (0)        &  1.41 (0.0)    &  6.8  (0.0)    &   0.35 (6.8)        &   2.71                  &   1.63 (0.07)              &  101             &  336 (336)             \\
SPH08-cJS-pp0  &  2 (1)        &  2.40 (1.45)   &  50.20(50.17)  &   0.080(0.25)       &   2.74                  &   0.86 (0.09)              &  25.2            &  84  (250)             \\
SPH09-cJS-pp0  &  3 (1)        &  2.41 (1.49)   &  53   (19.9)   &   0.15 (0.0034)     &   2.44                  &   1.09 (0.09)              &  33              &  98  (369)             \\
SPH10-cJS-pp0  &  3 (1)        &  2.44 (1.51)   &  35.6 (35.2)   &   0.30 (4.3)        &   2.66                  &   0.68 (0.08)              &  42              &  133 (217)             \\
               &               &  \bf{2.16 (1.13)}   &  \bf{34   (26)}     &   \bf{0.23 (2.8)}        &   \bf{2.6 (257)}             &   \bf{1.1}              &  \bf{51}              &  \bf{159 (296)}             \\
\hline                                                                                                                                                                                         
SPH11-eJS-pp1  &  3 (2)        &  2.00 (1.86)   &  0.52 (0.52)   &   0.67 (39)         &   2.38                  &   1.04 (0.15)              &  2.35            &  54  (350)             \\
SPH12-eJS-pp1  &  2 (1)        &  1.10 (1.00)   &  25.8 (25.8)   &   0.50 (10.3)       &   2.66                  &   1.82 (0.18)              &  7.3             &  379 (379)             \\
SPH13-eJS-pp1  &  2 (1)        &  1.07 (0.83)   &  19.7 (19.4)   &   1.14 (9.4)        &   2.42                  &   1.46 (0.11)              &  51              &  175 (175)             \\
SPH14-eJS-pp1  &  2 (0)        &  1.71 (0.0)    &  98   (0.0)    &   1.15 (9.6)        &   2.03                  &   1.20 (0.11)              &  9.2             &  183 (183)             \\
SPH15-eJS-pp1  &  3 (1)        &  1.48 (1.03)   &  2.38 (1.59)   &   0.89 (48)         &   2.41                  &   1.31 (0.09)              &  8.1             &  115 (253)             \\
               &               &  \bf{1.47 (0.94)}   &  \bf{29   (9.5)}    &   \bf{0.87 (23)}         &   \bf{2.4 (277)}             &   \bf{1.4}              &  \bf{16}              &  \bf{181 (268)}             \\
                                                                                                                                                                                         
SPH16-eJS-pp0  &  3 (1)        &  1.53 (1.28)   &  3.56 (3.41)   &   0.85 (50)         &   2.45                  &   1.25 (0.09)              &  1.8             &  174 (595)             \\
SPH17-eJS-pp0  &  3 (1)        &  1.55 (1.19)   &  33   (0.55)   &   0.73 (13.4)       &   2.38                  &   1.43 (0.13)              &  1.77            &  224 (451)             \\
SPH18-eJS-pp0  &  3 (1)        &  1.45 (1.04)   &  38.3 (3.16)   &   0.90 (50)         &   2.15                  &   1.58 (0.1)               &  6.1             &  416 (416)             \\
SPH19-eJS-pp0  &  2 (1)        &  1.52 (1.44)   &  33.3 (31.3)   &   0.80 (10.2)       &   2.50                  &   1.27 (0.16)              &  20.5            &  158 (237)             \\
SPH20-eJS-pp0  &  3 (1)        &  2.16 (0.89)   &  53.9 (11.3)   &   0.95 (39)         &   2.01                  &   0.96 (0.07)              &  31              &  89  (244)             \\
               &               &  \bf{1.64 (1.17)}   &  \bf{32   (9.9)}    &   \bf{0.85 (32)}         &   \bf{2.3 (269)}             &   \bf{1.3}              &  \bf{12}              &  \bf{214 (389)}             \\
\hline                                                                                                                                                                                                                                                                                                                                                                                  
PM01-cJS-pp1   &  4 (1)        &  3.90 (1.52)   &  154  (37.7)   &   0.15 (10.4)       &   2.03                  &   0                        &  0               &  91  (91)              \\
PM02-cJS-pp1   &  2 (1)        &  3.49 (1.85)   &  117  (75.3)   &   0.12 (4.2)        &   2.48                  &   0                        &  0               &  47  (198)             \\
PM03-cJS-pp1   &  3 (2)        &  3.41 (2.48)   &  115  (111)    &   0.12 (2.25)       &   2.55                  &   0                        &  0               &  104 (112)             \\
               &               &  \bf{3.6  (2.0)}    &  \bf{129  (75)}     &   \bf{0.13 (5.6)}        &   \bf{2.4 (210)}             &                            &                  &  \bf{81  (134)}             \\
                                                                                                                                                                                         
PM04-cJS-pp0   &  3 (1)        &  3.93 (1.87)   &  120  (41)     &   0.095(2.1)        &   2.06                  &   0                        &  0               &  112 (158)             \\
PM05-cJS-pp0   &  2 (1)        &  3.48 (2.51)   &  160  (113)    &   0.50 (4.3)        &   2.10                  &   0                        &  0               &  80  (80)              \\
PM06-cJS-pp0   &  3 (1)        &  3.42 (0.89)   &  117  (0.031)  &   0.083(2.1)        &   2.58                  &   0                        &  0               &  81  (144)             \\
               &               &  \bf{3.6  (1.76)}   &  \bf{132  (51)}     &   \bf{0.23 (2.8)}        &   \bf{2.2 (210)}             &                            &                  &  \bf{91  (127)}             \\
\hline                                                                                                                                                                                         
PM07-eJS-pp1   &  3 (1)        &  2.96 (1.83)   &  34   (0.064)  &   0.59 (10.7)       &   2.54                  &   0                        &  0               &  45  (55)              \\
PM08-eJS-pp1   &  2 (1)        &  2.76 (0.91)   &  2.8  (2.7)    &   0.52 (45)         &   2.81                  &   0                        &  0               &  161 (161)             \\
PM09-eJS-pp1   &  4 (2)        &  3.34 (2.30)   &  3.4  (3.32)   &   0.68 (84)         &   2.07                  &   0                        &  0               &  90  (225)             \\
               &               &  \bf{3.0  (1.68)}   &  \bf{13.4 (2.03)}   &   \bf{0.60 (47)}         &   \bf{2.5 (285)}             &                            &                  &  \bf{99  (147)}             \\
                                                                                                                                                                                         
PM10-eJS-pp0   &  2 (1)        &  2.79 (1.42)   &  37.4 (33.2)   &   0.73 (44)         &   2.56                  &   0                        &  0               &  55  (100)             \\
PM11-eJS-pp0   &  2 (1)        &  3.26 (1.05)   &  72   (40)     &   0.78 (11.3)       &   2.05                  &   0                        &  0               &  38  (117)             \\
PM12-eJS-pp0   &  3 (1)        &  3.07 (0.51)   &  32.2 (0.059)  &   0.65 (11.3)       &   2.36                  &   0                        &  0               &  170 (170)             \\
               &               &  \bf{3.0  (1.0)}    &  \bf{47   (25)}     &   \bf{0.72 (22)}         &   \bf{2.3 (276)}             &                            &                  &  \bf{88  (129)}             \\
\hline                                                                                                                                                                                                                                                                                                                                                                                 
\hline
EO-SPH21-cJS   &  3 (0)        &  1.36 (0.0)    &  22.8 (0.0)    &   0.38 (0.86)       &   2.36                  &   1.84 (0.11)              &  106             &  421 (421)             \\
EO-SPH22-cJS   &  4 (2)        &  2.01 (1.48)   &  72   (48)     &   0.090(1.13)       &   2.67                  &   1.17 (0.06)              &  52              &  294 (294)             \\
EO-SPH23-cJS   &  2 (1)        &  2.68 (1.81)   &  110  (0.55)   &   0.18 (0.018)      &   2.46                  &   0.63 (0.002)             &  1.05            &  156 (156)             \\
EO-SPH24-cJS   &  3 (1)        &  1.17 (0.16)   &  86.0 (1.17)   &   1.21 (1.04)       &   1.67                  &   1.89 (0.09)              &  42              &  764 (764)             \\
EO-SPH25-cJS   &  2 (1)        &  2.47 (1.01)   &  114  (66)     &   0.48 (0.56)       &   1.98                  &   1.02 (0.09)              &  69              &  259 (259)             \\
               &               &  \bf{1.94 (0.89)}   &  \bf{81   (23)}     &   \bf{0.47 (0.72)}       &   \bf{2.2 (239)}             &   \bf{1.3}              &  \bf{54}              &  \bf{379 (379)}             \\
\hline                                                                                                                                                                                         
EO-SPH26-eJS   &  3 (1)        &  2.08 (1.46)   &  62   (0.55)   &   0.60 (56)         &   2.43                  &   0.85 (0.08)              &  5.7             &  120 (120)             \\
EO-SPH27-eJS   &  2 (1)        &  1.59 (1.32)   &  100  (99)     &   1.21 (64)         &   2.39                  &   0.76 (0.07)              &  9.4             &  630 (630)             \\
EO-SPH28-eJS   &  1 (1)        &  0.71 (0.71)   &  7.3  (7.3)    &   0.67 (53)         &   2.82                  &   1.74 (0.11)              &  47              &  290 (290)             \\
EO-SPH29-eJS   &  2 (1)        &  2.03 (0.80)   &  53.0 (0.014)  &   0.86 (0.87)       &   2.34                  &   0.72 (0.02)              &  3.2             &  92  (92)              \\
EO-SPH30-eJS   &  3 (0)        &  0.78 (0.0)    &  18.3 (0.0)    &   0.74 (0.028)      &   2.79                  &   1.64 (0.13)              &  34              &  731 (731)             \\
               &               &  \bf{1.44 (0.86)}   &  \bf{48   (21)}     &   \bf{0.82 (35)}         &   \bf{2.6 (272)}             &   \bf{1.1}              &  \bf{20}              &  \bf{373 (373)}             \\
\hline                                                                                                                                                                                                                                                                                                                                                                                                                                                                                                                                                                           
EO-PM13-cJS    &  2 (0)        &  4.39 (0.0)    &  170  (0.0)    &   0.0  (0.0)        &   1.56                  &   0                        &  0               &  177 (177)             \\
EO-PM14-cJS    &  1 (1)        &  3.02 (3.02)   &  70.1 (70.1)   &   0.17 (0.006)      &   2.76                  &   0                        &  0               &  201 (201)             \\
EO-PM15-cJS    &  2 (1)        &  3.74 (2.65)   &  193  (65)     &   0.0  (0.0)        &   2.21                  &   0                        &  0               &  62  (62)              \\
               &               &  \bf{3.7  (1.9)}    &  \bf{144  (45)}     &   \bf{0.06 (0.002)}      &   \bf{2.2 (231)}             &                            &                  &  \bf{147 (147)}             \\
\hline                                                                                                                                                                                         
EO-PM16-eJS    &  2 (1)        &  2.19 (0.92)   &  1.00 (0.96)   &   0.76 (1.01)       &   3.00                  &   0                        &  0               &  84  (84)              \\
EO-PM17-eJS    &  2 (1)        &  1.98 (1.11)   &  0.069(0.039)  &   1.11 (1.02)       &   2.86                  &   0                        &  0               &  16  (16)              \\
EO-PM18-eJS    &  2 (1)        &  3.37 (2.67)   &  54.63(54.61)  &   0.76 (0.027)      &   1.82                  &   0                        &  0               &  65  (65)              \\
               &               &  \bf{2.5  (1.6)}    &  \bf{18.6 (18.5)}   &   \bf{0.88 (0.69)}       &   \bf{2.6 (356)}             &                            &                  &  \bf{55  (55)}              \\
	
	\hline
	\end{tabular}
	\tablefoot{All masses are given in $M_\oplus$, water masses always in $O_\oplus$, and time in Myr. Mean values of the respective subset are bold. `pot-hab' refers to the respective quantity only within the potentially habitable zone (0.75\,-\,1.5\,au), and `water' to the respective quantity of water (in $O_\oplus$). `outward-lost' is the sum of material that either collided with a giant planet or was ejected or removed from the system. `col-losses' refers to material not incorporated into either of the SPH post-collision bodies. $t_\mathrm{last-emb-col}$ and $t_\mathrm{last-col}$ are times of the last collision only among embryos and protoplanets and the last collision in general, also including planetesimals (but excluding the star and giant planets).}
	\end{table*}

Between one and four planets are formed in all scenarios, where between zero and two of them (one in the majority of runs) follow our definition of potentially habitable, that is they end up between 0.75 and 1.5\,au (see Sect.~\ref{sect:results_II}).
We count leftover embryos as planets, but not leftover planetesimals.
Most systems end up with either two or three planets -- and only rarely four -- which is somewhat lower than what was often found in previous studies where roughly three to four planets were formed.
A principal factor could be our relatively massive debris disk, with $\Sigma_0 =$ 10~g/cm$^2$ and extending from 0.5\,-\,4\,au (otherwise often $\Sigma_0 =$ 8~g/cm$^2$ and/or a cutoff at 2\,au), where it has been shown that lighter disks typically produce more planets \citep{Kokubo2006_Formation_of_terrestrial_planets_from_protoplanets_I, Izidoro2013_Compound_model_origin_water}.
Besides that, a better-resolved population of small bodies (planetesimals) can provide more efficient eccentricity damping by dynamical friction \citep{OBrien2006_Terrestrial_planet_formation_with_strong_dynamical_friction, Raymond2006_High_res_sims_planet_formation_I, Chambers2013_Late-stage_accretion_hit-and-run_fragmentation}, generally leading to a higher number of planets \citep{Chambers2001_Making_more_terrestrial_planets, Quintana2016_Frequency_giant_impacts_Earth-like_worlds}.
With our collision model a relatively large amount of small-scale collisional debris is often neglected, which would otherwise also enhance dynamical friction and lower eccentricities.
In contrast to that are results from \citet{Chambers2013_Late-stage_accretion_hit-and-run_fragmentation} and \citet{Quintana2016_Frequency_giant_impacts_Earth-like_worlds}, who use the semi-analytical collision outcome model by \citet{Leinhardt2012_Collisions_between_gravity-dominated_bodies_I, Leinhardt2015_Numerically_predicted_signatures_of_TPF}, which includes collisional debris as additional bodies (once sufficiently massive) that can contribute to dynamical friction.
With this model these latter authors obtain a considerably higher number of planets (up to 5) than with PM, unlike in our results.
A potential further influence in simulations with our model might be the substantially lengthened accretion times (see Sect.~\ref{sect:length_timing_of_accretion_phase}), which could lead to more ejections and removals of bodies due to longer chaotic interaction.
From the current data it is not clear which of these mechanisms is mainly responsible for the discrepancy in the number of formed planets and which results are more realistic.

The masses of the planets formed with our model are significantly lower than with PM (Table~\ref{tab:systems}), which is also clearly visible in the first panel of Fig.~\ref{fig:final_planets_III}. The latter plot also illustrates the strong decline of mass beyond $\sim$\,1.5\,au (presumably due to the disturbing influence of the giant planets), consistent with earlier studies \citep[e.g.,][]{Chambers2001_Making_more_terrestrial_planets, Chambers2013_Late-stage_accretion_hit-and-run_fragmentation, Fischer2014_Dynamics_terrestrial_planets_large_number_N-body_sims, Genda2017_Hybrid_code_Ejection_of_iron-bearing_fragments}, and the high number of planets formed around 0.5\,au, the inner edge of the initial disk.
These planets end up much less massive with our model than with simple PM, which is also clearly visible in Fig.~\ref{fig:final_planets_medN} and the first panel of Fig.~\ref{fig:masses_in_bins_over_t}, and often have highly elevated core mass fractions (CMF), as evident from the fourth panel of Fig.~\ref{fig:final_planets_III}.
This has to be viewed in light of our numerical model however, where small-scale debris and its potential re-accretion is not further tracked.
This interesting outcome, possibly a robust way to produce Mercury-like planets, is further discussed in Sect.~\ref{sect:discussion_system-wide_collisional_evolution_water_transport}.
Most other planets show only a moderate increase in CMF to around 0.3, consistent with earlier studies including fragmentation but complete mass conservation \citep[e.g.,][]{Chambers2013_Late-stage_accretion_hit-and-run_fragmentation}.

The mass and composition of the final systems generally depend on the initial inventory, which is reduced by material lost to the star or to the giant planets, removed or ejected from the system, or neglected as debris after SPH simulations.
With our model, from the initial $\sim$\,6\,$M_\oplus$, approximately 2\,$M_\oplus$ \changeslan{are accreted into} planets with cJS versus $\sim$\,1.5\,$M_\oplus$ with eJS.
In the PM comparison runs these figures are $\sim$\,3.5\,$M_\oplus$ for cJS versus $\sim$\,3\,$M_\oplus$ for eJS (Table~\ref{tab:systems}).
The decisive influence of the giant planet architecture has been studied comprehensively, mostly with the classical static orbits \citep[e.g.,][]{Raymond2004_Dynamical_sims_water_delivery, Raymond2006_High_res_sims_planet_formation_I, Raymond2007_High_res_sims_planet_formation_II, OBrien2006_Terrestrial_planet_formation_with_strong_dynamical_friction, Morishima2010_N-body_sims_including_gas_and_giant_planets, Izidoro2013_Compound_model_origin_water, Fischer2014_Dynamics_terrestrial_planets_large_number_N-body_sims, Quintana2014_Effect_of_planets_beyond_ice_line_on_volatile_accretion}, but also for migrating gas giants as in Grand-Tack scenarios \citep[e.g.,][]{OBrien2014_Water_delivery_and_giant_impacts_in_Grand_Tack}.
Our results confirm the detrimental influence of eccentric giant planets on the total surviving mass, also with our model.
On the other hand, the large variations in final water content found in earlier studies can be reproduced with PM, but are considerably relaxed with our model.
In the PM runs, the amount of surviving water (system-wide) is at most a few tens of $O_\oplus$ for eJS, but $\sim$\,100\,-\,150\,$O_\oplus$ with cJS, while with our model this figure is surprisingly similar for eJS and cJS with average values of $\sim$\,30\,-\,40\,$O_\oplus$ (cf.\ Table~\ref{tab:systems} and Fig.~\ref{fig:total_mass_and_water_over_t}), which is further discussed in Sect.~\ref{sect:water_inventories_of_terrestrial_planet_systems}.

\subsection{Radial mixing and fate of available building material}
\label{sect:fate_of_building_material}
	\begin{figure}
	\centering{\includegraphics[width=8.8cm]{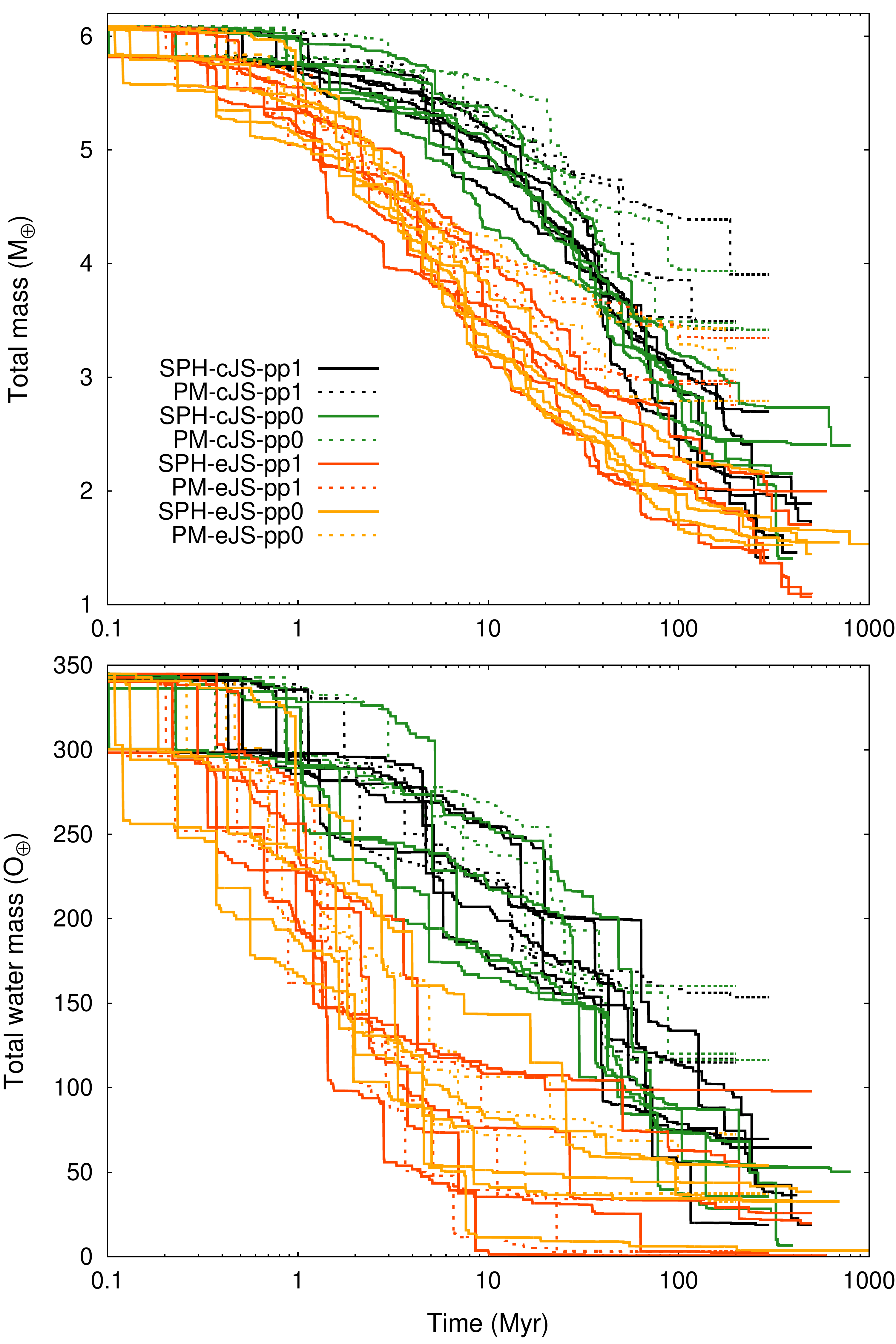}}
	\caption{Time evolution of the total mass (upper panel) and total water mass (lower panel) for all regular scenarios.}
	\label{fig:total_mass_and_water_over_t}
	\end{figure}
\begin{figure*}
	\centering{\includegraphics[width=0.98\hsize]{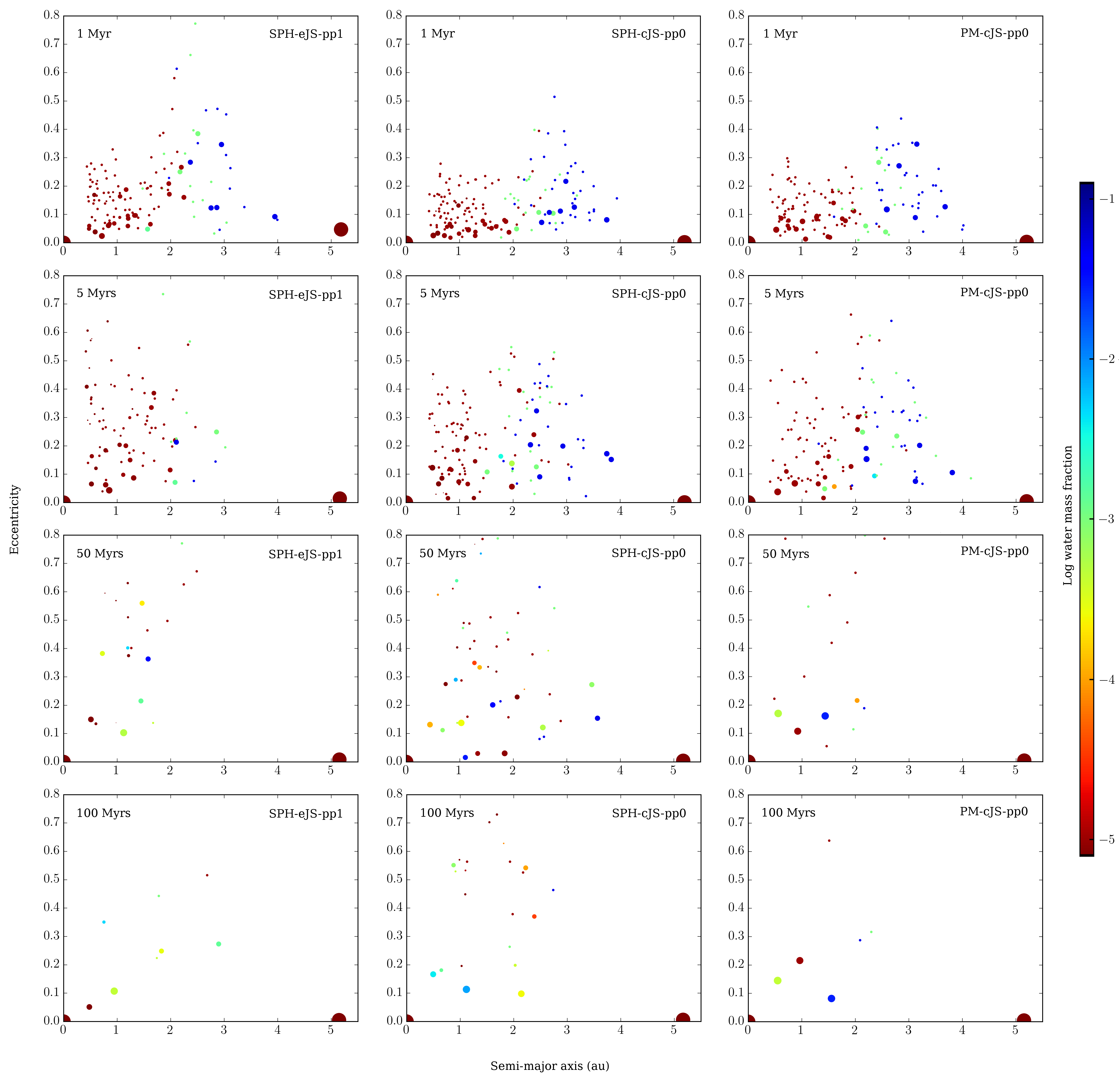}}
%	\centering{\includegraphics[width=0.66\columnwidth]{./graphics/N_body_snapshots/run63b_1015.png}}
%	\centering{\includegraphics[width=0.66\columnwidth]{./graphics/N_body_snapshots/run100b_1017.png}}
%	\centering{\includegraphics[width=0.66\columnwidth]{./graphics/N_body_snapshots/run84b_1012.png}} \\
%	\vspace{1.5mm}
%	\centering{\includegraphics[width=0.66\columnwidth]{./graphics/N_body_snapshots/run63b_1415.png}}
%	\centering{\includegraphics[width=0.66\columnwidth]{./graphics/N_body_snapshots/run100b_1421.png}}
%	\centering{\includegraphics[width=0.66\columnwidth]{./graphics/N_body_snapshots/run84b_1412.png}} \\
%	\vspace{1.5mm}
%	\centering{\includegraphics[width=0.66\columnwidth]{./graphics/N_body_snapshots/run63b_2327.png}}
%	\centering{\includegraphics[width=0.66\columnwidth]{./graphics/N_body_snapshots/run100b_2335.png}}
%	\centering{\includegraphics[width=0.66\columnwidth]{./graphics/N_body_snapshots/run84b_2318.png}} \\
%	\vspace{1.5mm}
%	\centering{\includegraphics[width=0.66\columnwidth]{./graphics/N_body_snapshots/run63b_2836.png}}
%	\centering{\includegraphics[width=0.66\columnwidth]{./graphics/N_body_snapshots/run100b_2836.png}}
%	\centering{\includegraphics[width=0.66\columnwidth]{./graphics/N_body_snapshots/run84b_2830.png}}
	\caption{Simulation snapshots in the a\,--\,e plane for three different runs (columns). Typical evolution with eJS in the left column (SPH15-eJS-pp1) vs.\ cJS in the middle (SPH06-cJS-pp0) vs.\ cJS with PM on the right (PM06-cJS-pp0). Symbol sizes are proportional to mass$^{1/3}$. Each row represents an equal time in all scenarios: Up to 1\,Myr the evolution appears similar, except some more dynamical excitation with eJS (left). Within only $\sim$\,5\,Myr however, the great majority of water-rich material is removed with eJS, but remains in the system with cJS and starts to diffuse inwards. At 50\,Myr, accretion is still actively ongoing with the SPH collision model (even though less so with eJS in this particular scenario), while with PM (right) it is almost over, except for some remaining planetesimals. We note that the cJS scenarios (middle and right columns) are pp0 runs, where some more leftover planetesimals are common. After $\sim$\,100\,Myr, body numbers are declining in the SPH scenarios (left and middle columns), but accretion and associated collisional evolution is still not over. We also note the damping effect on the eccentricity of the giant planet with eJS, consistent with similar studies \citep[e.g.,][]{OBrien2006_Terrestrial_planet_formation_with_strong_dynamical_friction}. These scenarios represent typical behavior and were not specifically selected, except that the left and middle runs are those where the potentially habitable planets in Figs.~\ref{fig:individual_planet_run63b_656} and \ref{fig:individual_planet_run100b_675} form.}
	\label{fig:N_body_snapshots}
	\end{figure*}
\begin{figure}
	\centering{\includegraphics[width=8.5cm]{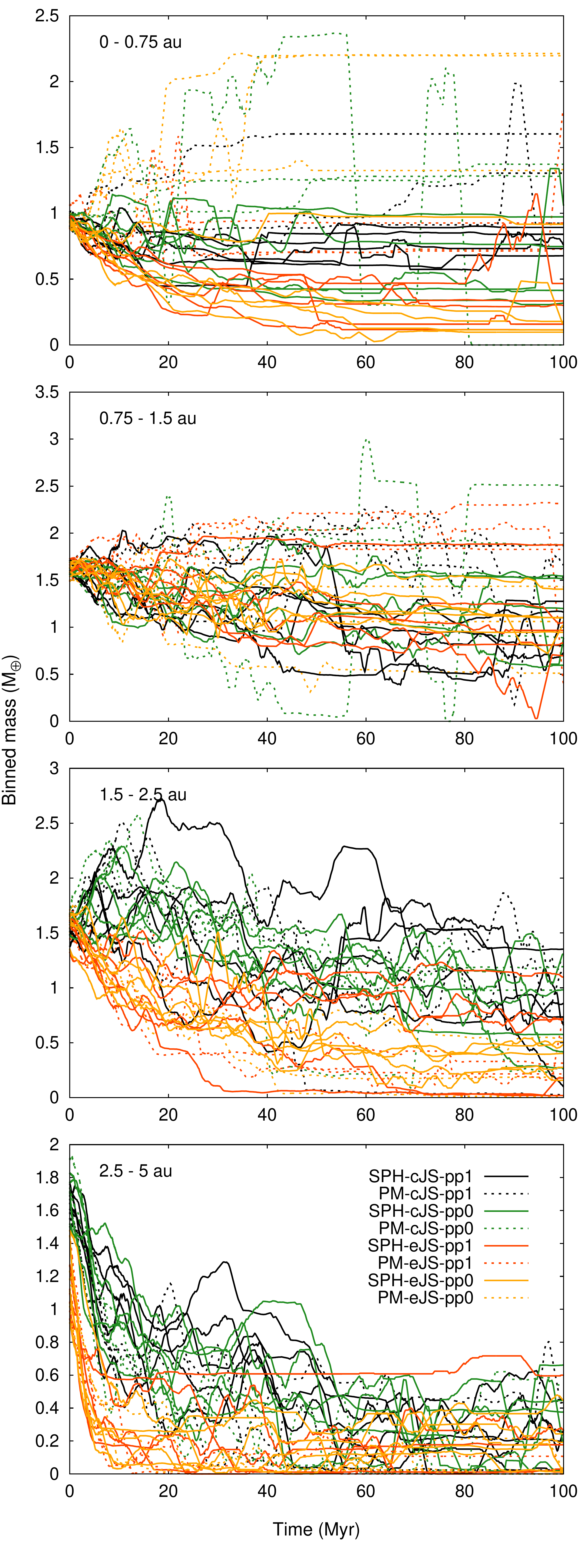}}
	\caption{Time evolution of the mass in four semi-major axis bins for all regular scenarios. The curves have been time-averaged (smoothed) over $\sim$\,3 Myr.}
	\label{fig:masses_in_bins_over_t}
	\end{figure}
\begin{figure}
	\centering{\includegraphics[width=8.5cm]{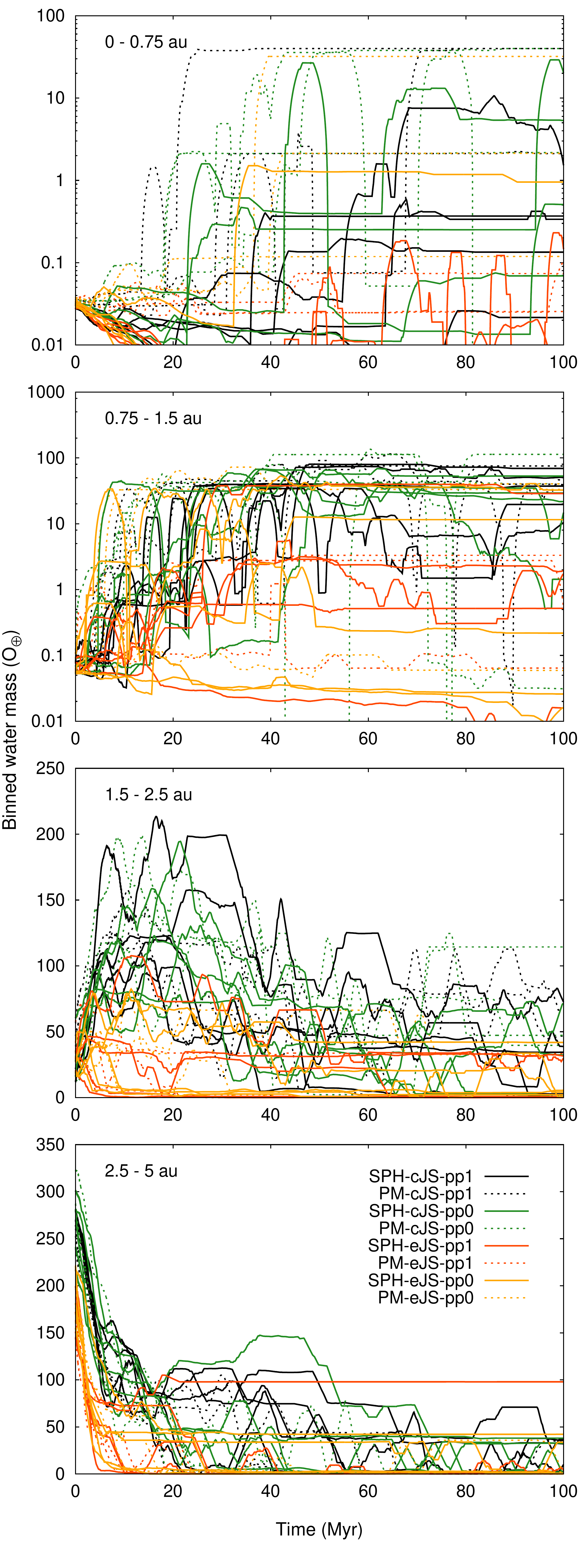}}
	\caption{Time evolution of the water mass (in $O_\oplus$) in four semi-major axis bins for all regular scenarios. We note the log y-axis in the upper two panels. The curves have been time-averaged (smoothed) over $\sim$\,3 Myr.}
	\label{fig:water_masses_in_bins_over_t}
	\end{figure}

The time evolution of the available building material ($\sim$\,6.1\,$M_\oplus$) and total water inventory ($\sim$\,345\,$O_\oplus$) is plotted in Fig.~\ref{fig:total_mass_and_water_over_t}. The total mass as well as water inventory decrease much more rapidly for eJS (red\,+\,orange) than for cJS (black\,+\,dark-green) for several Myr, which is also clearly visible in Fig.~\ref{fig:N_body_snapshots}. However, with our model especially the water contents converge to similar values eventually, further elaborated in Sect.~\ref{sect:water_inventories_of_terrestrial_planet_systems}.
To also track radial mixing, the time evolution of the mass in four semi-major axis bins is plotted in Fig.~\ref{fig:masses_in_bins_over_t}, and correspondingly for water in Fig.~\ref{fig:water_masses_in_bins_over_t}.
The second panels in these plots outline the region where potentially habitable planets might form.
Differences between cJS and eJS at the very beginning are clearly visible in the fourth panels, where strong dynamical forcing of the giant planets with eJS (red\,+\,orange) immediately ($\sim$\,5\,-\,10 Myr) removes the majority of the material from that region (cf.\ Fig.~\ref{fig:N_body_snapshots}).
The disturbing influence of the giant planets in our scenarios is mainly due to the 3:1 and 5:2 mean-motion resonances and the $\nu_6$ with the Saturn-like planet with eJS, which force bodies in the outer disk onto high-eccentricity orbits and often drive them into the star, a giant planet, or lead to ejection.
In the cJS simulations (black\,+\,dark-green) dynamical forcing acts as well, but is less pronounced and sets in more slowly, therefore considerably more water typically survives beyond 2.5\,au and slowly diffuses inwards.
With eJS, a general decline of material also between 1.5 and 2.5\,au can be observed within the first 10\,-\,20\,Myr, while with cJS it remains rather constant or even increases initially.
Radial mixing of water happens in all scenarios, and takes between a few and $\sim$\,10\,Myr to reach the 1.5\,-\,2.5\,au region, and on the order of 10\,-\,30\,Myr to arrive between 0.75 and 1.5\,au.
In earlier studies, similar timescales of around 10\,Myr for the onset of water accretion onto inner planets have been found \citep{Raymond2004_Dynamical_sims_water_delivery, Raymond2006_High_res_sims_planet_formation_I, Fischer2014_Dynamics_terrestrial_planets_large_number_N-body_sims}.
The later evolution of both mass and water content is generally marked by a slow and steady decline caused by further losses of bodies from the system, or collisional losses in SPH simulations for scenarios with our model.

The amount of material lost outwards, either by ejection or removal, or collision with a giant planet, is remarkably constant throughout all scenarios between 2.2 and 2.6\,$M_\oplus$ (Table~\ref{tab:systems}), which is more than a third of all available material.
Comparable results were found by \citet{OBrien2006_Terrestrial_planet_formation_with_strong_dynamical_friction} with strong dynamical friction, and \citet{Izidoro2013_Compound_model_origin_water} reported losses of more than 50\% in a similar dynamical environment (but perhaps including collisions with the star).
%\citet{Chambers2001_Making_more_terrestrial_planets}, using a disk that extends only out to 2\,au, found total losses of $\sim$\,25\%, mostly with the star, which indicates that the outwards-losses in our simulations originate mostly in water-rich material from beyond 2\,au.
For water it is even the majority of the initial inventory that is typically lost to ejections, removals, or the giant planets, about 250\,-\,300\,$O_\oplus$ in the regular scenarios.
Those losses are only considerably lower in the PM-cJS runs with $\sim$\,200\,$O_\oplus$, presumably due to the combination of less dynamical scattering and significantly faster accretion times (see Sect.~\ref{sect:length_timing_of_accretion_phase}), which prohibit further gradual losses (compared to SPH-cJS).
Collisions with the star account for about 0.5\,-\,1\,$M_\oplus$ with eJS (PM and our model alike), but only about a third of that with cJS, consistent with earlier studies \citep[e.g.,][]{Raymond2004_Dynamical_sims_water_delivery, OBrien2006_Terrestrial_planet_formation_with_strong_dynamical_friction}.
The difference is even larger for water, with an average of several tens of $O_\oplus$ with eJS, but typically only around 1/10 of that with cJS, presumably also caused by strong eccentricity forcing of outer disk material with eJS.

A third source of material loss is due to our collision model (Sect.~\ref{sect:collision_model}), where material that is not included in the further tracked bodies (gravitationally bound aggregates) is neglected. With cumulative amounts of between $\sim$\,1 and 1.5\,$M_\oplus$ (see Table~\ref{tab:systems}) these `losses' are surprisingly high, and while a considerable fraction of the refractory component may be re-accreted at some later point if it were tracked, the majority of the water content is likely lost.
Similar values have been found in comparable hybrid simulations by \citet{Genda2017_Hybrid_code_Ejection_of_iron-bearing_fragments}.
The background and rationale for ignoring this debris is that we are foremost interested in the fate of water, where small-scale fragments are probably often devolatilized to a large degree anyway after going through a high-energy collision.
The iron mass fraction of the neglected debris is approximately constant and around $\sim$\,0.1, clearly confirming that collisional erosion strips preferentially outer layers, while core material is lost only in the most destructive events.
The total amount of water lost in SPH collisions is on average tens of $O_\oplus$ with cJS, but several times less with eJS, probably because there is simply less water included in eJS-scenario collisions due to the high early losses of water-rich material (cf.\ Fig.~\ref{fig:total_mass_and_water_over_t}).
These higher collisional losses with cJS contribute significantly to eventually reduce the system-wide water content to similar levels as in eJS runs (see Sect.~\ref{sect:water_inventories_of_terrestrial_planet_systems}).

\subsection{Water inventories of final planetary systems}
\label{sect:water_inventories_of_terrestrial_planet_systems}
The water contents of all formed planets are in principle illustrated in Figs.~\ref{fig:final_planets_medN} and \ref{fig:final_planets_lowN} via their WMF.
The absolute amount of water (in $O_\oplus$) is plotted in the third panel of Fig.~\ref{fig:final_planets_III} and is given in Table~\ref{tab:systems}.
Here we discuss global water inventories of the formed systems, while potentially habitable planets in particular are addressed in Sect.~\ref{sect:results_II}.

All scenarios begin with a total of 345\,-\,375\,$O_\oplus$ ($\sim$\,0.1\,$M_\oplus$), located almost exclusively beyond 2\,au.
In the regular scenarios and with our collision model an average of $\sim$\,30\,-\,40\,$O_\oplus$ survives the accretion phase (system-wide), similar for cJS and eJS alike (cf.\ Table~\ref{tab:systems} and second panel in Fig.~\ref{fig:total_mass_and_water_over_t}).
Considering only the region of potentially habitable planets (0.75\,-\,1.5\,au), average water contents with cJS are still only about a factor of two above eJS.
\changeslan{The lack of a larger difference} is surprising and unexpected at first sight.
Previous studies with similar dynamical environments and initial conditions, but without any dedicated collision model (hence assuming PM), have consistently found that with giant planets on circular orbits (akin to cJS), planets emerge much more water-rich than in eJS-like scenarios \citep[e.g.,][]{Raymond2004_Dynamical_sims_water_delivery, OBrien2006_Terrestrial_planet_formation_with_strong_dynamical_friction, Quintana2014_Effect_of_planets_beyond_ice_line_on_volatile_accretion, Fischer2014_Dynamics_terrestrial_planets_large_number_N-body_sims}.
Our PM runs reflect these results perfectly (e.g., Fig.~\ref{fig:final_planets_III}), while the similar water contents resulting from a realistic collisional evolution appear to be an interplay of different loss mechanisms and presumably also lengthened accretion, as elaborated below.
However, the scatter in system-wide water content is much larger for eJS, ranging between $\sim$\,0.5 and 98\,$O_\oplus$, but only between $\sim$\,7 and 70\,$O_\oplus$ with cJS.
Strong resonant forcing by the giant planets, mainly of water-rich bodies in the outer disk, acts particularly with eJS (see Sect.~\ref{sect:fate_of_building_material}).
Nevertheless, the bulk of this effect is not to deliver water-rich material to the inner disk, but to efficiently (within $\sim$\,5\,-\,10 Myr) remove it from the system.
The consequences are that only a few water-rich bodies take part in the further evolution with eJS, which is clearly visible in Fig.~\ref{fig:N_body_snapshots} (left column), where especially the individual fate of water-rich embryos with their large inventories\footnote{In the regular scenarios there are initially 7 embryos and 37 planetesimals in the outer, water-rich region (with WMFs of 5\%), where each embryo contains 32\,-\,45, and each planetesimal $\sim$\,2\,$O_\oplus$.} can be decisive. This leads to high stochasticity and thus the large scatter in final water contents with eJS, and even more so for the subset of potentially habitable planets (Fig.~\ref{fig:water_masses_in_bins_over_t} and Table~\ref{tab:systems}).
With cJS, dynamical removal of outer disk material is much weaker and leaves more water-rich bodies to diffuse inwards (Fig.~\ref{fig:N_body_snapshots}, middle and right columns), making water delivery more robust.
One manifestation of these differences between cJS and eJS are water-rich embryos that remain stranded in the outer disk (see scenarios SPH01, SPH17, SPH18, PM07 and PM12 in Fig.~\ref{fig:final_planets_medN}).
Whether an individual embryo is transported inwards and delivers water to growing planets, survives as a stranded embryo, or is removed from the system often makes a tremendous difference with eJS, but usually has much less of an effect in cJS runs.
Another example are the highly differing (average) water contents of systems formed in the PM-eJS scenarios (PM07 to PM12, see Table~\ref{tab:systems}). Closer inspection revealed that the large apparent difference between the respective pp1 and pp0 scenarios is merely due to the high stochasticity of water delivery with eJS, without an actual systematic difference. Therefore, an average over all six scenarios would be more appropriate (and less misleading).

Initial dynamical removal of large quantities of water-rich material happens in all scenarios, and is particularly strong with eJS.
In PM simulations this point seems to be the decisive split between cJS and eJS, where for cJS accretion into typically large, water-rich planets proceeds quickly and prohibits much further losses, while for eJS only so much water-rich material is left to be distributed among the forming planets (cf.\ Figs.~\ref{fig:final_planets_medN} and \ref{fig:final_planets_lowN}).
While the initial removal phase is similar to PM, realistic treatment of collisions changes the subsequent evolution drastically, where several tens of $O_\oplus$ are removed in the form of continuous collisional losses with cJS, but only a fraction of it with eJS (Table~\ref{tab:systems}), presumably simply because less water is available (after initial removal).
The considerably prolonged growth phase compared to PM (Sect.~\ref{sect:length_timing_of_accretion_phase}) can be assumed to also allow for more losses of water-rich material.
This can readily explain the increased amount of outward losses (collisions with the giant planets or ejections and removals) with SPH-cJS, which is much closer to the high values of SPH-eJS runs than to PM-cJS scenarios (Table~\ref{tab:systems}).

A general observation is that WMFs never increase above the initial maximum value of 0.05 (second panel in Fig.~\ref{fig:final_planets_III}), confirming earlier results \citep{Marcus2010_Icy_Super_Earths_max_water_content, Burger2018_Hit-and-run}. A decrease below the initial minimum of $10^{-5}$ is of course possible (with our model), but actually happens only in relatively few cases, presumably also because at least some water-rich material is frequently accreted at some point and not necessarily due to inefficient collisional water losses.
Another interesting outcome is that absolute water masses are rather unaffected by semi-major axis, except for the innermost part of the disk around 0.5\,au (third panel in Fig.~\ref{fig:final_planets_III}). This indicates efficient mixing of water also into the region of potentially habitable planets (see Sect.~\ref{sect:fate_of_building_material}).
For the WMFs of the final planets on the other hand, the situation is more complicated (second panel in Fig.~\ref{fig:final_planets_III}), where the decreasing masses of planets formed in the outer disk lead to higher WMFs, also for similar absolute inventories \citep[see also, e.g.,][]{Raymond2007_High_res_sims_planet_formation_II}.
This is where one has to be careful with notions of `water-rich', `water-poor', and so on, as discussed at the beginning of Sect.~\ref{sect:results_I}.

\subsection{Length of the accretion phase}
\label{sect:length_timing_of_accretion_phase}
Our model allows for \changeslan{0, 1, or 2} further-tracked post-collision bodies, therefore their number is still monotonically declining, \changeslan{but considerably slower than with PM}.
This suggests longer accretion times in general, and indeed, when the time of the last collision (excluding star and giant planets) is used as a measure, accretion times roughly double compared to PM (Table~\ref{tab:systems}). Interestingly, the same holds if the last collision only among embryos and protoplanets is considered, which is probably a more robust measure since it excludes the (in principle indefinite) accretionary tail and rather signifies the last giant collision.
The mean time of the last collision only among embryos and protoplanets is between 80 and 100\,Myr in PM runs, but between 150 and well over 200\,Myr with our model.
These differences are even more pronounced in the embryo-only scenarios, presumably due to the absence of damping by dynamical friction, which results in more energetic collisions with a lower likelihood of accretion.
Looking at the growth curves of individual (potentially habitable) planets (Fig.~\ref{fig:pot_hab_planets_over_t}, lower panel) confirms that planets formed with PM finish accretion considerably sooner than those assembled with our model.
Another aspect is the frequency of collisions, which is naturally declining over time and will be further discussed in Sect.~\ref{sect:timing_accretion_water_delivery_pot_hab_planets}.

% N_bodies over t medN
	\begin{figure}
	\centering{\includegraphics[width=8.8cm]{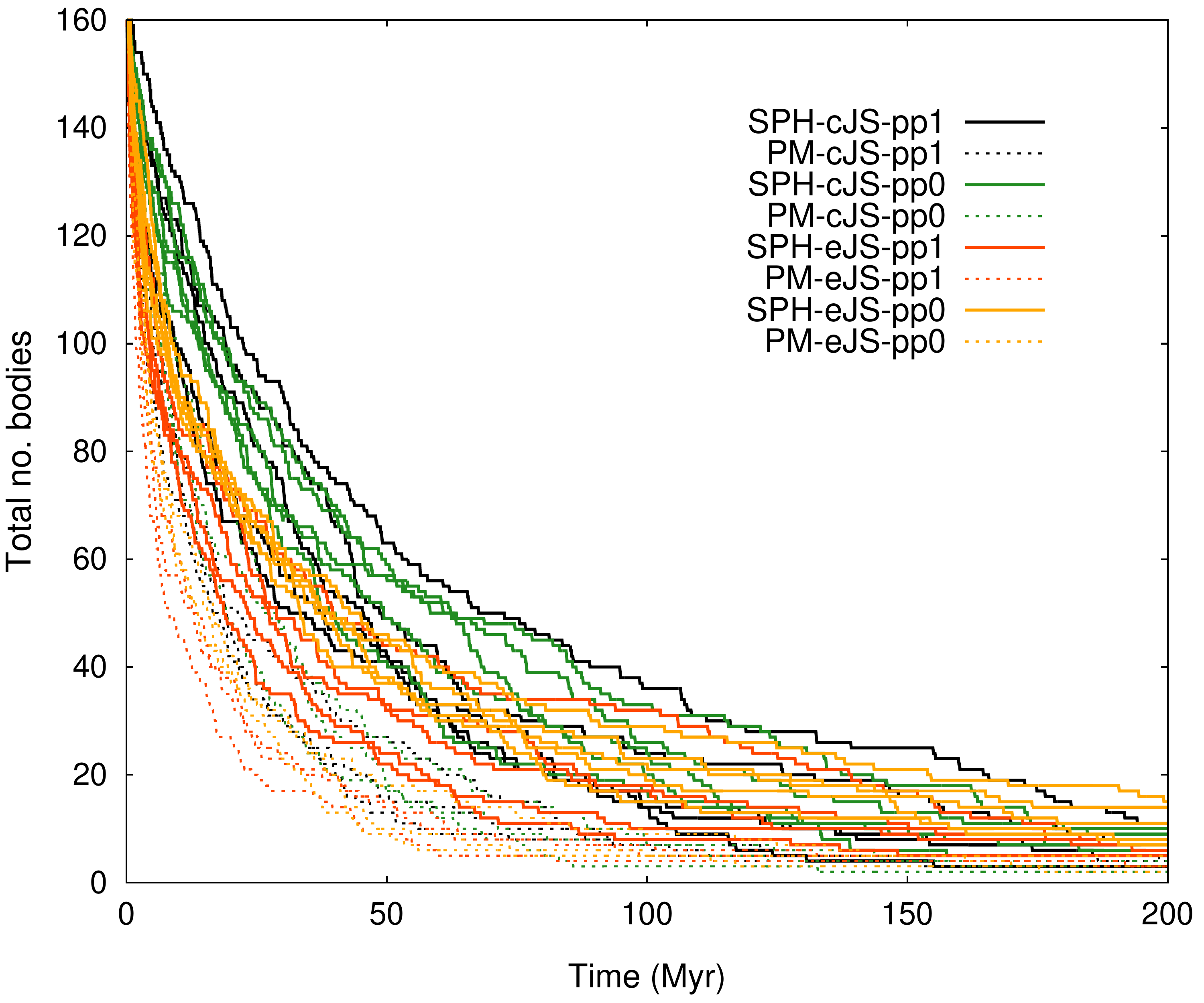}}
	\caption{Total number of bodies (embryos or protoplanets + planetesimals) over time for all regular scenarios.}
	\label{fig:body_numbers_over_t}
	\end{figure}
% N_embryos over t medN
	\begin{figure}
	\centering{\includegraphics[width=8.8cm]{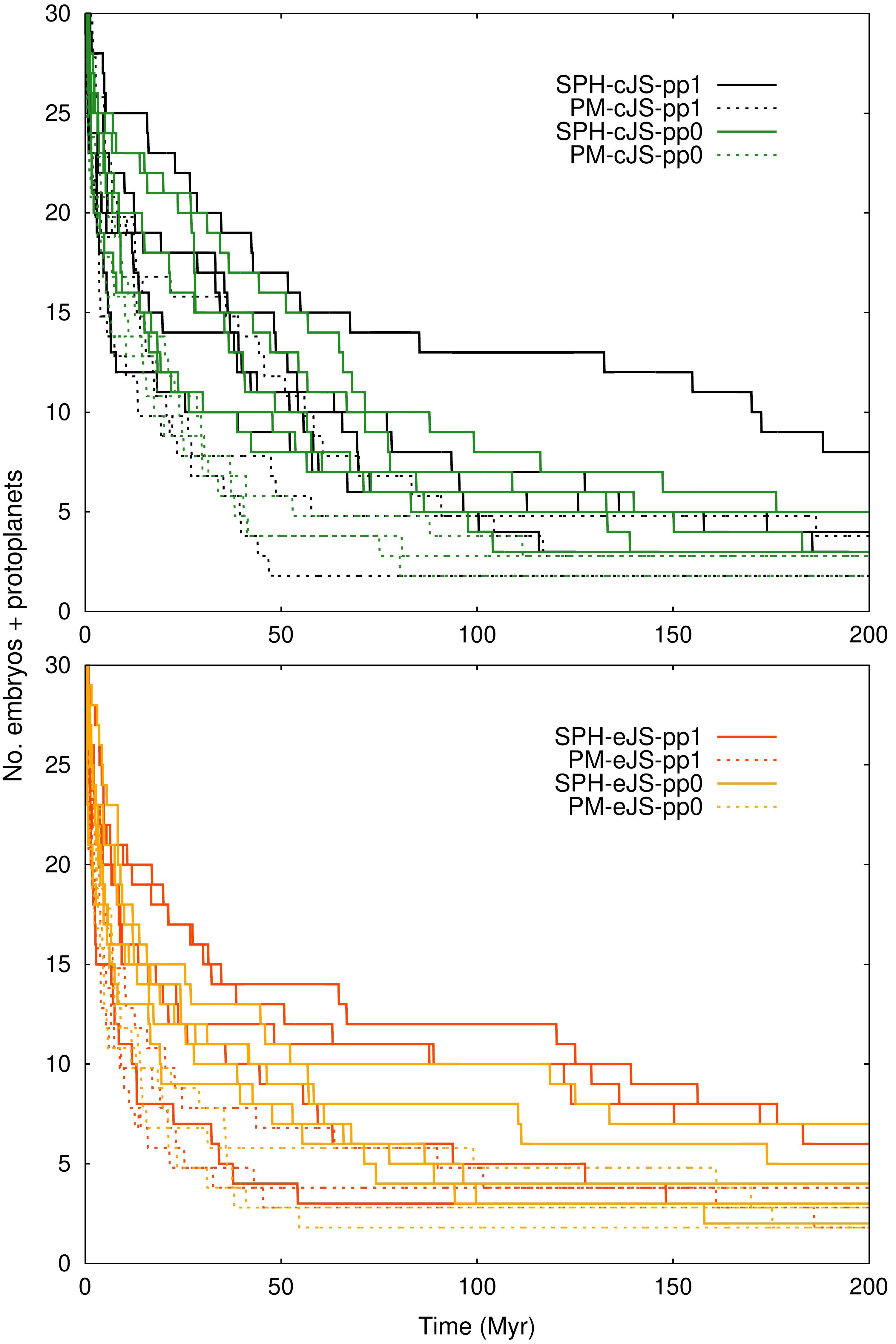}}
	\caption{Number of embryos\,+\,protoplanets (i.e., excluding planetesimals) over time for all regular scenarios, split into cJS (upper panel) and eJS (lower panel) for better visibility. The respective `PM' curves have a slight y-offset of -\,0.2 for the same reason.}
	\label{fig:embryo_numbers_over_t}
	\end{figure}
% N_embryos over t lowN
%	\begin{figure}
%	\centering{\includegraphics[width=\columnwidth]{./graphics/embryo_numbers_over_t_lowN.eps}}
%	\caption{...}
%	\label{fig:embryo_numbers_over_t_lowN}
%	\end{figure}
Figure~\ref{fig:body_numbers_over_t} shows the time evolution of the total number of bodies, where accretion times with our model seem to be clearly lengthened and body numbers are roughly twice those in PM runs over long periods of time.
Similar results have also been found in previous studies where collisional fragmentation was included \citep{Chambers2013_Late-stage_accretion_hit-and-run_fragmentation, Quintana2016_Frequency_giant_impacts_Earth-like_worlds}, based on the semi-analytical model by \citet{Leinhardt2012_Collisions_between_gravity-dominated_bodies_I, Leinhardt2015_Numerically_predicted_signatures_of_TPF}.
However, in these studies the number of embryos (protoplanets) declined approximately equally, and the authors concluded that hit-and-run and fragmentation do not actually lengthen the bulk growth of planets, but that higher body numbers are mostly due to low-mass planetesimals (collision fragments).
Our results indicate a different behavior.
Figure~\ref{fig:embryo_numbers_over_t} shows that the number of embryos\,+\,protoplanets (i.e., excluding planetesimals) is also significantly higher compared to PM for most of the time, and a similar plot (not included here) for the embryo-only scenarios shows the same behavior.
The growth times for potentially habitable planets  to reach 50\% ($t_\mathrm{50}$) and 90\% of their final mass ($t_\mathrm{90}$), summarized in Table~\ref{tab:pot_hab_planets}, also confirm that our collision model indeed results in roughly doubled accretion times compared to simple PM. \citet{Chambers2013_Late-stage_accretion_hit-and-run_fragmentation} on the other hand found similar values for $t_\mathrm{50}$ when comparing their fragmentation model to PM, possibly because of differences of the applied collision and fragmentation models. While we include two post-collision bodies at most and neglect the remaining material not gravitationally bound to them, \citet{Chambers2013_Late-stage_accretion_hit-and-run_fragmentation} and \citet{Quintana2016_Frequency_giant_impacts_Earth-like_worlds} conserve mass by adding additional fragments on (slightly) escaping trajectories if necessary.
The additional dynamical friction of those fragments might lead to more gentle subsequent collisions and therefore explain their faster decline in embryo\,+\,protoplanet numbers \citep[see also][]{OBrien2006_Terrestrial_planet_formation_with_strong_dynamical_friction}.
However, the lowest allowed planetesimal\,/\,fragment mass in their model is $\sim$\,$5\times10^{-3}$\,$M_\oplus$ (for lower masses they simply assume PM), while in our model this is $\sim$\,$2\times10^{-4}$\,$M_\oplus$, which allows fragment formation in a wide mass range where theirs already assumes PM.
The reason for their faster decline in embryo\,+\,protoplanet numbers may also be found somewhere else in the differing methodology, where a priori our actual hydrodynamic simulations for each individual collision scenario are probably more reliable than semi-analytical scaling laws, for example in predicting hit-and-run outcomes. Eventually it remains unclear which results are more accurate, and further work is required to gain a better understanding of this important issue.

\subsection{Global collision characteristics}
\label{sect:global_collision_characteristics}
The regular scenarios each include about 150\,-\,200 collisions among embryos, protoplanets, and planetesimals with our model (30\,-\,45 among embryos and protoplanets alone), and roughly 100 collisions in the PM runs (about 40\,-\,50 in the embryo-only scenarios, and only $\sim$\,15\,-\,20 in their PM versions).
These numbers are relatively robust and show only little variation on average (even when the different planetesimal--planetesimal interaction models pp1 and pp0 are compared).
A detailed list of collision number statistics specifically for the growth histories of potentially habitable planets is included in Table~\ref{tab:pot_hab_planets_collisions} and discussed in Sect.~\ref{sect:collision_characteristics_pot_hab}.

In the regular scenarios about 100 collisions (out of the 150\,-\,200) result in hit-and-run, more than half of all events.
In the embryo-only runs even more than 75\% are hit-and-run, presumably because this outcome is more likely among more similar-sized bodies, and due to the lack of dynamical friction (to reduce impact velocities). The latter is also supported by the fact that in the regular runs, where dynamical friction is at work, the hit-and-run rate also in collisions among embryos and protoplanets alone is not substantially above 50\%.
Studies in similar dynamical environments found results in broad agreement: \citet{Kokubo2010_Formation_under_realistic_accretion} reported 49\% hit-and-run (embryo-only simulations), \citet{Chambers2013_Late-stage_accretion_hit-and-run_fragmentation} found that 42\% of collisions among embryos and protoplanets are hit-and-run (they also had planetesimals), \citet{Quintana2016_Frequency_giant_impacts_Earth-like_worlds} found that 31\% of impacts that lead to Earth-like planets are hit-and-run (they also included planetesimals), and \citet{Genda2017_Hybrid_code_Ejection_of_iron-bearing_fragments} reported 52\% in their hybrid code (embryos only).
We note that the definition of hit-and-run and its demarcation to other outcome regimes is not unambiguous, and here and in the following we use the term for all collisions that resulted again in two bodies according to our model (Sect.~\ref{sect:collision_model}), independently of their mass ratio.
This can result in (post-collision) mass ratios that would no longer be deemed hit-and-run by most definitions \citep{Asphaug2010_Similar_sized_collisions_diversity_of_planets, Genda2012_Merging_criteria_for_giant_impacts}, and can therefore somewhat overstate its occurrence rate.

	\begin{figure}
	\centering{\includegraphics[width=8.8cm]{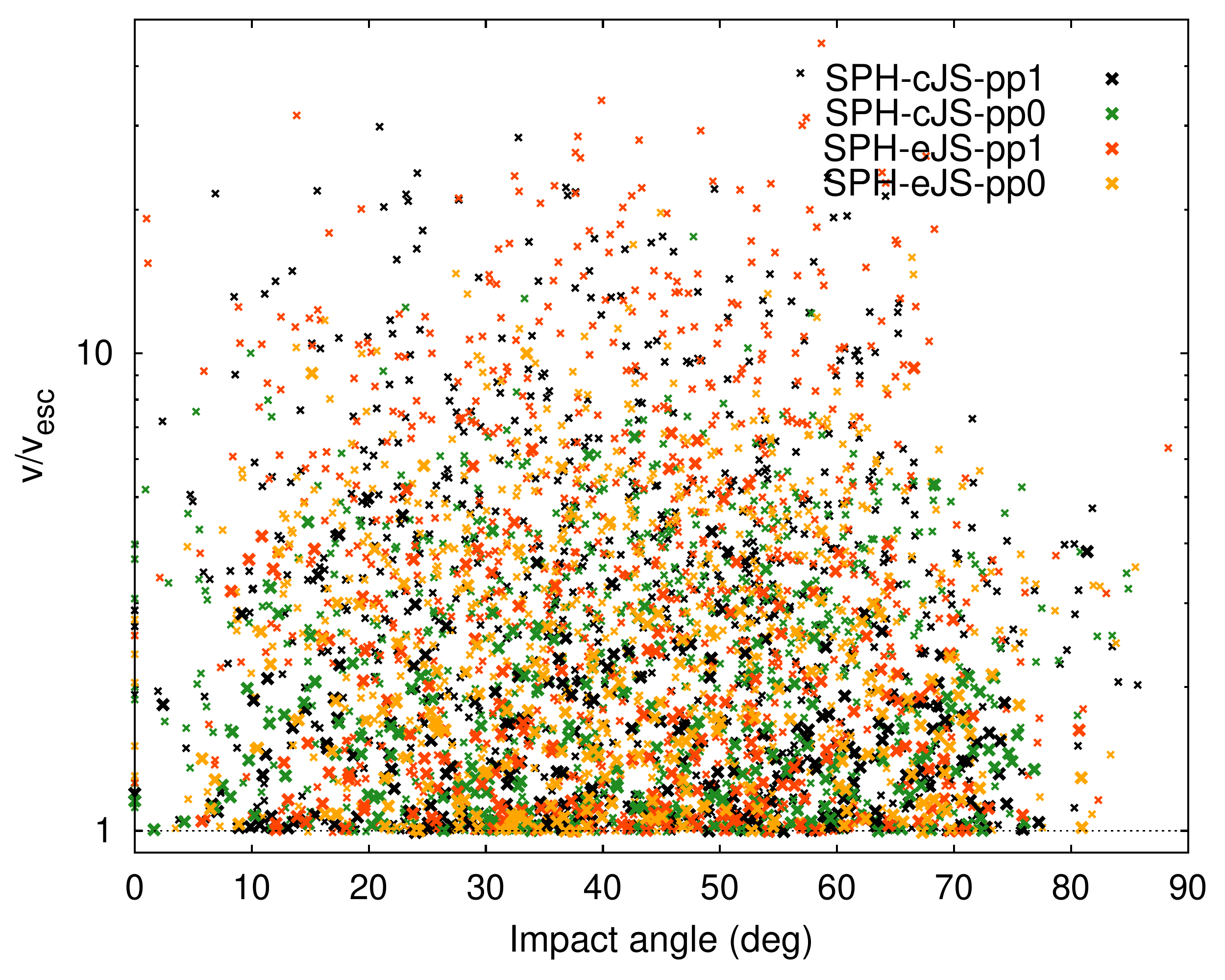}}
	\caption{Impact velocity in units of $v_\mathrm{esc}$ plotted vs.\ impact angle for all collisions in the regular scenarios with our collision model. Encounters exclusively among embryos and protoplanets (i.e., excluding planetesimals) are highlighted as larger symbols.}
	\label{fig:v_vesc_over_alpha}
	\end{figure}
Figure~\ref{fig:v_vesc_over_alpha} shows the collision phase space of impact velocity versus impact angle. Impact velocity is normalized to the mutual escape velocity $v_\mathrm{esc} = \sqrt{2GM/r}$, with the combined mass $M$ and the sum of the bodies' radii $r$. The impact angle $\alpha$ is defined as illustrated in Fig.~\ref{fig:collision_geometry_differentiated}, with $\alpha = 0^\circ$ for head-on.
As expected, impact velocities are always around or greater than $v_\mathrm{esc}$ and are rarely above $10\times v_\mathrm{esc}$. The highest values (and therefore the most destructive\,/\,erosive events) occur preferentially in the pp1 simulations, and are likely encounters between planetesimals, which are more likely on dynamically hot orbits. This is also the reason why pure embryo\,/\,protoplanet collisions (larger symbols) occur more frequently at lower velocities.
Besides that, there seem to be no significant differences between the parameter setups (colors).
Due to the limited number and fraction of planetesimals, dynamical friction is likely stronger in reality \citep[see, e.g.,][]{OBrien2006_Terrestrial_planet_formation_with_strong_dynamical_friction, Raymond2006_High_res_sims_planet_formation_I, Chambers2013_Late-stage_accretion_hit-and-run_fragmentation}, which would further damp eccentricities of embryos and protoplanets, presumably leading to somewhat gentler collisions (lower velocities).

	\begin{figure}
	\centering{\includegraphics[width=8.8cm]{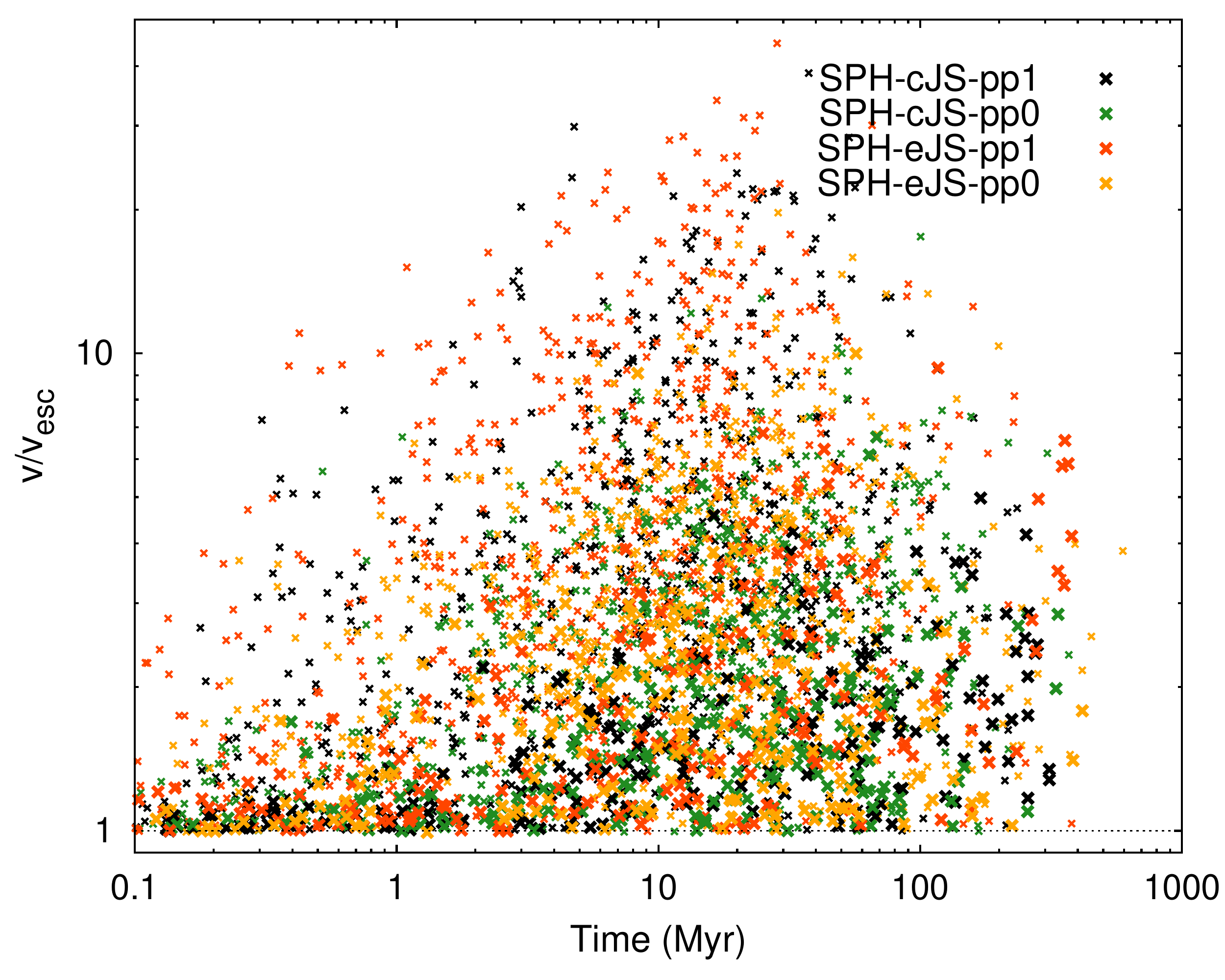}}
	\caption{Impact velocity in units of $v_\mathrm{esc}$ over time for all regular scenarios with our collision model. Larger symbols mark collisions exclusively among embryos and protoplanets.}
	\label{fig:v_vesc_over_t}
	\end{figure}
Figure~\ref{fig:v_vesc_over_t} illustrates the impact velocity as a function of time.
The initially dynamically cold systems remain calm within the first $\sim$\,1\,Myr, and collisions preferentially result in accretion.
This could be considered an artifact of the initial configuration, but also as the natural result of residual damping by the former gas disk and small body population.
After the initial phase, orbits quickly become excited within the first few Myr, and appear to reach maximum excitation by $\sim$\,10\,Myr.
The apparent decline at later times is probably rather due to the lower number of bodies, before indeed most highly excited planetesimals are eventually accreted or lost at the latest times.
The last or at least one of the latest protoplanet impacts during Earth's accretion is often linked to the formation of the Moon, where besides the classical (canonical) scenario \citep[e.g.,][]{Canup2013_Lunar-forming_impacts_High-resolution_SPH_and_AMR-CTH_simulations}, various different collision configurations are discussed, including formation of moons by hit-and-run \citep{Reufer2012_Hit-and-run_Moon_formation}.
Even though we cannot make any particular predictions, our data suggest that encounters between protoplanet-sized objects are very common in the later phases of the accretion process (Fig.~\ref{fig:v_vesc_over_t}), covering a wide range of impact velocities and angles in parameter space.

	\begin{figure}
	\centering{\includegraphics[width=8.8cm]{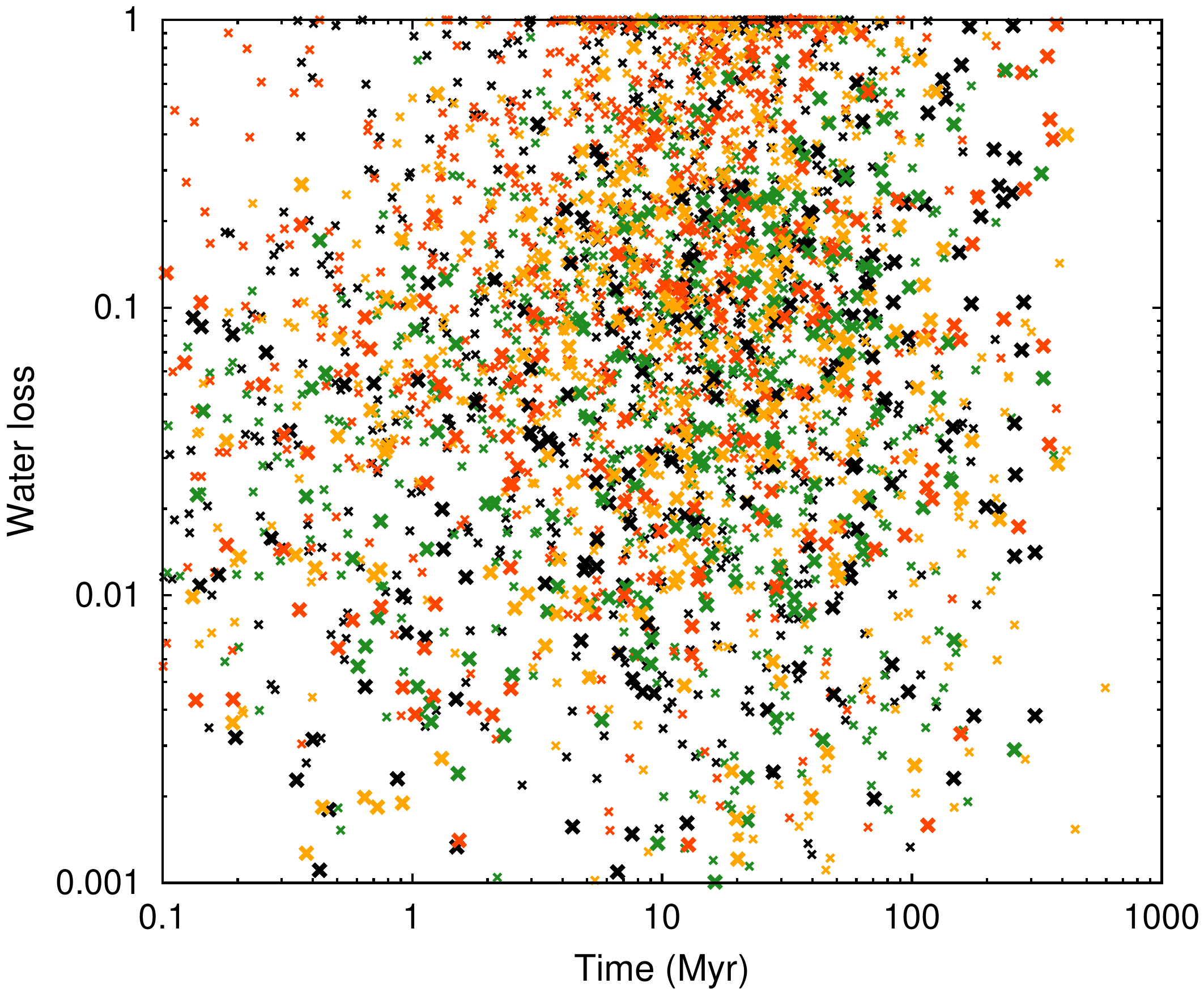}}
	\caption{Water loss in individual collisions over time for all regular scenarios with our model. Color coding is identical to Figs.~\ref{fig:v_vesc_over_alpha} and \ref{fig:v_vesc_over_t}. Water loss is defined as the fraction of the (combined) pre-collision water inventory that is not incorporated in either of the post-collision bodies. Larger symbols mark collisions exclusively among embryos and protoplanets.}
	\label{fig:total_water_loss_over_t}
	\end{figure}
Again for the same sample of collisions, Fig.~\ref{fig:total_water_loss_over_t} illustrates (combined) water losses over time. This quantity is frequently found in the literature, but does not account for more complex processes like water transfer in hit-and-run, for example from a water-rich to a dry body.
Not surprisingly, losses increase with time, with a naturally strong correlation to the increase of impact velocities in Fig.~\ref{fig:v_vesc_over_t}.
There appears to be a noticeable clustering of water losses around 1, corresponding to encounters where the colliding bodies are either completely gravitationally dispersed, or at least highly erosive events which desiccate them entirely.
The majority of collisions however, end up with combined water losses below $\sim$\,20\,-\,30\%, with a sizeable fraction even below 10\%.
Therefore, (average) losses are actually relatively low, but it is only the cumulative effect -- perhaps including low-probability chance events -- on the way to final planets that is of true significance.
Following these evolutionary tracks for potentially habitable planets is the subject of Sect.~\ref{sect:results_II}.

\section{Results II: Water transport to potentially habitable planets}
\label{sect:results_II}
	\begin{figure}
	\centering{\includegraphics[width=8.8cm]{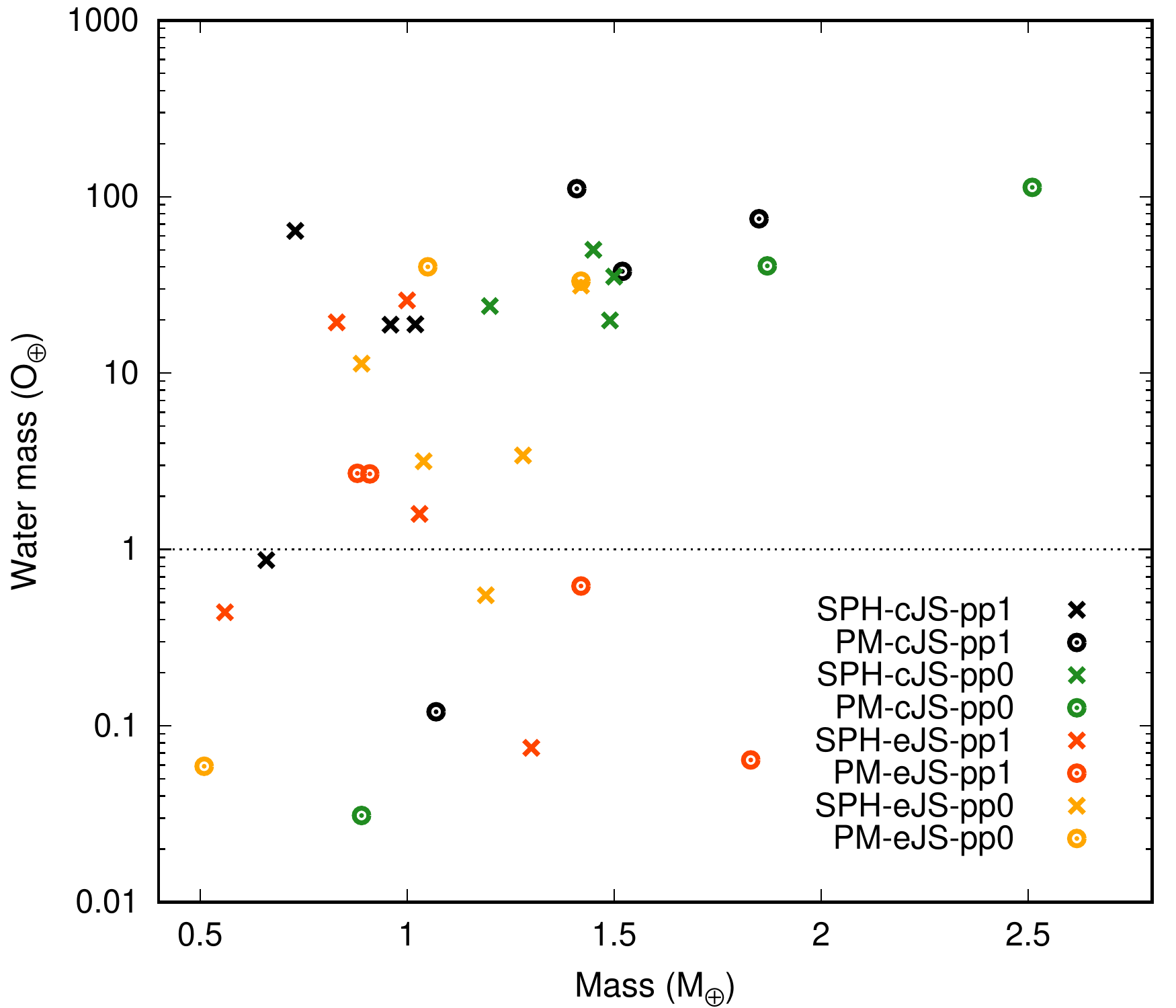}}
	\caption{All potentially habitable planets formed in the regular scenarios in the water-inventory vs.\ mass space. We note the log y-axis.}
	\label{fig:pot_hab_planets_water_over_mass}
	\end{figure}
	\begin{figure*}
	\centering{\includegraphics[width=0.93\hsize]{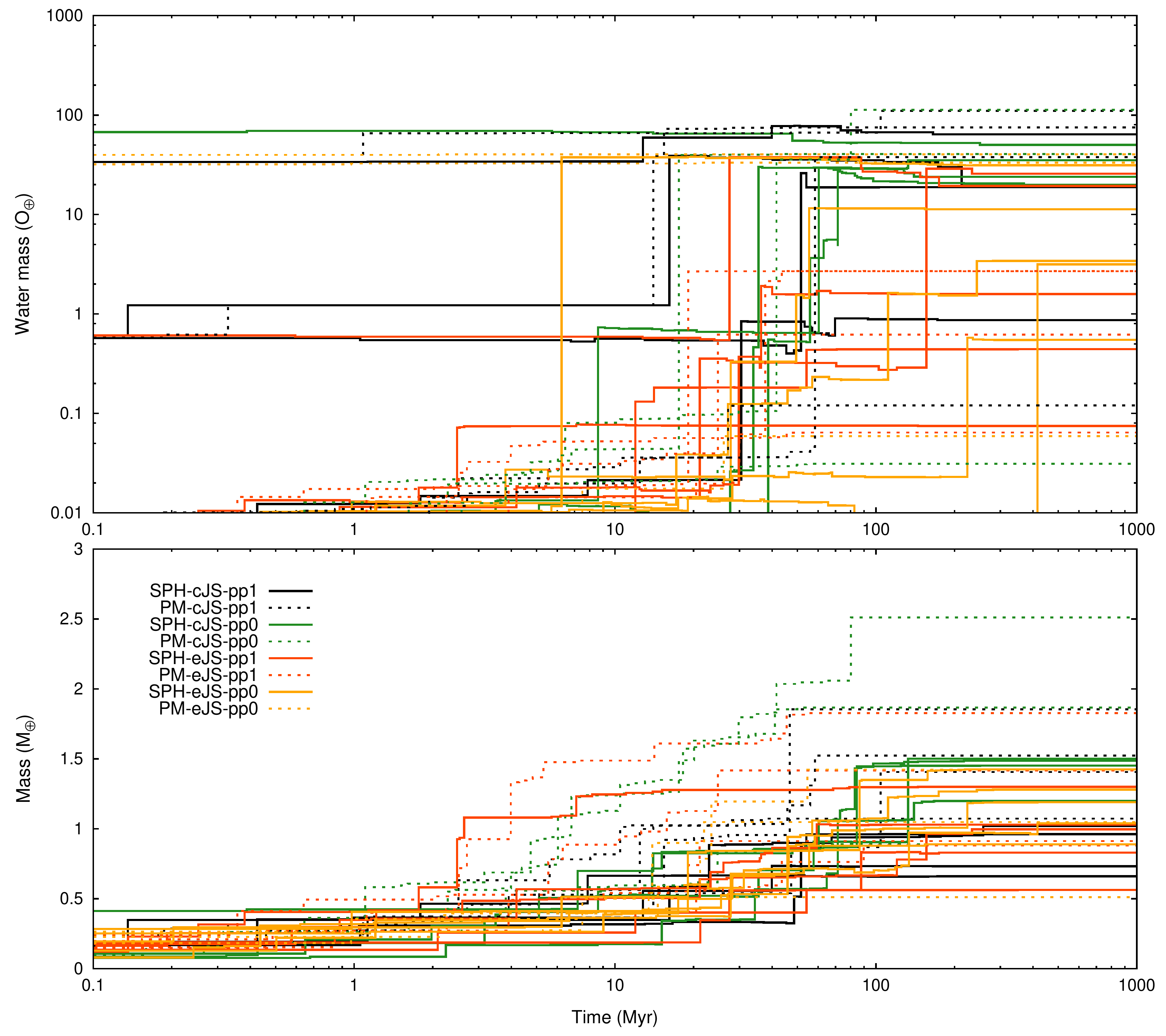}}
	\caption{Growth curves of all potentially habitable planets formed in the regular scenarios, with water inventories in the top panel (note the log y-axis) and mass in the bottom panel. The precise definition according to which evolutionary tracks are followed (especially for hit-and-run) is discussed in the text.}
	\label{fig:pot_hab_planets_over_t}
	\end{figure*}
A central aspect of our work, and one that is crucial for assessing habitability, is to constrain initial water inventories of potentially habitable planets.
We therefore analyzed (water) accretion of this subset in more detail.
Due to the inherent stochasticity of long-term dynamical simulations and the uncertainties related to the extent of habitable zones themselves \citep[see, e.g.,][]{Kasting1993_Habitable_zones, Kopparapu2013_Habitable_zones}, we use a semi-major axis range of between 0.75 and 1.5\,au to define `potentially habitable planets' for our purposes -- broadly the space between Venus and Mars in the Solar System.
This region is also highlighted in Figs.~\ref{fig:final_planets_medN}, \ref{fig:final_planets_lowN}, and \ref{fig:final_planets_III}, showing the final systems and major properties of formed planets.

\subsection{Main characteristics of potentially habitable planets}
\label{sect:potentially_habitable_planets}
In all simulations, between zero and two planets form between 0.75 and 1.5\,au, with one planet in most scenarios (Table~\ref{tab:systems}).
A complete list of these planets and their properties is provided in Table~\ref{tab:pot_hab_planets}, and specific data on collisions related to our model are additionally summarized in Table~\ref{tab:pot_hab_planets_collisions}.
The overview in Fig.~\ref{fig:pot_hab_planets_water_over_mass} reveals a distinctive twofold behavior of cJS and eJS simulations. For cJS, our model leads to significantly lower mass planets (around $\sim$\,1\,$M_\oplus$) compared to PM, akin to the system-wide trend. Average water contents are $\sim$\,30\,$O_\oplus$, compared to roughly doubled values for PM (Table~\ref{tab:pot_hab_planets}).
For eJS the situation is more complicated, and while masses are around 1\,$M_\oplus$ for both our model and PM, final water inventories are highly scattered.
The reason is again the high stochasticity of water transport with eJS, where often only a few water-rich bodies survive an initial removal phase of water-rich material due to forcing by the giant planets, as already detailed in Sect.~\ref{sect:water_inventories_of_terrestrial_planet_systems}.
Water delivery to potentially habitable planets is even more stochastic than the global inventory, because it additionally depends on inward transport of water-rich material (instead of, e.g., remaining as/on stranded embryos in the outer disk).
One manifestation of this is seen in the highly different (average) water contents of the PM-eJS-pp0 and PM-eJS-pp1 planets (Table~\ref{tab:pot_hab_planets}), where closer inspection revealed that this is simply a result of the high stochasticity (as found also system wide, see Sect.~\ref{sect:water_inventories_of_terrestrial_planet_systems}).
Therefore, it is justified and more reasonable to average over all PM-eJS planets, which leads to similar values with our model and PM of around 10\,$O_\oplus$ with eJS.
The highly stochastic water transport with eJS thus disguises the actual differences between our model and simple PM in those scenarios.
Stochastic variations aside, potentially habitable planets formed with cJS are likely more water-rich than with eJS (by a factor of two or more), while average masses are roughly equal and around 1\,$M_\oplus$.

\subsection{Accretion and water delivery to potentially habitable planets}
\label{sect:accretion_water_delivery_pot_hab_planets}
In this section we discuss the accretion tracks of potentially habitable planets. Two particularly insightful examples are analyzed in Sect.~\ref{sect:example_pot_hab_planets}.
This is unfortunately only possible for a small, selected sample, but we provide comprehensive data on accretion histories of all planets to interested researchers\footnote{\label{footnote:offer_our_data}If you are interested simply contact the lead author via e-mail at \texttt{christoph.burger@uni-tuebingen.de}.}, in order to serve as a possible basis for in-depth modeling of further aspects and physical processes.
	\begin{figure*}
	\centering{\includegraphics[width=0.245\hsize]{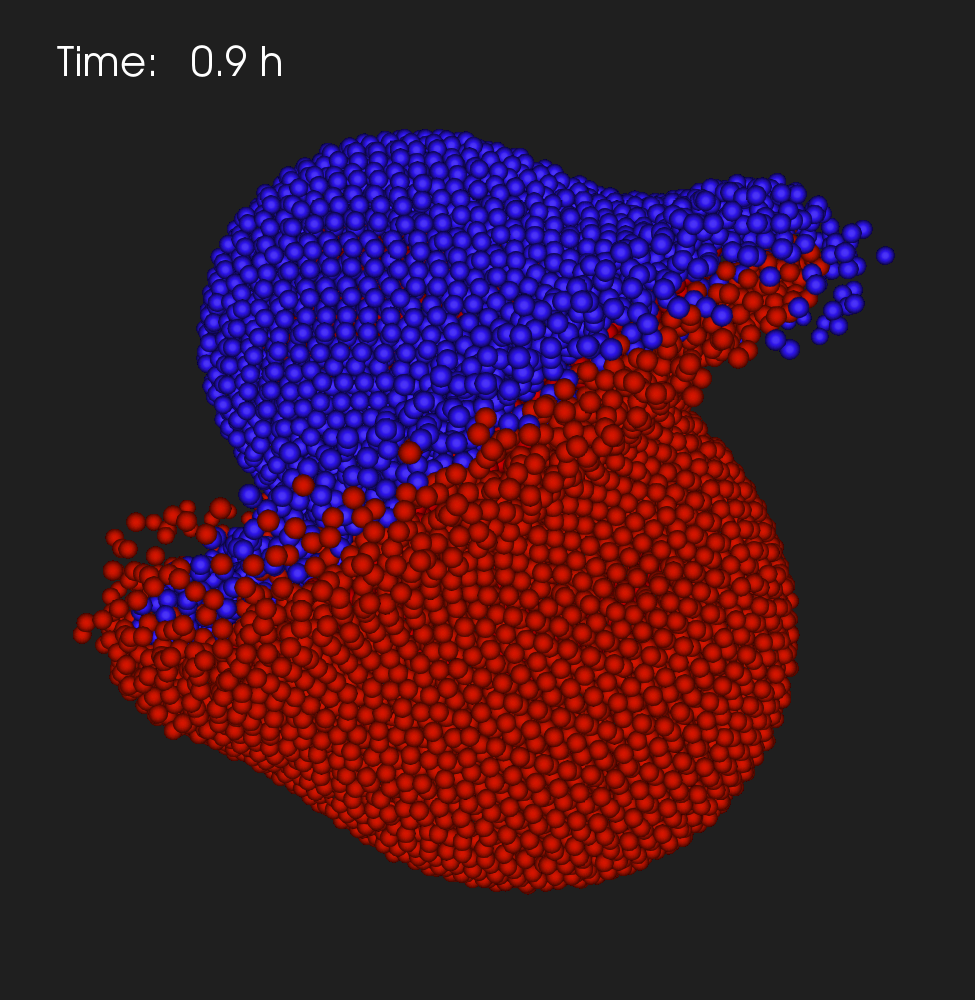}}
	\centering{\includegraphics[width=0.245\hsize]{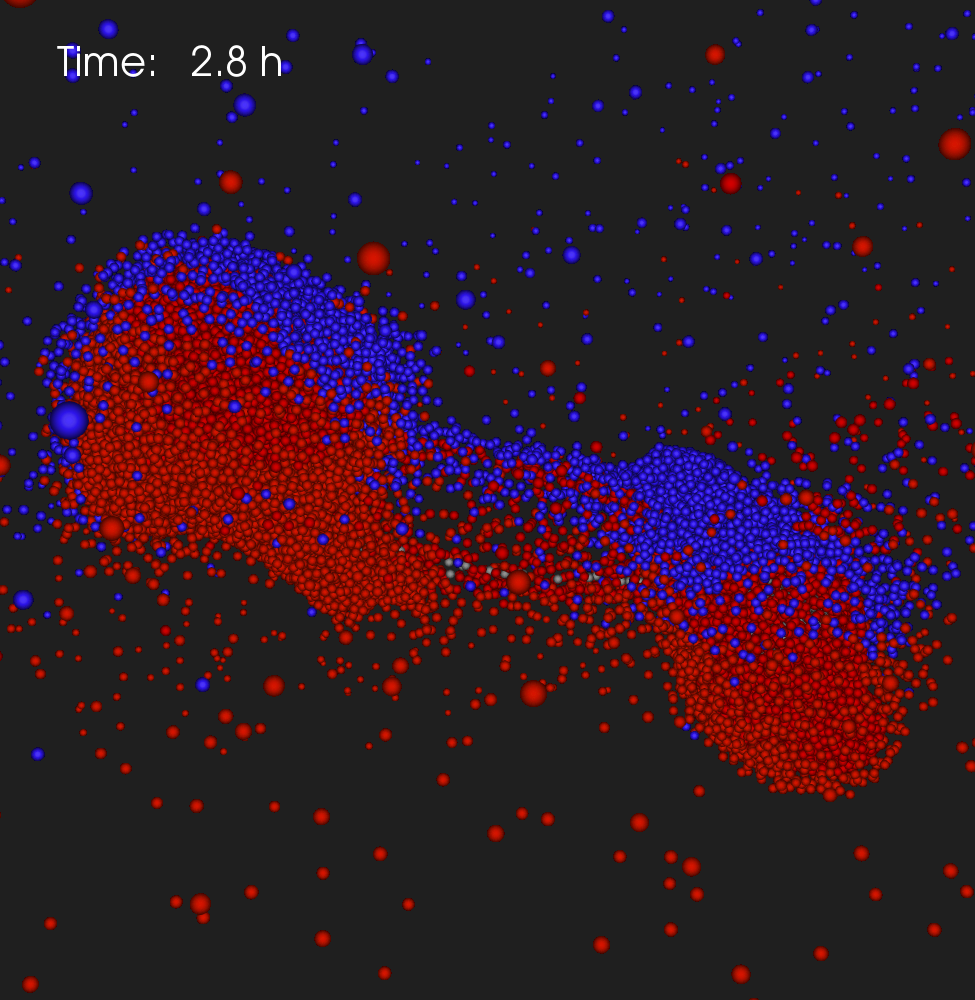}}
	\centering{\includegraphics[width=0.245\hsize]{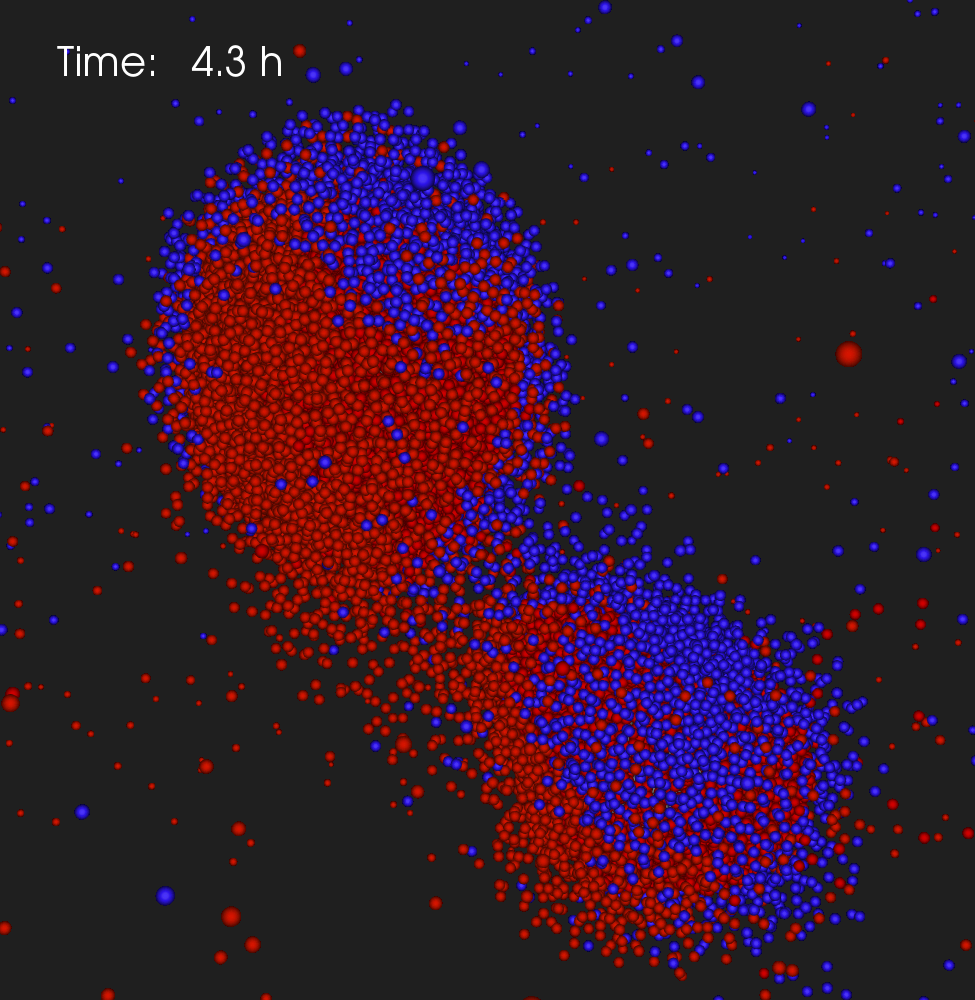}}
	\centering{\includegraphics[width=0.245\hsize]{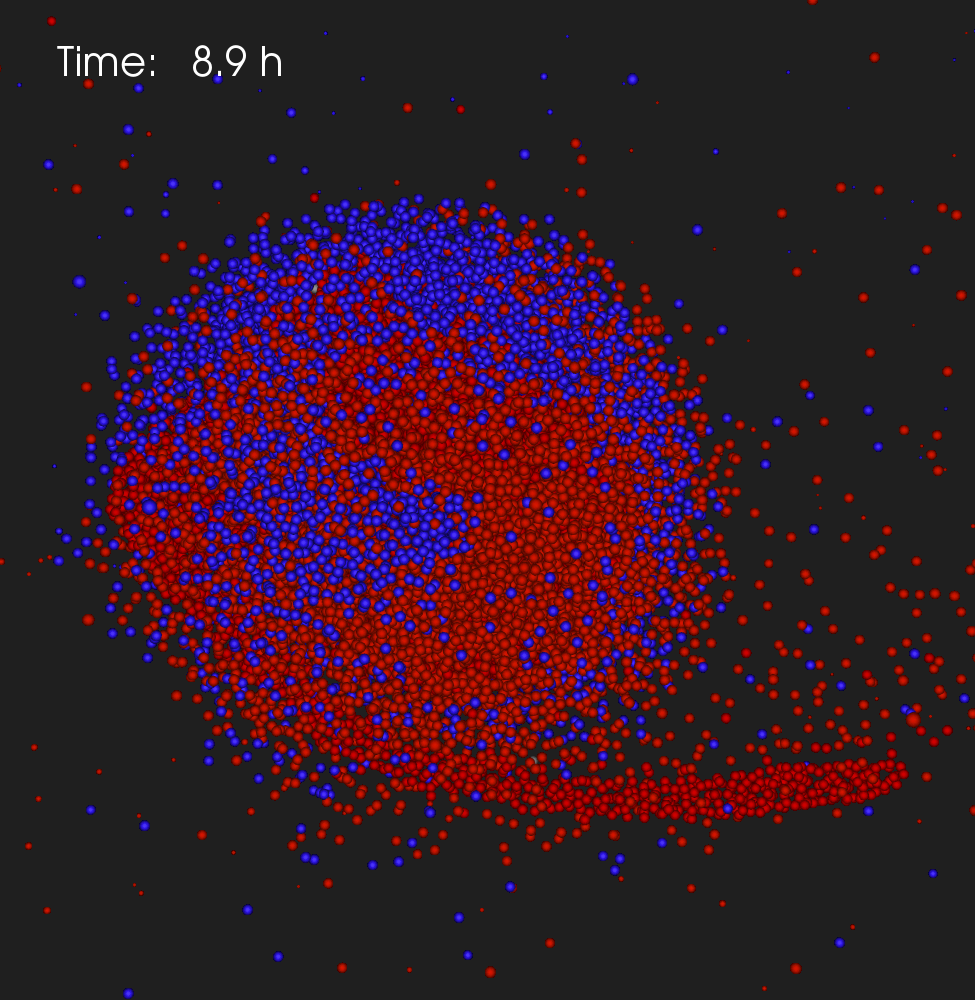}} \\
	\vspace{1.5mm}
	\centering{\includegraphics[width=0.245\hsize]{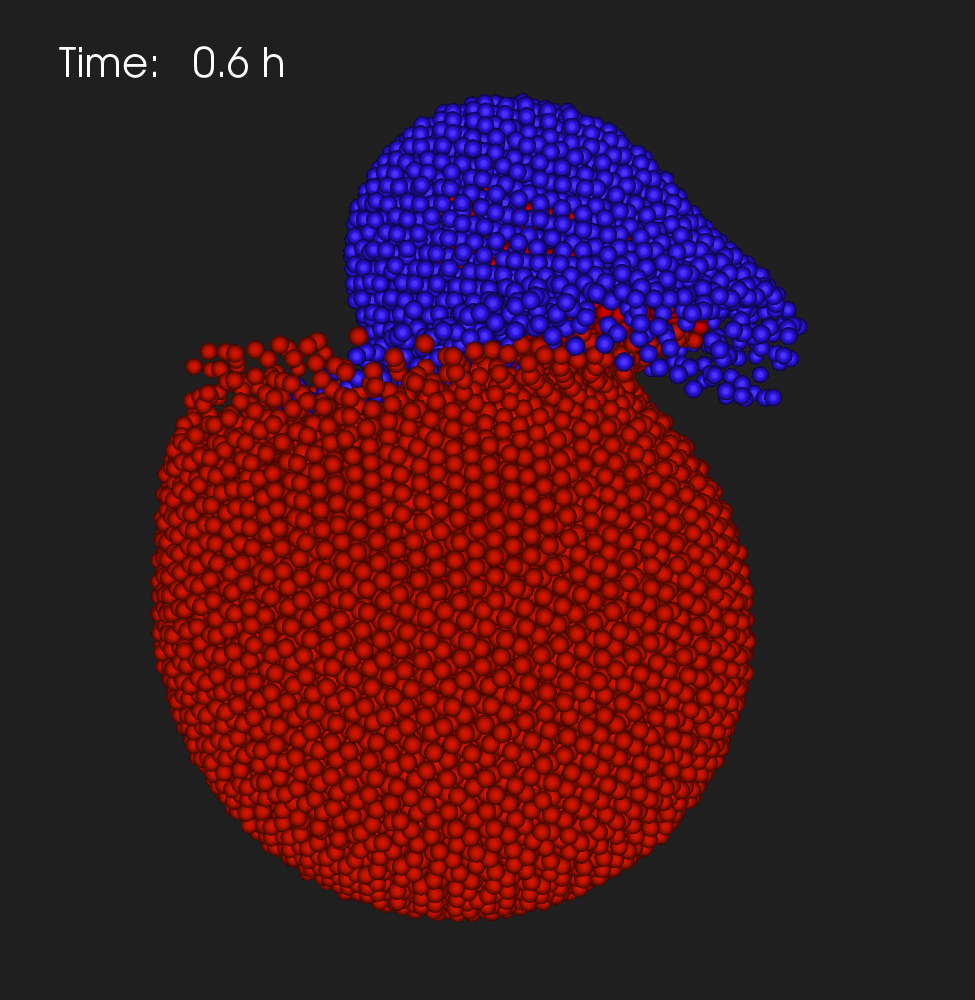}}
	\centering{\includegraphics[width=0.245\hsize]{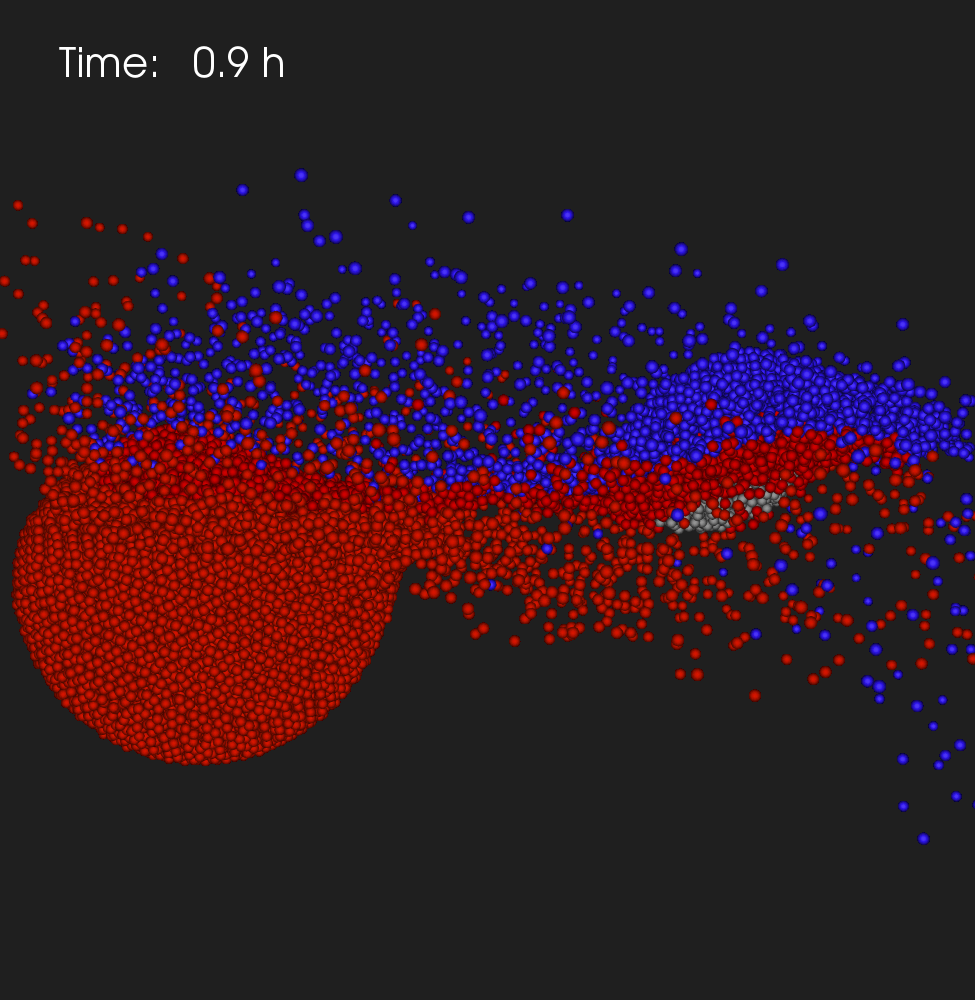}}
	\centering{\includegraphics[width=0.4955\hsize]{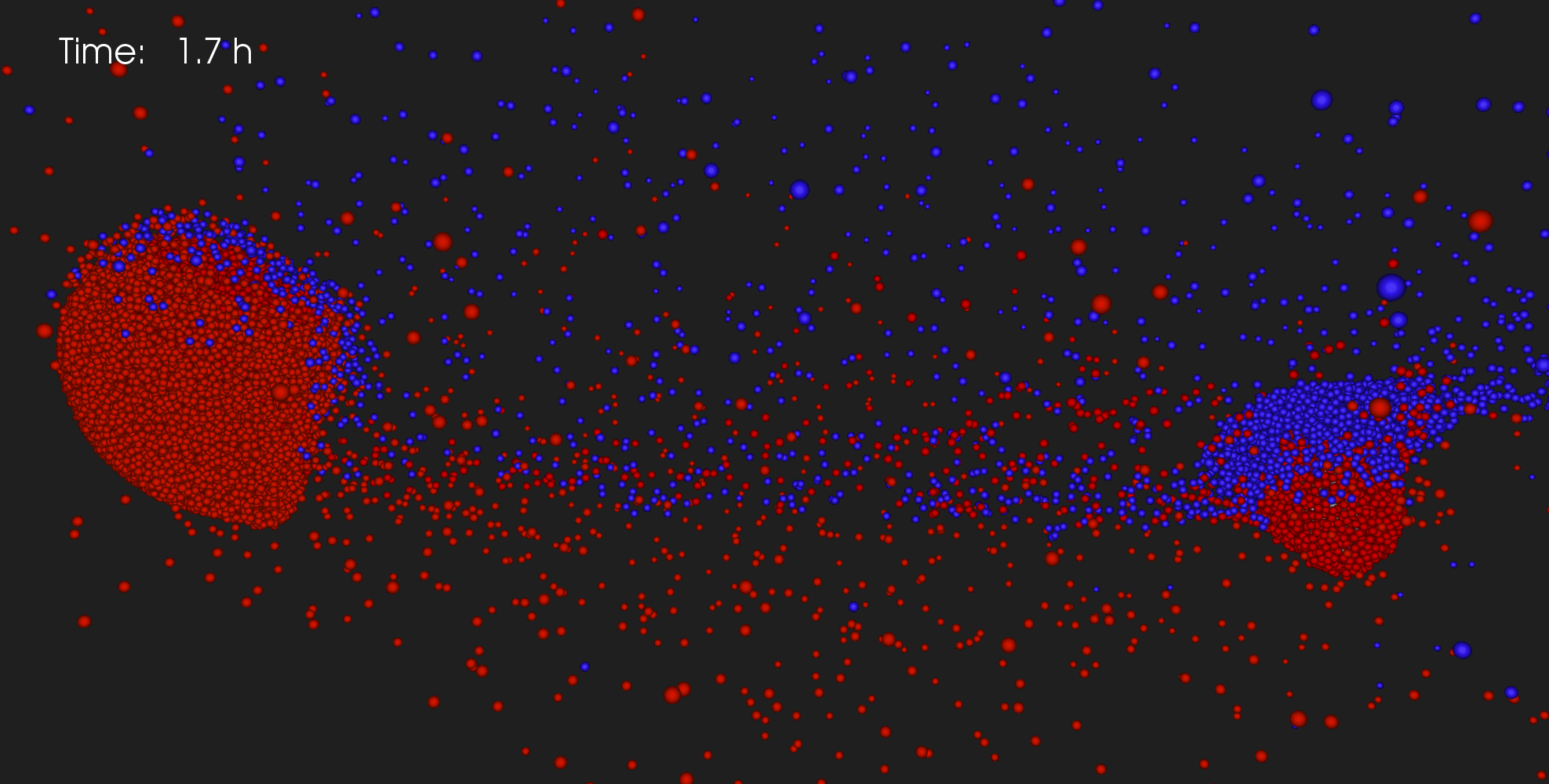}}
	\caption{Snapshots of SPH collision simulations of two different scenarios (top and bottom row). The impactor (smaller body, in blue) approaches the target from the left. The color coding shows water on the impactor in blue and silicate material (basalt) in red. (Iron core material is hardly exposed and is light gray.) The target body was actually also simulated with a water shell, but is marked in red too in order to highlight water transfer from a water-rich impactor to a dry target. The top row shows an accretionary collision (more precisely `graze-and-merge', where after a brief hit-and-run-like period the bodies fall back and merge), with $v_\mathrm{imp}/v_\mathrm{esc} = 1.2$, impact angle $\alpha = 28^\circ$ and impactor-to-target mass ratio $\gamma$\,$\sim$\,1:2. The bottom row shows a typical hit-and-run encounter with only relatively little water transfer, where $v_\mathrm{imp}/v_\mathrm{esc} = 1.8$, $\alpha = 47^\circ$ and $\gamma$\,$\sim$\,1:6. The collision in the top panels is the last (at $\sim$\,70\,Myr) decisive water-delivering event for the planet in Fig.~\ref{fig:individual_planet_run100b_675}, the one in the bottom panels shows the first (at $\sim$\,30\,Myr) of two such decisive collisions for the planet in Fig.~\ref{fig:individual_planet_run63b_656} (see Sect.~\ref{sect:example_pot_hab_planets} for details).}
	\label{fig:SPH_snapshots}
	\end{figure*}
Figure~\ref{fig:pot_hab_planets_over_t} illustrates growth of water content and mass for all potentially habitable planets formed in the regular scenarios.
Finding the growth path of a planet is no longer entirely trivial once hit-and-run collisions come into play. Our results are based on the following algorithm: Starting from the final planet, the evolution is traced backwards, where in collisions with a single survivor the more massive pre-collision body is followed.
For hit-and-run, where the larger (smaller) pre-collision body practically always ends up as the larger (smaller) post-collision object, this characteristic is used to trace the respective pre-collision body further.
A possible alternative would be to continue always with the more massive pre-collision body also for hit-and-run, which would not reliably trace the evolution passed these encounters however.
Figure~\ref{fig:pot_hab_planets_over_t} makes it immediately apparent that our collision model (solid lines) frequently results in slower growth (see Sect.~\ref{sect:timing_accretion_water_delivery_pot_hab_planets}) and considerably less massive planets than simple PM (lower panel), and also a more complex evolution of their water inventories (upper panel), including frequent collisional erosion as well.
In addition, the substantial scatter of final water contents and also the generally higher values for cJS (black\,+\,dark-green) than eJS (red\,+\,orange), are well visible.
Final planets all end up between 0.5 and 1.5\,$M_\oplus$ with our model, while PM runs can result in considerably higher masses.

When one follows the evolutionary tracks for water content in Fig.~\ref{fig:pot_hab_planets_over_t} it may be surprising that often very few large steps -- corresponding to single collisions -- seem to decisively determine the final inventory.
For the growth histories of all 18 potentially habitable planets formed with our model in the regular runs, this is indeed practically always the case, which means that the great majority of their final water content is delivered by a few -- frequently only one or two -- collisions.
Furthermore, the water-delivering impactor is almost always an embryo or larger protoplanet, and is only very rarely a small (planetesimal) body (also in the pp0 scenarios).
For the 18 planets, only twice did such a body deliver any appreciable amount of water (relative to the total accreted amount).
This indicates that water-rich planetesimals are either removed from the system or incorporated into growing protoplanets before those can deliver water, thus showing a different behavior in terms of inward transport of these two types of source materials\,/\,bodies \citep{Morbidelli2000_Source_regions_for_water}.
Therefore, the picture of inwardly scattered small bodies that directly deliver water to growing planets is rather not supported, at least not if the majority of mass has already accreted into embryos throughout the whole disk initially.
We note that these considerations include all possibilities for the origin or formation of small bodies in our model -- either pristine planetesimals from the outer disk, collisionally modified ones, or planetesimal-sized hit-and-run survivors.

This outlined mode of water delivery is basically the same in cJS and eJS scenarios. Exceptions are cases where the planet originates in a water-rich embryo from the outer disk beyond 2.5\,au, which already start with 32\,-\,45\,$O_\oplus$, but have to be transported inwards over large distances. This happens for two of the eight planets with cJS (runs SPH03 and SPH08), and for none out of the ten planets with eJS. These planets are the most water-rich formed, with 64 and 50\,$O_\oplus$, respectively. The evolution of their water inventory can be easily followed in Fig.~\ref{fig:pot_hab_planets_over_t}.
All other 16 planets start formation inside 2.5\,au, most from dry material within 2\,au and some from embryos in the intermediate range between 2 and 2.5\,au with WMF\,=\,0.001 (but their initial water content is also modest $\sim$\,0.5\,$O_\oplus$).
The few water-rich impactors either originate directly from outer-disk material or have accreted considerable quantities from other bodies during their inward transport or diffusion.
A significant difference between cJS and eJS is that the water-delivering embryos or protoplanets are often more water-rich with cJS, which explains the larger final water contents of those planets.
A noticeable feature in several cases is that a first collision delivers a relatively small but significant amount (typically below 1\,$O_\oplus$), and a later one eventually delivers the majority of the final inventory.
If the first collision does not happen the planet still ends up water-rich, but if the second impactor does not arrive, a relatively dry planet emerges.
A general point to consider is that the few decisive events in the \emph{direct} growth history do not take into account the impactors' histories, which have their own, often complicated accretion paths, and are only rarely pristine bodies from the initial configuration.
The decisive water-delivering collisions are sometimes accretionary and sometimes hit-and-run, with their typically low water-transfer efficiencies (see Sect.~\ref{sect:collision_characteristics_pot_hab} and Fig.~\ref{fig:SPH_snapshots}).

\subsection{Two examples for accretion of potentially habitable planets}
\label{sect:example_pot_hab_planets}
	\begin{table*}
	\centering
	\small
%	\scriptsize
	\caption{Summarized data on all potentially habitable planets.}
	\label{tab:pot_hab_planets}
	\begin{tabular}[b]{p{2.3cm} p{0.8cm} p{0.8cm} p{1.1cm} p{0.8cm} p{2.5cm} p{1.8cm} p{2.2cm} p{2.2cm}}
	\hline
	\hline
	    Scenario name        &   $M_\mathrm{planet}$     &  $M_\mathrm{water}$    &  WMF      &   CMF    &   a (au) / ecc / inc($^\circ$)     &   $N_\mathrm{col}$\hspace{1mm}/\hspace{1mm}$N_\mathrm{emb-col}$  &   $t_\mathrm{50}$\hspace{1mm}/\hspace{1mm}$t_\mathrm{90}$\hspace{1mm}/ $t_\mathrm{last-emb-col}$  &    $t_\mathrm{25,w}$ / $t_\mathrm{50,w}$ / $t_\mathrm{90,w}$  \\
\hline
SPH01-cJS-pp1   &   0.66    &  0.87   &  3.7e-4   &   0.26   &   0.86 / 0.09 / 4.9   &   27 / 10         &   0.4  / 8   / 93      &    31   / 31   / 31       \\
SPH03-cJS-pp1   &   0.73    &  64     &  0.025    &   0.27   &   1.10 / 0.13 / 10.5  &   9  / 4          &   13   / 40  / 49      &    1    / 1    / 13       \\
SPH04-cJS-pp1   &   1.02    &  18.9   &  0.0053   &   0.31   &   1.08 / 0.12 / 3.1   &   24 / 11         &   16   / 68  / 258     &    16   / 16   / 16       \\
SPH05-cJS-pp1   &   0.96    &  18.8   &  0.0056   &   0.31   &   1.39 / 0.17 / 5.4   &   14 / 6          &   49   / 54  / 54      &    52   / 52   / 52       \\
                &   \bf{0.84}    &  \bf{26}     &  \bf{0.0088}   &      &                       &   \bf{19 / 8}          &   \bf{20   / 43  / 114}     &    \bf{25   / 25   / 28}       \\

$^1$SPH06-cJS-pp0   &   1.20    &  24     &  0.0057   &   0.32   &   1.11 / 0.08 / 6.3   &   39 / 15         &   65   / 140 / 140     &    71   / 71   / 71       \\
SPH08-cJS-pp0   &   1.45    &  50     &  0.0098   &   0.28   &   1.20 / 0.13 / 2.0   &   13 / 6          &   14   / 84  / 84      &    0.01 / 0.01 / 0.01     \\
SPH09-cJS-pp0   &   1.49    &  20     &  0.0038   &   0.30   &   0.95 / 0.08 / 4.5   &   45 / 13         &   15   / 83  / 98      &    60   / 60   / 60       \\
SPH10-cJS-pp0   &   1.50    &  35     &  0.0067   &   0.28   &   0.92 / 0.07 / 6.1   &   35 / 15         &   58   / 133 / 133     &    36   / 36   / 133      \\
                &   \bf{1.4}     &  \bf{32}     &  \bf{0.0065}   &      &                       &   \bf{33 / 12}         &   \bf{38   / 110 / 114}     &    \bf{42   / 42   / 66}       \\
\hline                                                                                                                                                              
SPH11-eJS-pp1   &   0.56    &  0.44   &  2.2e-4   &   0.28   &   0.84 / 0.19 / 2.4   &   11 / 5          &   12   / 54  / 54      &    12   / 54   / 54       \\
                &   1.30    &  0.075  &  1.6e-5   &   0.26   &   1.31 / 0.06 / 1.8   &   26 / 7          &   3    / 7   / 7       &    2    / 2    / 2        \\
SPH12-eJS-pp1   &   1.00    &  26     &  0.0074   &   0.27   &   1.23 / 0.07 / 9.7   &   28 / 11         &   9    / 156 / 233     &    156  / 156  / 156      \\
SPH13-eJS-pp1   &   0.83    &  19.4   &  0.0067   &   0.31   &   0.89 / 0.33 / 7.8   &   16 / 9          &   28   / 88  / 175     &    28   / 28   / 28       \\
$^2$SPH15-eJS-pp1   &   1.03    &  1.59   &  4.4e-4   &   0.30   &   0.89 / 0.09 / 5.1   &   31 / 16         &   4    / 59  / 115     &    36   / 36   / 36       \\
                &   \bf{0.94}    &  \bf{9.5}    &  \bf{0.0029}   &      &                       &   \bf{22 / 10}         &   \bf{11   / 73  / 117}     &    \bf{47   / 55   / 55}       \\

SPH16-eJS-pp0   &   1.28    &  3.4    &  7.6e-4   &   0.32   &   1.00 / 0.08 / 3.8   &   28 / 12         &   46   / 111 / 174     &    111  / 244  / 244      \\
SPH17-eJS-pp0   &   1.19    &  0.55   &  1.3e-4   &   0.29   &   1.18 / 0.10 / 2.3   &   38 / 11         &   28   / 224 / 224     &    224  / 224  / 224      \\
SPH18-eJS-pp0   &   1.04    &  3.2    &  8.7e-4   &   0.38   &   0.89 / 0.12 / 1.9   &   46 / 16         &   19   / 134 / 416     &    416  / 416  / 416      \\
SPH19-eJS-pp0   &   1.42    &  31     &  0.0062   &   0.28   &   1.04 / 0.04 / 8.8   &   26 / 7          &   19   / 87  / 158     &    6    / 6    / 6        \\
SPH20-eJS-pp0   &   0.89    &  11.3   &  0.0036   &   0.30   &   1.26 / 0.07 / 5.4   &   13 / 7          &   28   / 55  / 55      &    55   / 55   / 55       \\
                &   \bf{1.2}     &  \bf{9.9}    &  \bf{0.0023}   &      &                       &   \bf{30 / 11}         &   \bf{28   / 122 / 205}     &    \bf{162  / 189  / 189}      \\
\hline                                                                                                                                                              
PM01-cJS-pp1    &   1.52    &  38     &  0.0070   &   0.25   &   1.01 / 0.09 / 8.2   &   18 / 9          &   12   / 58  / 58      &    58   / 58   / 58       \\
PM02-cJS-pp1    &   1.85    &  75     &  0.0116   &   0.25   &   1.31 / 0.04 / 5.2   &   8  / 4          &   20   / 47  / 47      &    14   / 14   / 15       \\
PM03-cJS-pp1    &   1.07    &  0.12   &  3.2e-5   &   0.25   &   0.92 / 0.12 / 13    &   13 / 7          &   3    / 11  / 27      &    11   / 27   / 27       \\
                &   1.41    &  111    &  0.022    &   0.25   &   1.33 / 0.05 / 7.3   &   5  / 4          &   47   / 104 / 104     &    1    / 1    / 104      \\
                &   \bf{1.5}     &  \bf{56}     &  \bf{0.011}    &      &                       &   \bf{11 / 6}          &   \bf{21   / 55  / 59}      &    \bf{21   / 25   / 51}       \\

PM04-cJS-pp0    &   1.87    &  41     &  0.0062   &   0.25   &   1.07 / 0.07 / 2.6   &   26 / 10         &   6    / 41  / 41      &    18   / 18   / 18       \\
PM05-cJS-pp0    &   2.51    &  113    &  0.0128   &   0.25   &   0.80 / 0.18 / 11    &   39 / 11         &   11   / 80  / 80      &    42   / 80   / 80       \\
PM06-cJS-pp0    &   0.89    &  0.031  &  1.0e-5   &   0.25   &   0.96 / 0.21 / 5.4   &   15 / 5          &   4    / 25  / 25      &    0.6  / 4    / 25       \\
                &   \bf{1.8}     &  \bf{51}     &  \bf{0.0081}   &      &                       &   \bf{27 / 9}          &   \bf{7    / 49  / 49}      &    \bf{20   / 34   / 41}       \\
\hline
PM07-eJS-pp1    &   1.83    &  0.064  &  1.0e-5   &   0.25   &   0.92 / 0.13 / 2.3   &   26 / 10         &   3    / 40  / 45      &    2.5  / 2.7  / 40       \\
PM08-eJS-pp1    &   0.91    &  2.7    &  8.4e-4   &   0.25   &   1.33 / 0.08 / 5.8   &   9  / 4          &   4    / 21  / 21      &    19   / 19   / 19       \\
PM09-eJS-pp1    &   1.42    &  0.62   &  1.3e-4   &   0.25   &   0.81 / 0.12 / 5.4   &   18 / 10         &   6    / 25  / 25      &    25   / 25   / 25       \\
                &   0.88    &  2.7    &  8.7e-4   &   0.25   &   1.27 / 0.13 / 6.5   &   8  / 5          &   21   / 90  / 90      &    38   / 38   / 44       \\
                &   \bf{1.3}     &  \bf{1.5}    &  \bf{3.3e-4}   &      &                       &   \bf{15 / 7}          &   \bf{9    / 44  / 45}      &    \bf{21   / 21   / 32}       \\
                                                                                                                                                              
PM10-eJS-pp0    &   1.42    &  33     &  0.0066   &   0.25   &   1.22 / 0.20 / 4.2   &   8  / 5          &   21   / 55  / 55      &    1    / 1    / 1        \\
PM11-eJS-pp0    &   1.05    &  40     &  0.011    &   0.25   &   1.40 / 0.24 / 9.3   &   5  / 4          &   13   / 22  / 22      &    2    / 2    / 2        \\
PM12-eJS-pp0    &   0.51    &  0.059  &  3.3e-5   &   0.25   &   1.34 / 0.02 / 11    &   4  / 2          &   0.05 / 8   / 8       &    8    / 26   / 26       \\
                &   \bf{1.0}     &  \bf{24}     &  \bf{0.0068}   &      &                       &   \bf{6  / 4}          &   \bf{11   / 28  / 28}      &    \bf{4    / 10   / 10}       \\
\hline
\hline
EO-SPH22-cJS    &   0.63    &  4.6    &  0.0021   &   0.50   &   0.89 / 0.21 / 7.8   &   21 / 21         &   8    / 11  / 288     &    288  / 288  / 288      \\
                &   0.85    &  44     &  0.0146   &   0.27   &   1.42 / 0.10 / 4.9   &   6  / 6          &   0.2  / 27  / 294     &    27   / 27   / 27       \\
EO-SPH23-cJS    &   1.81    &  0.55   &  8.7e-5   &   0.29   &   0.86 / 0.28 / 4.3   &   14 / 12         &   8    / 46  / 156     &    6.5  / 6.5  / 6.6      \\
EO-SPH24-cJS    &   0.162   &  1.17   &  0.0021   &   0.67   &   1.03 / 0.59 / 14    &   12 / 12         &   16   / 16  / 764     &    580  / 580  / 764      \\
EO-SPH25-cJS    &   1.01    &  66     &  0.0185   &   0.29   &   0.96 / 0.18 / 12    &   6  / 6          &   1    / 105 / 105     &    1    / 1    / 1        \\
                &   \bf{0.89}    &  \bf{23}     &  \bf{0.0074}   &      &                       &   \bf{12 / 11}         &   \bf{7    / 41  / 321}     &    \bf{181  / 181  / 217}      \\
\hline                                                                                                                                                              
EO-SPH26-eJS    &   1.46    &  0.55   &  1.1e-4   &   0.30   &   1.06 / 0.08 / 3.3   &   17 / 14         &   0.5  / 51  / 97      &    25   / 25   / 25       \\
EO-SPH27-eJS    &   1.32    &  99     &  0.021    &   0.25   &   0.77 / 0.10 / 3.6   &   7  / 7          &   16   / 17  / 73      &    0.4  / 0.4  / 0.4      \\
EO-SPH28-eJS    &   0.71    &  7.3    &  0.0029   &   0.35   &   0.95 / 0.20 / 7.1   &   11 / 11         &   158  / 273 / 290     &    40   / 40   / 40       \\
EO-SPH29-eJS    &   0.80    &  0.014  &  4.9e-6   &   0.37   &   0.81 / 0.24 / 14    &   16 / 16         &   28   / 92  / 92      &    0.6  / 0.6  / 49       \\
                &   \bf{1.1}     &  \bf{27}     &  \bf{0.007}    &      &                       &   \bf{13 / 12}         &   \bf{51   / 108 / 138}     &    \bf{17   / 17   / 29}       \\
\hline                                                                                                                                                                                                                                                                                                                            
EO-PM14-cJS     &   3.02    &  70     &  0.0066   &   0.25   &   0.81 / 0.40 / 7.8   &   6  / 6          &   100  / 201 / 201     &    100  / 100  / 100      \\
EO-PM15-cJS     &   2.65    &  65     &  0.0069   &   0.25   &   0.78 / 0.07 / 3.4   &   12 / 12         &   23   / 59  / 59      &    25   / 25   / 25       \\
                &   \bf{2.8}     &  \bf{68}     &  \bf{0.0069}   &      &                       &   \bf{9  / 9}          &   \bf{62   / 130 / 130}     &    \bf{63   / 63   / 63}       \\
\hline                                                                                                                                                              
EO-PM16-eJS     &   0.92    &  0.96   &  3.0e-4   &   0.25   &   1.17 / 0.10 / 24    &   4  / 4          &   1    / 84  / 84      &    84   / 84   / 84       \\
EO-PM17-eJS     &   1.11    &  0.039  &  1.0e-5   &   0.25   &   1.15 / 0.10 / 5.1   &   4  / 4          &   10   / 16  / 16      &    0.3  / 10   / 16       \\
EO-PM18-eJS     &   2.67    &  55     &  0.0058   &   0.25   &   0.78 / 0.10 / 6.7   &   11 / 11         &   18   / 27  / 65      &    18   / 18   / 18       \\
                &   \bf{1.6}     &  \bf{19}     &  \bf{0.0034}    &      &                       &   \bf{6  / 6}          &   \bf{10   / 42  / 55}      &    \bf{34   / 37   / 39}       \\
	\hline
	\end{tabular}
	\tablefoot{All masses are given in $M_\oplus$, water masses always in $O_\oplus$, and time in Myr. Mean values of subsets are bold. $t_{25}$, $t_{50}$ and $t_{90}$ denote growth times to the respective mass percentage, where the additional subscript `w' indicates water inventories alone (see the text for a detailed definition). $N_\mathrm{col}$ and $N_\mathrm{emb-col}$ are numbers of all collisions and of those only among embryos and protoplanets (i.e., excluding planetesimals), respectively. $t_\mathrm{last-emb-col}$ is the time of the last pure embryo\,/\,protoplanet collision. Note that mean WMFs are computed from mean $M_\mathrm{water}$ and $M_\mathrm{planet}$. Superscript $^1$ indicates the planet in Fig.~\ref{fig:individual_planet_run100b_675}, $^2$ the one in Fig.~\ref{fig:individual_planet_run63b_656}.}
	\end{table*}
	
	\begin{figure}
	\centering{\includegraphics[width=8.8cm]{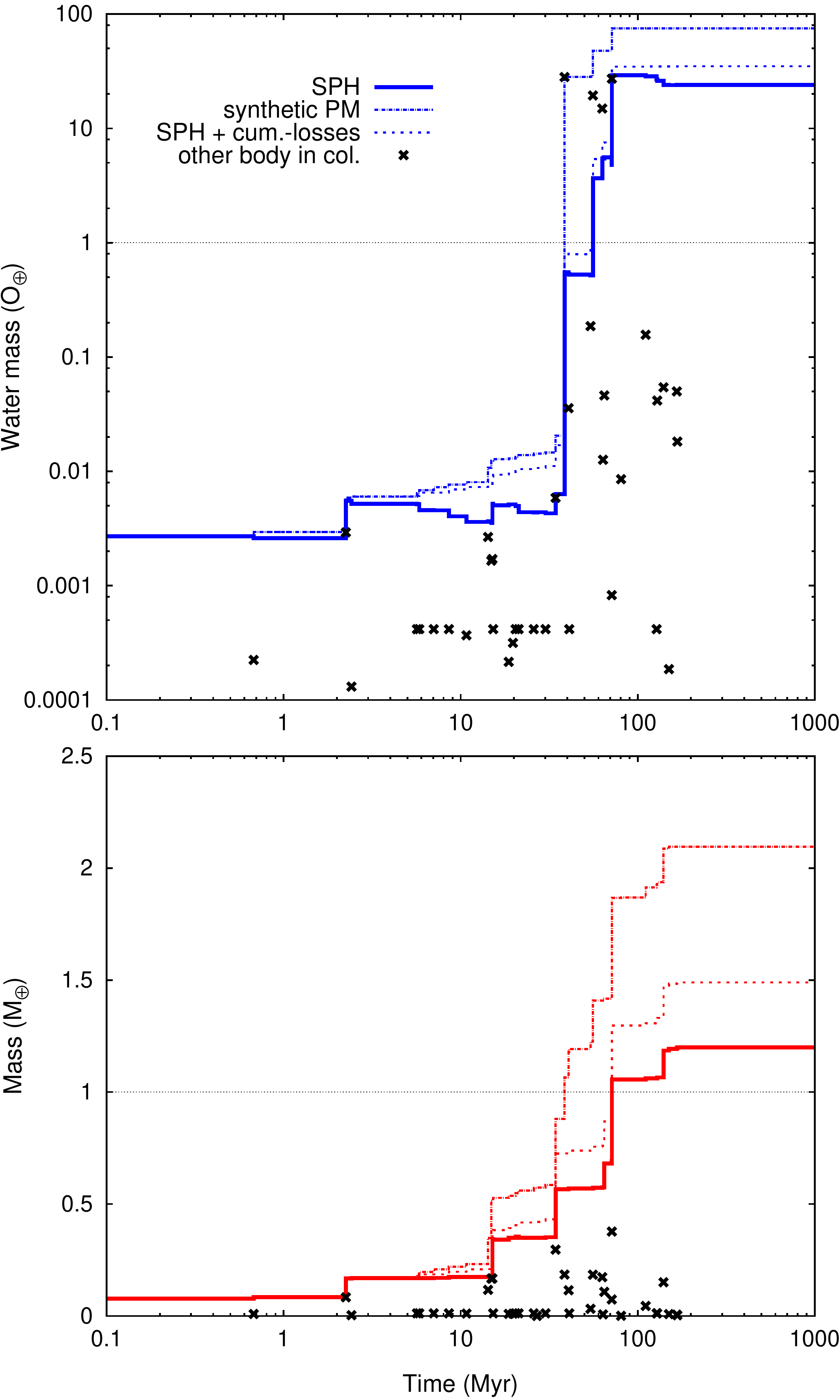}}
	\caption{Accretion history of the potentially habitable planet formed in run SPH06-cJS-pp0 (marked with superscript $^1$ in Tables~\ref{tab:pot_hab_planets} and \ref{tab:pot_hab_planets_collisions}, and also illustrated in Fig.~\ref{fig:N_body_snapshots}). It forms at $\sim$\,1.1\,au, with 1.2\,$M_\oplus$ and 24\,$O_\oplus$ of water. Besides the actual results with our SPH collision model (solid lines), hypothetical growth curves with synthetic PM, and also with added cumulative losses are plotted -- both defined and discussed in the text (Sect.~\ref{sect:example_pot_hab_planets}). The crosses mark the respective property of the other (second) body in collisions, besides the traced planet.}
	\label{fig:individual_planet_run100b_675}
	\end{figure}

	\begin{figure}
	\centering{\includegraphics[width=8.8cm]{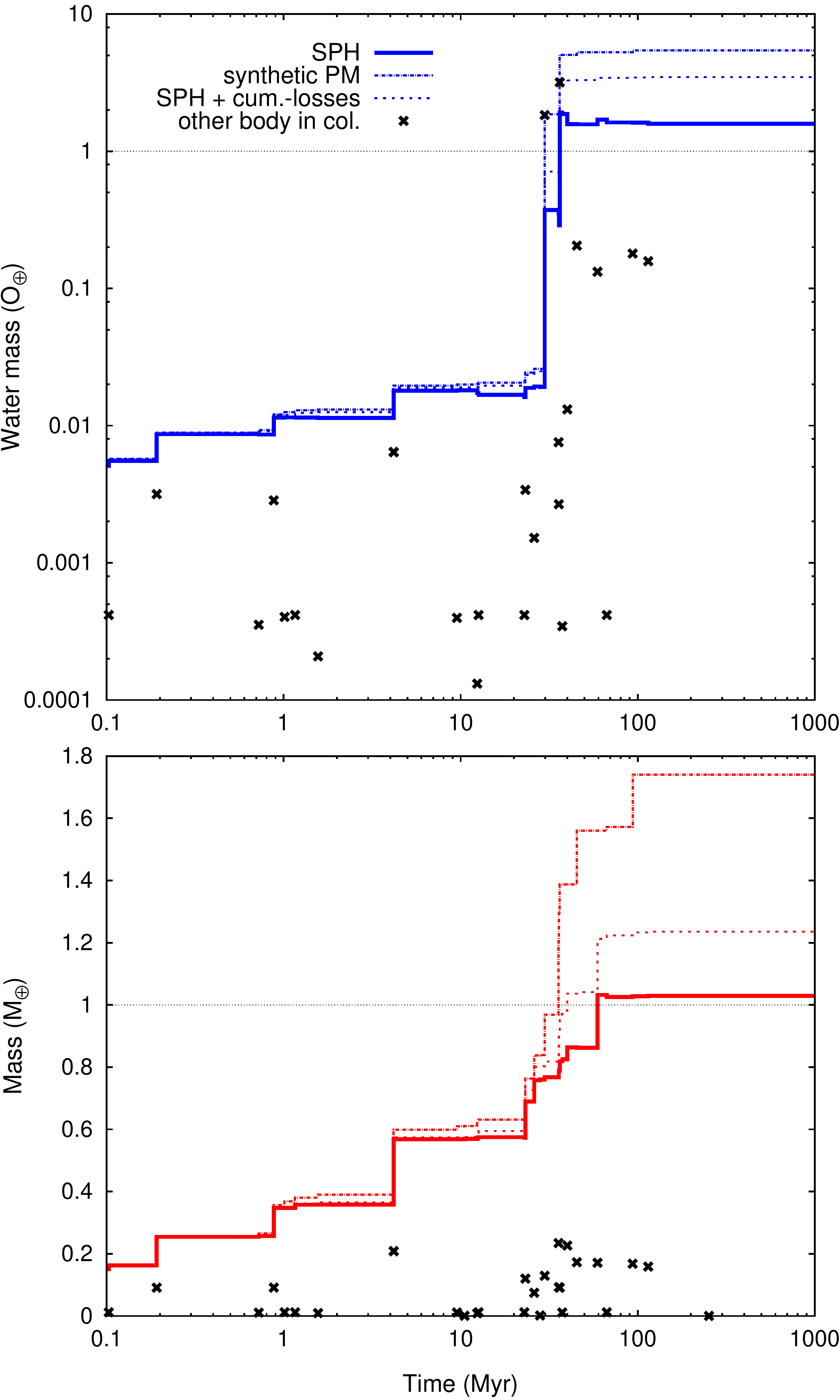}}
	\caption{Accretion history of the potentially habitable planet formed in run SPH15-eJS-pp1 (marked with superscript $^2$ in Tables~\ref{tab:pot_hab_planets} and \ref{tab:pot_hab_planets_collisions}, and also illustrated in Fig.~\ref{fig:N_body_snapshots}). It forms at $\sim$\,0.9\,au, with 1.03\,$M_\oplus$ and 1.59\,$O_\oplus$ of water. Lines and symbols have the same meanings as in Fig.~\ref{fig:individual_planet_run100b_675}.}
	\label{fig:individual_planet_run63b_656}
	\end{figure}

To illustrate the accretion of potentially habitable planets in detail, two typical examples are presented in Figs.~\ref{fig:individual_planet_run100b_675} and \ref{fig:individual_planet_run63b_656}, where growth of their mass as well as water content are plotted (thick solid lines).
The planet in Fig.~\ref{fig:individual_planet_run100b_675} forms in a cJS architecture (run SPH06) and ends up with $\sim$\,24\,$O_\oplus$.
It experiences four decisive collisions that deliver significant amounts of water, which is the highest number in all considered scenarios, but they are nevertheless representative and it is an interesting illustrative example.
The first of these four collisions, a very grazing hit-and-run, happens after $\sim$\,40\,Myr and transfers less than 1\,$O_\oplus$ to the growing planet (note the log y-axis).
After that, two further hit-and-run collisions deliver considerable amounts of water, where the impactor in the second one is actually the second hit-and-run survivor from the first one of these two events ($\sim$\,7\,Myr before), which hits again and transfers some more of its water inventory.
Finally, accretion of a water-rich protoplanet delivers $\sim$\,24\,$O_\oplus$ at $\sim$\,70\,Myr. Snapshots of this latter graze-and-merge event are included in Fig.~\ref{fig:SPH_snapshots} (top row), illustrating water delivery by the impacting body (the target is color-coded as dry).
Before the first of those events, \changeslan{practically} only dry accretion occurs, and after the last one several collisions with dry bodies result in clearly visible losses of more than 5\,$O_\oplus$.
The second example planet (Fig.~\ref{fig:individual_planet_run63b_656}) forms in an eJS architecture (run SPH15), and ends up with $\sim$\,1.6\,$O_\oplus$, shaped by two decisive events.
After 30 Myr of dry accretion, a hit-and-run encounter with a moderately water-rich protoplanet transfers $\sim$\,0.35\,$O_\oplus$ to the growing planet, which is also illustrated as snapshots in Fig.~\ref{fig:SPH_snapshots} (bottom row), where the rather low water transfer efficiency can be clearly discerned.
Only 6 Myr later, another (quite accretionary) hit-and-run with a more water-rich protoplanet delivers $\sim$\,1.6\,$O_\oplus$.

For the planets in Figs.~\ref{fig:individual_planet_run100b_675} and \ref{fig:individual_planet_run63b_656} we also include two additional growth curves, which are intended to provide some measure of the theoretical evolution without proper treatment of collisions.
The first one gives the actual results with added cumulative collision losses, meaning that all material that is not included in either post-collision body is simply added (cumulatively).
\changeslan{Final water and total masses are then the sum of the actual values (resulting from our model) plus all these losses during the accretion of the planet} (also given in Table~\ref{tab:pot_hab_planets_collisions}).
In some sense this gives a lower limit for theoretical growth without proper collision treatment: while for events with a single post-collision body all material neglected as debris is now included (thus equivalent to PM), only a (typically small) fraction of the remaining material is included in hit-and-run.

The second theoretical growth curve, labeled `synthetic PM', merges all material also in hit-and-run into the planet's original growth path (again cumulative).
In cases where the two survivors of a hit-and-run collision subsequently collide again, they are merged only once.
Interestingly it turned out that defining such synthetic PM is not as straightforward as one might think, particularly due to hit-and-run in the original results, leading to convoluted accretion tracks. Eventually we decided to use the back-traced history of planets, that is, the regular results from our collision model since that is their natural dynamical way to evolve, and to trace this path again forward in time while applying synthetic PM (instead of some direct and only forward-in-time approach).
The obtained theoretical water contents are also included in Table~\ref{tab:pot_hab_planets_collisions}.
We note that even with synthetic PM the collision histories of the bodies impacting the growing planet are not also followed and replaced by PM, where complex accretion paths due to hit-and-run would complicate matters, for example avoiding that bodies or material are counted several times.

These two approaches yield growth curves that are significantly above the actual ones, clearly visible in Figs.~\ref{fig:individual_planet_run100b_675} and \ref{fig:individual_planet_run63b_656}.
Due to the abundance of hit-and-run, it is furthermore not surprising that synthetic PM results in considerably larger values than counting only cumulative losses.
A comparison of the average values for synthetic PM (Table~\ref{tab:pot_hab_planets_collisions}) with the actual outcomes of our PM runs (Table~\ref{tab:pot_hab_planets}) shows roughly similar results and indicates that synthetic PM provides quite reliable predictions for water contents if collisional effects were ignored.
For the planet in Fig.~\ref{fig:individual_planet_run100b_675} one obtains final water inventories of 35 and 75\,$O_\oplus$ for the two described approaches, respectively (vs.\ 24\,$O_\oplus$ as the actual result).
For the planet in Fig.~\ref{fig:individual_planet_run63b_656} these values are 3.5 and 5.4\,$O_\oplus$ (vs.\ 1.6\,$O_\oplus$ as the actual result).

\subsection{Collision characteristics of potentially habitable planets}
\label{sect:collision_characteristics_pot_hab}
	\begin{table*}
	\centering
	\small
%	\scriptsize
	\caption{Data on collisions of all potentially habitable planets formed with our model.}
	\label{tab:pot_hab_planets_collisions}
	\begin{tabular}[b]{p{2.5cm} p{1.8cm} p{1.8cm} p{2.6cm} p{2.2cm} p{2.6cm} p{2.0cm}}
	\hline
	\hline
Scenario name        &   $M_\mathrm{planet}$\newline \scriptsize{(col-losses)} &  $M_\mathrm{water}$\newline \scriptsize{(col-losses)} &  $N_\mathrm{col}$\newline ($N_\mathrm{hit-and-run}$ / $N_\mathrm{re-col}$) &  $t_\mathrm{re-col,median}$ \scriptsize{(range)} &  $N_\mathrm{emb-col}$\newline ($N_\mathrm{hit-and-run}$ / $N_\mathrm{re-col}$) &  $M_\mathrm{water,synthetic-PM}$ \\
\hline
\hline
SPH01-cJS-pp1   &   0.66  (0.14)    &  0.87  (2.1)     &  27 (17/4)            &  7 (0.45...23)               &  10 (6/2)		&	37                 \\
SPH03-cJS-pp1   &   0.73  (0.09)    &  64    (45)      &  9  (3/0)             &  n.a.                        &  4  (2/0)		&	112                 \\
SPH04-cJS-pp1   &   1.02  (0.19)    &  19    (22)      &  24 (13/5)            &  8.5 (0.03...40)             &  11 (6/4)	&	42                 \\
SPH05-cJS-pp1   &   0.96  (0.15)    &  19    (18)      &  14 (6/0)             &  n.a.                        &  6  (4/0)		&	38                 \\
                &   \bf{0.84  (0.14)}    &  \bf{26    (22)}      &  \bf{19 (10/2)}            &                              &  \bf{8  (5/1.5)}		&	\bf{57}               \\
                                                                                                                                          
$^1$SPH06-cJS-pp0   &   1.20  (0.29)    &  24    (11)      &  39 (23/6)            &  11 (0.2...26)               &  15 (11/4)	&	75                \\
SPH08-cJS-pp0   &   1.45  (0.08)    &  50    (24)      &  13 (5/0)             &  n.a.                        &  6  (2/0)		&	76                 \\
SPH09-cJS-pp0   &   1.49  (0.18)    &  20    (15)      &  45 (21/8)            &  18 (0.01...70)              &  13 (4/2)		&	35                 \\
SPH10-cJS-pp0   &   1.50  (0.18)    &  35    (14)      &  35 (23/7)            &  0.7 (0.2...6.4)             &  15 (11/4)		&	56                \\
                &   \bf{1.4   (0.18)}    &  \bf{32    (16)}      &  \bf{33 (18/5)}            &                              &  \bf{12 (7/2.5)}		&	\bf{61}               \\
\hline                                                                                                                                          
SPH11-eJS-pp1   &   0.56  (0.014)   &  0.44  (0.04)    &  11 (4/0)             &  n.a.                        &  5  (2/0)	&	0.82                 \\
                &   1.30  (0.03)    &  0.075 (0.012)   &  26 (7/0)             &  n.a.                        &  7  (1/0)		&	0.088                 \\
SPH12-eJS-pp1   &   1.00  (0.14)    &  26    (5.2)     &  28 (14/1)            &  20                          &  11 (5/0)		&	31                 \\
SPH13-eJS-pp1   &   0.83  (0.17)    &  19    (18)      &  16 (11/3)            &  37 (1.5...54)               &  9  (7/3)		&	39                 \\
$^2$SPH15-eJS-pp1   &   1.03  (0.21)    &  1.6   (1.9)     &  31 (18/5)            &  19 (6e-3...28)              &  16 (11/4)		&	5.4                \\
                &   \bf{0.94  (0.11)}    &  \bf{9.4   (5.0)}     &  \bf{22 (11/2)}            &                              &  \bf{10 (5/1.4)}		&	\bf{15}               \\
                                                                                                                                          
SPH16-eJS-pp0   &   1.28  (0.09)    &  3.4   (0.62)    &  28 (16/2)            &  59 (46...71)                &  12 (6/0)		&	4.8                 \\
SPH17-eJS-pp0   &   1.19  (0.14)    &  0.55  (0.07)    &  38 (16/7)            &  35 (7e-5...86)              &  11 (3/1)		&	0.63                 \\
SPH18-eJS-pp0   &   1.04  (0.42)    &  3.2   (2.2)     &  46 (29/6)            &  1.5 (0.02...17)             &  16 (11/3)		&	5.4                \\
SPH19-eJS-pp0   &   1.42  (0.06)    &  31    (7.3)     &  26 (11/6)            &  13 (0.5...140)              &  7  (1/1)		&	39                 \\
SPH20-eJS-pp0   &   0.89  (0.07)    &  11    (1.3)     &  13 (5/2)             &  33 (6...61)                 &  7  (2/1)		&	13                 \\
                &   \bf{1.2   (0.16)}    &  \bf{9.8   (2.3)}     &  \bf{30 (15/5)}            &                              &  \bf{11 (5/1)}	&	\bf{13}                 \\
\hline                                                                                                                                          
EO-SPH22-cJS    &   0.63  (0.93)    &  4.6   (14)      &  21 (17/8)            &  22 (0.2...69)               &  21 (17/8)		&	31                \\
                &   0.85  (0.12)    &  44    (26)      &  6  (3/1)             &  11                          &  6  (3/1)	&	101                 \\
EO-SPH23-cJS    &   1.81  (0.19)    &  0.55  (0.5)     &  14 (9/4)             &  0.9 (0.16...16)             &  12 (7/4)	&	1.06                 \\
EO-SPH24-cJS    &   0.16  (0.48)    &  1.2   (10)      &  12 (11/2)            &  101 (18...184)              &  12 (11/2)		&	38                \\
EO-SPH25-cJS    &   1.01  (0.15)    &  66    (51)      &  6  (4/0)             &  n.a.                        &  6  (4/0)		&	124                 \\
                &   \bf{0.89  (0.37)}    &  \bf{23    (20)}      &  \bf{12 (9/3)}             &                              &  \bf{11 (8/3)}	&	\bf{59}                 \\
\hline                                                                                                                                          
EO-SPH26-eJS    &   1.46  (0.25)    &  0.55  (0.51)    &  17 (9/4)             &  3.6 (0.8...18)              &  14 (9/4)	&	1.2                 \\
EO-SPH27-eJS    &   1.32  (0.03)    &  99    (9.3)     &  7  (4/0)             &  n.a.                        &  7  (4/0)		&	110                 \\
EO-SPH28-eJS    &   0.71  (0.32)    &  7.3   (41)      &  11 (9/3)             &  1.4 (0.8...13)              &  11 (9/3)	&	61                 \\
EO-SPH29-eJS    &   0.80  (0.23)    &  0.014 (0.08)    &  16 (13/3)            &  17 (0.1...56)               &  16 (13/3)		&	0.96                \\
                &   \bf{1.1   (0.21)}    &  \bf{27    (13)}      &  \bf{13 (9/2.5)}           &                              &  \bf{12 (9/2.5)}		&	\bf{43}               \\
	\hline
	\end{tabular}
	\tablefoot{All masses are given in $M_\oplus$, water masses always in $O_\oplus$, and time in Myr. Mean values of subsets are bold. `col-losses' refers to the respective quantity (mass or water mass) not incorporated in either of the SPH post-collision bodies, summed over all collisions. $N_\mathrm{col}$ and $N_\mathrm{emb-col}$ are numbers of all collisions and of those only among embryos and protoplanets (i.e., excluding planetesimals), respectively. `re-col' refers to a re-collision of the growing planet with the second hit-and-run survivor of a previous encounter. For the time-span between hit-and-run collision and re-collision ($t_\mathrm{re-col}$), median and range are given where applicable. The theoretical water contents labeled `synthetic PM' are defined and discussed in Sect.~\ref{sect:example_pot_hab_planets}. Superscript $^1$ indicates the planet in Fig.~\ref{fig:individual_planet_run100b_675}, $^2$ the one in Fig.~\ref{fig:individual_planet_run63b_656}.}
	\end{table*}
Table~\ref{tab:pot_hab_planets} provides overall collision numbers for all potentially habitable planets, while Table~\ref{tab:pot_hab_planets_collisions} summarizes more detailed collision information for the subset formed with our model.
In the regular scenarios and with our model the average number of collisions in the growth history of a potentially habitable planet is $\sim$\,20 with pp1 and $\sim$\,30 with pp0.
This is not surprising since with pp1 planetesimals can accrete mutually before hitting the growing planet, but they cannot with pp0.
However, the total number of collisions is only of secondary importance anyway, since it depends strongly on the initial number of bodies, which is determined by computational limitations, not physics.
More interesting is the number of impacts by large bodies (embryos and protoplanets), because the initial embryo population is accurately resolved according to our physical model.
This number averages to approximately ten collisions, with pp1 and the same with pp0.
In comparison, potentially habitable planets formed with PM experience roughly one-third less impacts by embryos or protoplanets, presumably due to the absence of hit-and-run.
With our model, the prevalence of hit-and-run in the accretion histories of potentially habitable planets is similar to what was found and discussed in a system-wide context in Sect.~\ref{sect:global_collision_characteristics}.
\changeslan{In the regular scenarios, about 50\% of all collisions are hit-and-run, also if only impacts by embryos and protoplanets are considered, and even $\sim$\,75\% in embryo-only simulations.}

Hit-and-run collisions where a water-rich target is hit by a considerably drier projectile can strip water from the target body, even though this becomes only really transformative if hit by an approximately equally massive or larger impactor \citep[e.g.,][]{Asphaug2010_Similar_sized_collisions_diversity_of_planets, Burger2018_Hit-and-run}.
In the accretion histories of all 18 potentially habitable planets formed with our model in the regular scenarios, only once is the growing planet hit by a more massive impactor, and collisions with equally massive bodies are also relatively rare. Most impacts are by significantly lower mass objects (see Figs.~\ref{fig:individual_planet_run100b_675} and \ref{fig:individual_planet_run63b_656}).
However, energetic individual events as well as a large number of smaller impactors can lead to considerable collisional water losses.
Especially interesting for the delivery of water on the other hand are encounters where a drier target body is hit by a water-rich projectile, with potential transfer to the target. However, for hit-and-run this transfer is often not very efficient, especially compared to accretionary (e.g., more head-on) collisions \citep{Burger2018_Hit-and-run, Burger2019_IAU_proceedings_water_delivery_to_dry_protoplanets_hit_and_run}.
For commonly occurring collision parameters, typically less than $\sim$\,25\% of the projectile water is transferred in hit-and-run, but 50\% or more is transferred in accretionary impacts (cf.\ Fig.~\ref{fig:SPH_snapshots}).
The rationale of past studies behind still assuming PM (in spite of usually recognizing the high frequency of hit-and-run) was often that the two hit-and-run survivors remain on similar orbits and re-collide soon after the first encounter.
Our results support this assumption only to a low degree, as was also found in a recent, dedicated study by \citet{Emsenhuber2019_Fate_of_Runner_in_har}.
Table~\ref{tab:pot_hab_planets_collisions} lists the total number of collisions, along with the number of hit-and-run and how many of those led to re-collision at a later time in brackets (and the same for purely embryo\,/\,protoplanet encounters).
Re-collisions happen on average in only $\sim$\,25\% of cases, thus the majority of bodies that go into hit-and-run do not directly meet again.
For comparison, \citet{Chambers2013_Late-stage_accretion_hit-and-run_fragmentation} found that 54\% of hit-and-run events (considering only those among embryos and protoplanets) lead to re-collision of the same bodies later.
A follow-up question that presents itself pertains to the nature of the re-collisions. Even though this analysis suffers from low sample sizes, our results indicate (not included in Table~\ref{tab:pot_hab_planets_collisions}) that about half of all re-collisions after hit-and-run are again hit-and-run \citep[][found a somewhat reduced probability for repeated hit-and-run]{Emsenhuber2019_Fate_of_Runner_in_har}.
Therefore our data suggest no special trend in re-collision outcomes, and do not support the hypothesis that the majority of hit-and-run encounters are followed by subsequent accretion.
The time between hit-and-run and re-collision varies over a very wide range (Table~\ref{tab:pot_hab_planets_collisions}).
For those planets where relatively good statistics (high hit-and-run and re-collision numbers) exist, median values are on the order of 1 to 10~Myr, and re-collision times can be as high as several 10~Myr \citep[][found a mean time of 1.3~Myr]{Chambers2013_Late-stage_accretion_hit-and-run_fragmentation}.
Very immediate re-collisions seem to be rare -- in our data (on potentially habitable planets) only one such event occurs after just 70\,years, where the next shortest time to re-collision is $\sim$\,6000\,years.
The high frequency of hit-and-run and the often low water transfer efficiency, combined with low re-collision rates, makes it particularly important to accurately model this type of encounter, especially if water delivery is dominated by few decisive collisions with large bodies (embryos or protoplanets), which naturally tend to be more similar sized (and thus more likely hit-and-run) than delivery by smaller bodies.

	\begin{figure}
	\centering{\includegraphics[width=8.8cm]{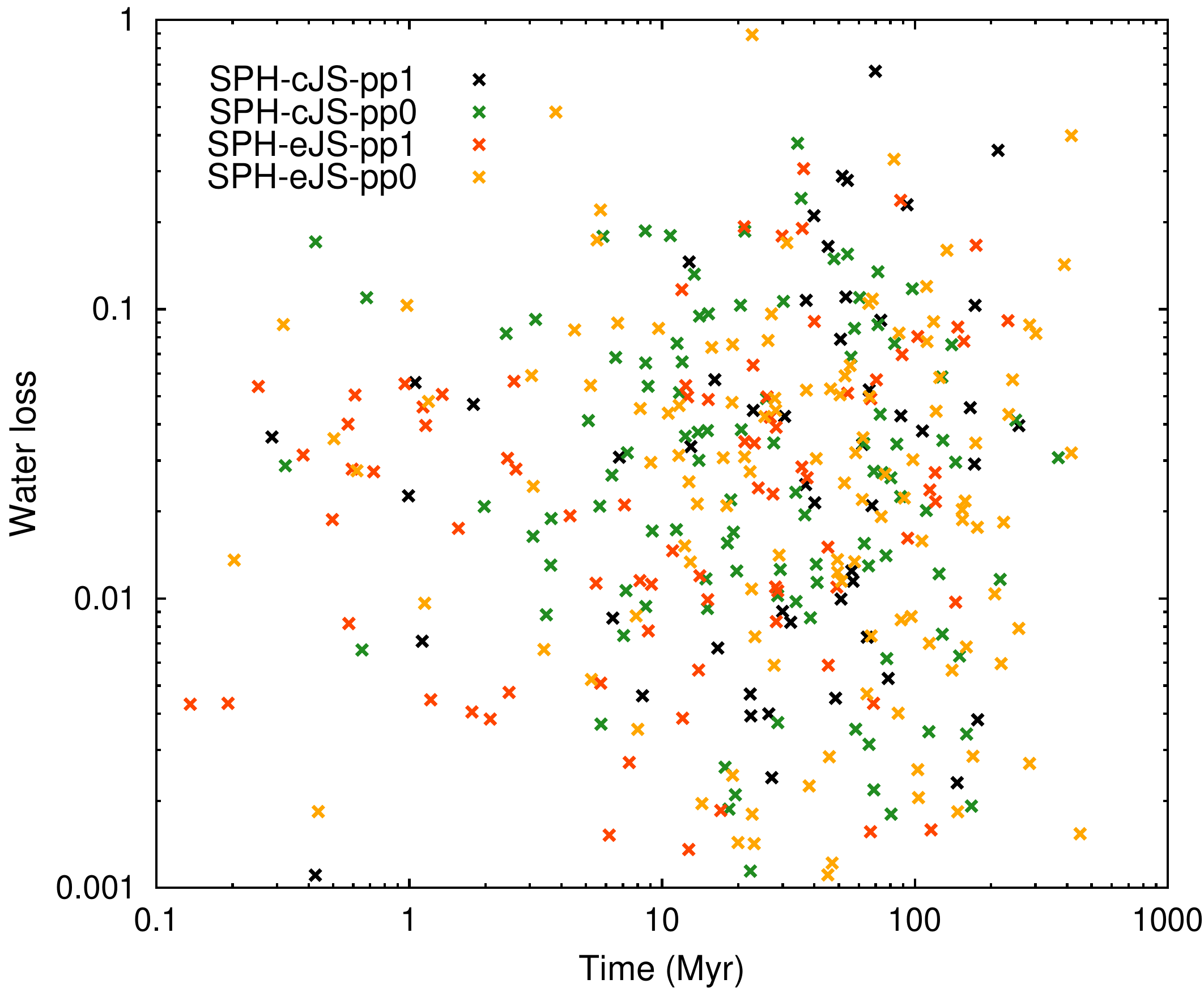}}
	\caption{Water losses in individual collisions over time during accretion of all 18 potentially habitable planets in the regular scenarios. Water loss is defined as the fraction of the (combined) pre-collision water inventory that is not incorporated into either of the post-collision bodies.}
	\label{fig:pot_hab_planets_water_loss_over_t}
	\end{figure}
Figure~\ref{fig:pot_hab_planets_water_loss_over_t} shows water losses over time for collisions in the growth histories of potentially habitable planets.
This is a subset of Fig.~\ref{fig:total_water_loss_over_t}, where all collisions in the regular scenarios are plotted.
One apparent difference is the practical absence of very high water losses in Fig.~\ref{fig:pot_hab_planets_water_loss_over_t}, which are presumably often either very destructive events (thus not directly part of planetary growth), or happen in the innermost part of the disk (see Sect.~\ref{sect:discussion_system-wide_collisional_evolution_water_transport}).
Otherwise there are no clear differences between either the giant planet architecture (cJS vs.\ eJS) or the planetesimal--planetesimal collision model (pp1 vs.\ pp0) in Fig.~\ref{fig:pot_hab_planets_water_loss_over_t}.
Water losses are mostly between 1 and 10\%, while very erosive events, and even losses significantly above 10\% are rare.
	\begin{figure}
	\centering{\includegraphics[width=8.7cm]{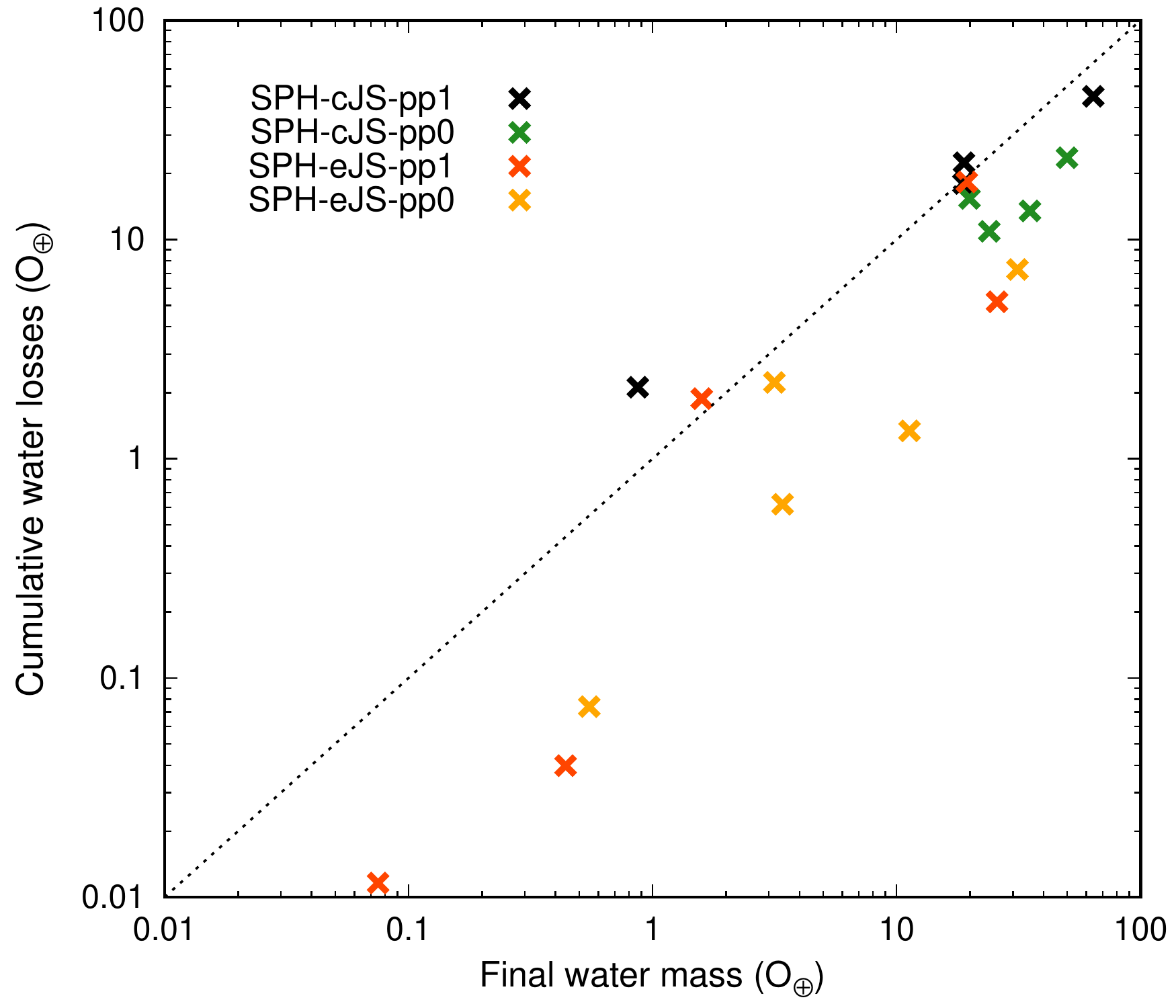}}
	\caption{Cumulative water losses (from their individual collision histories) vs.\ final water inventory for all 18 potentially habitable planets formed with our model in the regular scenarios. The diagonal line indicates cumulative losses equal to the final water content.}
	\label{fig:pot_hab_planets_cum_col_water_losses_over_water}
	\end{figure}
Knowing water losses in all collisions on the way to the final planet offers the possibility to compare their cumulative sum to the planet's final water inventory. This is listed in Table~\ref{tab:pot_hab_planets_collisions} (third column) and plotted in Fig.~\ref{fig:pot_hab_planets_cum_col_water_losses_over_water}.
Cumulative water losses are typically somewhat less, but make up a sizeable fraction of the final inventory, despite considerable scatter especially with eJS.
Considering also the histories of the impactors and associated losses (which are not included in Fig.~\ref{fig:pot_hab_planets_cum_col_water_losses_over_water} and Table~\ref{tab:pot_hab_planets_collisions}), a reasonable first-order estimate might be that total losses in the whole accretion tree of a planet are roughly equal to its final water content (the diagonal line in Fig.~\ref{fig:pot_hab_planets_cum_col_water_losses_over_water}).
For comparison with simple PM, the addition of cumulative losses likely still gives lower results, since in actual PM runs the whole impactor (including a potential second hit-and-run survivor) is always merged into the growing planet.
\changeslan{This is also confirmed by the theoretical growth curves in Figs.~\ref{fig:individual_planet_run100b_675} and \ref{fig:individual_planet_run63b_656}, where synthetic PM (intended to emulate actual PM outcomes) gives considerably larger water content (and mass) than merely adding cumulative losses (see Sect.~\ref{sect:example_pot_hab_planets}).}
We note that cumulative mass losses are only 10\,-\,20\% in the regular scenarios (Table~\ref{tab:pot_hab_planets_collisions}), while cumulative water losses are usually much higher, a clear indication that (surface) water is stripped more easily in collisions than other material.

In order to asses the actual efficiency of water delivery in individual collisions we use a modified version of the water accretion efficiency \citep{Burger2017_hydrodynamic_scaling},
\begin{equation}
\Xi_w =
\left\{
\begin{array}{lc}
\frac{\alpha}{M_\mathrm{w,p}} & \qquad \mathrm{if}\quad \alpha > 0 \\
\rightarrow 0 & \qquad \mathrm{if} \quad \alpha \rightarrow 0 \\
\frac{\alpha}{M_\mathrm{w,t}} & \qquad \mathrm{if}\quad \alpha < 0
\end{array}
\right.
\end{equation}
where $\alpha = M_\mathrm{w,tf} - M_\mathrm{w,t}$.
$M_\mathrm{w,t}$ and $M_\mathrm{w,p}$ denote the (absolute) pre-collision water masses on target and projectile, where `target' always means the body that is followed (even if it is the smaller one). $M_\mathrm{w,tf}$ denotes the mass of water on the post-collision body that originated from the target (relevant only for hit-and-run, because otherwise there is only one surviving body).
The interpretation of $\Xi_w$ is the following: If the target body has increased its (absolute) water inventory during a collision ($\alpha > 0$) then $\Xi_w \in (0,1]$ and the amount of increase in units of the (pre-collision) projectile inventory is given.
Conversely, if the water inventory of the target decreases ($\alpha < 0$) then $\Xi_w \in [-1,0)$ and the amount of decrease in units of the (pre-collision) target inventory is given.
For practically unchanged water contents $\Xi_w$ is close to zero.
This definition provides an easy-to-interpret measure of the efficiency of water accretion and simultaneously also erosion, in all cases, independent of whether both bodies are water-rich or if either of them is dry.
	\begin{figure}
	\centering{\includegraphics[width=8.8cm]{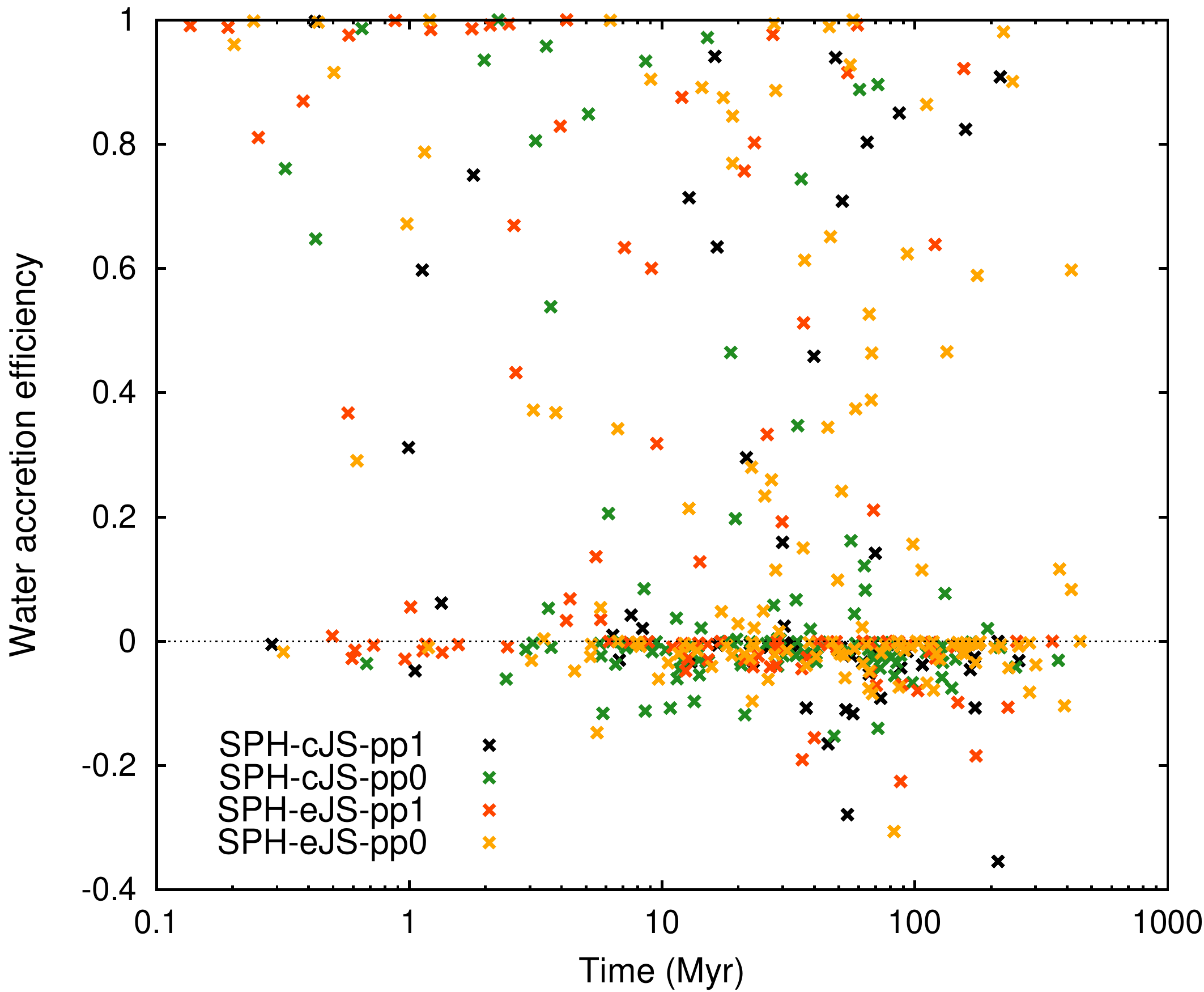}}
	\caption{Water accretion efficiency $\Xi_w$ in individual collisions over time during accretion of all 18 potentially habitable planets in the regular scenarios. Positive values indicate an increase in water inventory (in units of the projectile's water content), while negative ones indicate a decrease (in units of the target's water content). See the text for a precise definition of $\Xi_w$ and further discussion.}
	\label{fig:pot_hab_planets_XI_w_over_t}
	\end{figure}
A plot of $\Xi_w$ over time in Fig.~\ref{fig:pot_hab_planets_XI_w_over_t} shows that in the majority of collisions it remains close to or below zero, which indicates losses of water inventory, albeit only by a few percent in most cases.
Only a handful of collisions erode more than 10\% of the target's water content ($\Xi_w$ below -0.1), and not a single one erodes more than 40\%.
This is likely a simple consequence of the fact that impactors are mostly smaller than the growing planets, especially in their later evolution (cf.\ Figs.~\ref{fig:individual_planet_run100b_675} and \ref{fig:individual_planet_run63b_656}), and it is difficult for impactors to erode larger targets, even more so for hit-and-run \citep{Asphaug2010_Similar_sized_collisions_diversity_of_planets, Burger2018_Hit-and-run}.
A comparatively small fraction of collisions results in clearly positive $\Xi_w$, but these are distributed rather uniformly up to $\Xi_w = 1$.
Results by \citet{Burger2019_IAU_proceedings_water_delivery_to_dry_protoplanets_hit_and_run} indicate that the larger positive values are likely from accretionary collisions, while for hit-and-run they found values of $\sim$\,0.3 at most, and frequently considerably less.
Positive values of $\Xi_w$ occur not only in the few decisive water-delivering events (see Sects.~\ref{sect:accretion_water_delivery_pot_hab_planets} and \ref{sect:example_pot_hab_planets}), but also whenever more water is delivered than removed.
However, this becomes increasingly less likely for drier impactors, and it is the prevalence of significantly drier impactors in the growth histories of potentially habitable planets (clearly visible in Figs.~\ref{fig:individual_planet_run100b_675} and \ref{fig:individual_planet_run63b_656}) that cause $\Xi_w$ to be at least slightly negative in the majority of collisions.
We note that with simple PM, $\Xi_w$ is always 1, and none of this diverse behavior is modeled at all.

\subsection{Timing of accretion and water delivery to potentially habitable planets}
\label{sect:timing_accretion_water_delivery_pot_hab_planets}
Mass accretion timescales of potentially habitable planets are indicated by $t_\mathrm{50}$, $t_\mathrm{90}$ and $t_\mathrm{last-emb-col}$ in Table~\ref{tab:pot_hab_planets}, and for their water inventories by $t_\mathrm{25,w}$, $t_\mathrm{50,w}$, and $t_\mathrm{90,w}$.
These represent the accretion of 25, 50, and 90\% of their final mass and water content and the time of the last pure embryo\,/\,protoplanet collision (i.e., excluding planetesimals), and are all computed from the final simulation time backwards (not from t\,=\,0 forwards).
There appears to be no obvious strong difference between cJS and eJS in the regular scenarios, and they are also considerably lengthened by roughly a factor of two compared to PM, in concordance with accretion times of the entire systems (Sect.~\ref{sect:length_timing_of_accretion_phase}).
However, the formation of individual (potentially habitable) planets generally proceeds faster, which is not surprising since only a single object is considered, and additionally accretion in the inner part of the disk is generally faster than for planets farther out.
With our collision model it takes potentially habitable planets on the order of 10\,-\,50\,Myr to accumulate 50\% of their final mass, and 40\,-\,140\,Myr to accrete 90\%. In some cases $t_\mathrm{90}$ and $t_\mathrm{last-emb-col}$ coincide, but frequently there are still further collisions with other embryos or protoplanets ahead even after a planet has accreted almost all of its mass, presumably often (not very accretionary) hit-and-run encounters.
For comparison, PM simulations by \citet{Chambers2001_Making_more_terrestrial_planets} found that accretion of 50\% (90\%) of Earth analogs is completed in about 20 (50)\,Myr, and \citet{OBrien2006_Terrestrial_planet_formation_with_strong_dynamical_friction} found similar values, even though they considered all formed planets in the system.
In very high-resolution simulations, \citet{Fischer2014_Dynamics_terrestrial_planets_large_number_N-body_sims} found mean values of $\sim$\,60\,-\,80\,Myr for the last collision with an embryo or larger body for their habitable-zone planets.
These results are consistent with our PM runs (albeit often somewhat larger), but are considerably lower than those of our model.
There is a potential conflict especially with Hf-W measurements which seem to indicate shorter timescales for core formation on Earth.
However, this is presumably connected rather to the last accretionary giant collision, and the influence of large-scale hit-and-run interaction (as happens regularly with our collision model) is unclear.
\citet{OBrien2006_Terrestrial_planet_formation_with_strong_dynamical_friction} also found somewhat shorter accretion times if strong (and thus probably more realistic) dynamical friction is included, which would indicate that our results have not yet reached resolution convergence.

The water accretion times $t_\mathrm{25,w}$, $t_\mathrm{50,w}$ and $t_\mathrm{90,w}$ and particularly their frequent equality (Table~\ref{tab:pot_hab_planets}), mainly show what has been described in Sects.~\ref{sect:accretion_water_delivery_pot_hab_planets} and \ref{sect:example_pot_hab_planets} already, namely that the vast majority of the final planets' water inventory is delivered in a few -- often only one or two -- decisive collisions, while all other impactors deliver only small amounts, or remove water.
Radial mixing of water-rich material to the region of potentially habitable planets was found to take on the order of 10\,-\,30\,Myr (see second panel in Fig.~\ref{fig:water_masses_in_bins_over_t}).
This is confirmed by their accretion histories, where the decisive water-delivering collisions rarely happen prior to 10\,Myr, and often also much later, especially with eJS (see also Fig.~\ref{fig:total_mass_and_water_over_t}).
Exceptions are the planets whose accretion starts from a water-rich embryo (those formed in runs SPH03 and SPH08), which are water-rich from the beginning but nevertheless have to move inwards from their initial orbits.
Once accreted, the chance of long-term water retention is certainly a strong function of the planet's current mass, but also depends on its final mass, which determines how much collisional accretion (or erosion) is still ahead.
Our results indicate that the bulk of final water inventories is often delivered late, between $t_\mathrm{50}$ and $t_\mathrm{90}$ or after $t_\mathrm{90}$ (see Table~\ref{tab:pot_hab_planets}).
Delivery after $t_\mathrm{90}$ seems more frequent in eJS scenarios than with cJS, and water retention is therefore more likely.

	\begin{figure}
	\centering{\includegraphics[width=8.8cm]{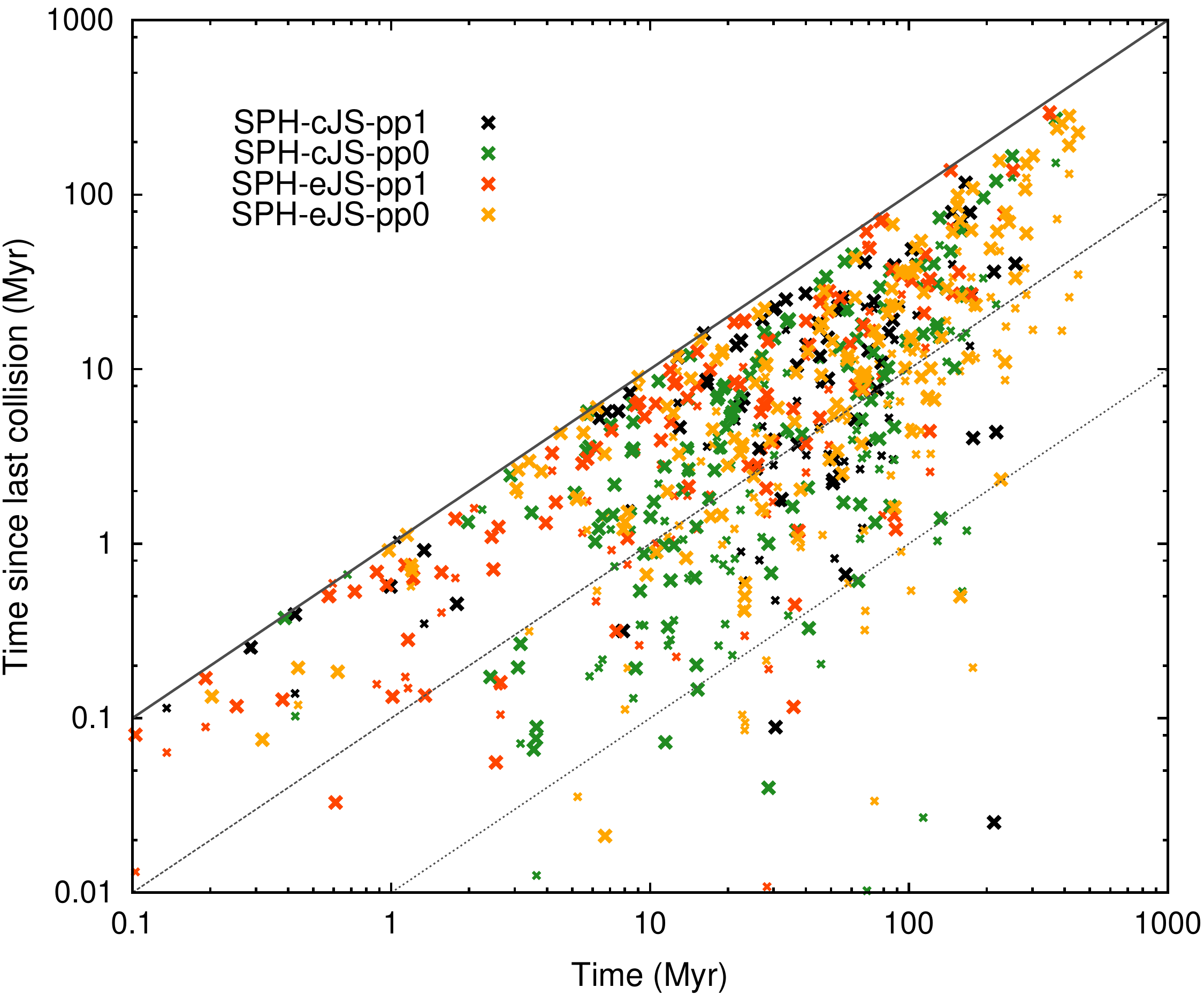}}
	\caption{Time since the last collision (evaluated each time a collision happens) from the perspective of a growing planet as a function of time, for all 18 potentially habitable planets in the regular scenarios with our model. Small crosses count all collisions, large ones only impacts by embryos and protoplanets (i.e., excluding planetesimals). Each collision is assigned two data points in general -- a small and a large cross -- and if the last impacting body was an embryo or protoplanet then they coincide. The continuous and dashed lines indicate elapsed time (since the last collision): since simulation start (upper line), 1/10 of that time (middle line), and 1/100 of that time (lower line).}
	\label{fig:pot_hab_planets_time_since_last_col_over_t}
	\end{figure}
Another aspect of the timing of collisions is the intervals between those events for a growing planet, especially impacts by large bodies (embryos or protoplanets).
The elapsed time since the last collision is plotted in Fig.~\ref{fig:pot_hab_planets_time_since_last_col_over_t} for the (direct) growth paths of potentially habitable planets.
Giant collisions with embryos or protoplanets (large crosses) generally deposit enough energy to not only vaporize surface water and ice, but can also produce local or even global magma oceans.
Therefore, they determine the physical state of water for typically a couple of Myr (until re-solidification of the surface at least), and thus not only the evolution until then but also how subsequent impactors encounter the growing planet, depending on the elapsed time since the last large event (see Sect.~\ref{sect:discussion_physical_state_water_further_losses} for further discussion).
The average time since the last collision naturally increases as planets grow and body numbers decrease.
The diagonal lines in Fig.~\ref{fig:pot_hab_planets_time_since_last_col_over_t} indicate that intervals between collisions are almost always larger than 1/100 of the current system age (above lower line) and for the majority at least 1/10 of that age (above middle line).
Typical intervals between collisions are therefore above 1\,Myr at t\,=\,10\,Myr, and already above 10\,Myr at t\,=\,100\,Myr.
The relative scarcity of planetesimal impacts (small crosses not over-plotted by large ones in Fig.~\ref{fig:pot_hab_planets_time_since_last_col_over_t}) is essentially a resolution effect, where a better-resolved planetesimal population would certainly lead to an increased number of planetesimal impacts between collisions involving only protoplanets and embryos.
A closer look reveals that in the very end phase of accretion, considerably more planetesimal impacts occur with pp0 (dark-green and orange) than with pp1 (black and red), which indicates that a sizeable small body population survives longer with pp0 (see also Fig.~\ref{fig:N_body_snapshots}).

\section{Discussion and conclusions}
\label{sect:discussion_and_conclusions}
The direct combination of long-term N-body integrations and dedicated hydrodynamical simulations of all occurring collisions is basically a simple and obvious approach, but has nevertheless been applied only very few times to self-consistently model the collisional evolution in N-body planet formation simulations \citep{Genda2011_Direct_coupling_N-body_SPH, Genda2017_Hybrid_code_Ejection_of_iron-bearing_fragments}.
Moreover, to our knowledge, this approach has never been used to also trace volatile constituents like water.
The obviously high computational demands may have been one reason, the intricacies involved in the realization of such a method might have been another.
Although it appears rather straightforward, we found it to be a formidable task to self-consistently implement the interface between the two worlds of N-body computations and SPH collision simulations, which resulted in more than 7\,000 lines of additional code, so far.

\subsection{System-wide collisional evolution and water transport}
\label{sect:discussion_system-wide_collisional_evolution_water_transport}
The results of our perfect merging (PM) comparison simulations are generally in agreement with existing studies.
In particular the influence of giant planets on eccentric orbits (eJS) leads to vastly reduced water contents compared to circular gas giants (cJS), where (system-wide) not more than a few tens of $O_\oplus$ survive with eJS, but $\sim$\,100\,-\,150\,$O_\oplus$ survive with cJS (Table~\ref{tab:systems}).
With eJS, strong dynamical forcing by the giant planets rapidly ($\sim$\,5\,-\,10\,Myr) removes the majority of water-rich material from the system, leaving only a few water-rich bodies (cf.\ Fig.~\ref{fig:N_body_snapshots}), which additionally renders final water content and distribution highly stochastic (with large scatter between scenarios).
With cJS, an initial removal of water-rich material occurs too, but a substantial amount remains, which is the main reason for the much larger final inventories in PM runs with cJS and makes them also more robust and less stochastic.
With our collision model on the other hand, final (system-wide) water contents with cJS are not much higher than with eJS (for both $\sim$\,30\,-\,40\,$O_\oplus$ on average), where mainly two mechanisms appear to further reduce water amounts with cJS (see also Sect.~\ref{sect:water_inventories_of_terrestrial_planet_systems}):
First, several tens of $O_\oplus$ are lost to collisional erosion, but only a fraction of that with eJS, presumably simply because eJS systems contain much less water after the initial removal phase, while there are no large differences in collision characteristics (especially impact velocities) between cJS and eJS (Sect.~\ref{sect:global_collision_characteristics}).
Second, our results indicate that the accretion phase is truly lengthened, by roughly a factor of two with realistic collision outcomes and especially accurate treatment of hit-and-run encounters (see Sect.~\ref{sect:length_timing_of_accretion_phase}, and also Sect.~\ref{sect:timing_accretion_water_delivery_pot_hab_planets}).
The associated slower growth allows for further removal of water from the system, considerably more so with cJS, presumably again due to higher abundances after the initial removal phase.

Unlike with cJS, the resolution (number of bodies) in the regular scenarios appears to be still too low to obtain robust results on water transport with eJS, at least for initial configurations like ours, where embryos have already also formed in the outer disk, and with a rather sharp transition between the water-poor inner and water-rich outer regions.
In contrast to that would be more smooth initial water distributions \citep{Quintana2014_Effect_of_planets_beyond_ice_line_on_volatile_accretion}, possibly as a result of earlier radial mixing \citep[e.g.,][]{Carter2015_Compositional_evolution_during_accretion}, which would likely also make results with eJS less stochastic.
Eventually a higher number and/or fraction of small bodies (planetesimals) might be required to model dynamical friction more realistically and better resolve the water-rich outer disk.
However, eJS simulations by \citet{OBrien2006_Terrestrial_planet_formation_with_strong_dynamical_friction} with such higher resolution found that even less water-rich material remains to later diffuse inwards (or remains as `stranded embryos') compared to our lower-resolution simulations.
This would suggest that a better-resolved planetesimal population can robustly model their radial mixing and water delivery behavior, but inwards water transport by embryos (if they form in the outer disk) ultimately remains a stochastic process in eJS-like environments.
We note that inward diffusion of water-rich material itself is likely a robust process dominated by interaction with the debris disk and not the giant planets, where water is reliably transported inwards provided it is not removed before. The same behavior has also been observed without any giant planets at all \citep{Quintana2014_Effect_of_planets_beyond_ice_line_on_volatile_accretion}.
In embryo-only scenarios the quick removal of water-rich material\footnote{\label{footnote:embryo_only_water}In the embryo-only scenarios, initially only six embryos comprise the outer, water-rich region (with WMFs of 5\%), where each embryo contains 53\,-\,73\,$O_\oplus$.} can have an even profounder influence, again more so with eJS, where the (chaotic) fate of individual water-rich embryos is often decisive.
At least in one scenario (EO-PM17-eJS, see Fig.~\ref{fig:final_planets_lowN}) all water-rich material is removed before delivering any water and both formed planets are composed exclusively of dry material from inside 2\,au.

A central part of our model is the accurate treatment of material and water transfer and loss in hit-and-run encounters, which are prevalent in collisions among embryos and protoplanets with a roughly 50\% occurrence rate (see Sect.~\ref{sect:global_collision_characteristics} and Fig.~\ref{fig:SPH_snapshots}).
The often-stated claim that the two hit-and-run survivors mostly remain on similar orbits and re-collide (and accrete) soon after the first encounter is not supported by our results (Sect.~\ref{sect:collision_characteristics_pot_hab}), and therefore accurate treatment of hit-and-run appears to play a decisive role in water transport and late-stage accretion in general.
Hit-and-run can be transformative especially for the smaller body of the colliding pair and strip it of mantle material and water \citep{Burger2018_Hit-and-run}, but still let the colliding bodies further evolve as individual objects.
Therefore, it has been suggested that unaccreted second-largest bodies might often be shaped by hit-and-run events with the largest body in the region \citep{Asphaug2010_Similar_sized_collisions_diversity_of_planets}.
This is not supported by the accretion histories of the 18 potentially habitable planets formed with our model in the regular runs, where merely once the growing planet is hit by a larger body, and only rarely by an approximately equally massive one (cf.\ Figs.~\ref{fig:individual_planet_run100b_675} and \ref{fig:individual_planet_run63b_656}).
However, these planets are often the largest formed (Fig.~\ref{fig:final_planets_medN}), so we also checked all other planets in those systems, with interesting results.
In most scenarios there is no collisional interaction at all between the planets in their later growth phases, that is, they form independently, and also mostly by impactors that are smaller or equally massive at most.
There are some instances where growing planets have hit-and-run encounters with larger bodies which are then removed from the system or accreted elsewhere, but very transformative hit-and-run collisions with larger impactors appear to be rare exceptions rather than the rule.
In 5 out of 20 scenarios growing protoplanets do collide late in their evolution but still end up as separate planets, but in all except for a single case they have very grazing (and thus not particularly transformative) hit-and-run encounters.
This happens for rather unsurprising configurations, like the two close inner planets in the SPH01 run (see Fig.~\ref{fig:final_planets_medN}), but also for planets that eventually end up on widely separated orbits, like in SPH02 or the middle and outer planets in SPH10.
In only one instance, the SPH07 scenario, is a smaller planet decisively shaped by survived (hit-and-run) collisions with a larger one, where the outer planet has two late hit-and-run encounters with the roughly three times more massive inner planet.
These collisions lead to considerable water losses on the more water-rich inner planet, but especially strip the outer planet of almost half its mass -- preferentially rocky mantle material -- and its CMF increases from 0.27 to 0.43.
This behavior might be a simple consequence of the rather narrow allowed parameter range for such encounters: Too grazing and they undergo no significant interaction, too central and they are more likely to merge and only a single planet forms instead of two (perhaps along with a smaller hit-and-run survivor). It is only in between, and with sufficient $v_\mathrm{imp}/v_\mathrm{esc}$, where highly transformative hit-and-run collisions are possible, which still leave two planet-sized objects. In addition, the protoplanets have to eventually achieve separated, long-time stable orbits again after those events.
While the mostly larger mass (gravity) of growing planets aids in water accretion from the very few water-rich impactors (in accretionary as well as hit-and-run collisions), this factor also hinders erosion of existing water by smaller impactors in hit-and-run \citep{Burger2018_Hit-and-run}.
In this sense, collisional water delivery is commonly more efficient than collisional erosion, especially during the later evolution when impactors are typically smaller than their targets.

A related observation is the high abundance of close-in (around 0.5\,au) low-mass planets with high CMF, which form with our collision model but not with PM. Examples can be easily identified, for example in all scenarios in the first row of Fig.~\ref{fig:final_planets_medN} (runs SPH01, SPH06, SPH11 and SPH16), and are formed in roughly half of all systems (see also first and fourth panels in Fig.~\ref{fig:final_planets_III}).
As evident from the discussion above, not individual mantle-stripping events (e.g., by neighboring planets), but rather a long sequence of energetic collisions gradually erode mantle material and increase their CMF, regularly up to 0.5 or more.
The dynamical environment close to the star leads to high relative velocities, and planets that experience long chains of often erosive collisions with several times $v_\mathrm{esc}$.
The impactors frequently also have high CMFs, which suggests that they have undergone erosive encounters of their own, and that altogether mutual erosion towards ever more iron-rich bodies is typical for that region of the disk.
This appears to be not only a possible natural explanation for Mercury's large CMF -- without the necessity of some individual chance event -- but also suggests that similar planets might be common in extrasolar systems.
We want to emphasize that our model was designed primarily to track collisional water transport and its limitations especially with respect to small-scale refractory debris (see Sect.~\ref{sect:discussion_collision_model}) make it hard to assess the accuracy of these results.
Nevertheless, the possibility that the outlined collisional evolution might be a regular and robust pathway to small, close-in, iron-rich planets certainly merits further investigation.

\subsection{Water delivery to potentially habitable planets}
\label{sect:discussion_water_delivery_pot_hab_planets}
A main focus of this study is collisional growth and water delivery to potentially habitable planets, which we define as lying between 0.75 and 1.5\,au.
In the regular scenarios and with cJS, potentially habitable planets with $\sim$\,1\,$M_\oplus$ and average water contents of $\sim$\,30\,$O_\oplus$ form with our model, while PM runs result in masses mostly above $\sim$\,1.5\,$M_\oplus$ and roughly doubled water inventories.
Therefore, even with realistic collision treatment these planets end up still more water-rich than Earth, but are considerably closer than the very high final water contents in PM simulations.
In the respective eJS scenarios, potentially habitable planets appear more Earth-like on average, with $\sim$\,1\,$M_\oplus$ and water contents of $\sim$\,10\,$O_\oplus$, but substantial scatter, typical for our eJS scenarios as discussed above (Sect.~\ref{sect:discussion_system-wide_collisional_evolution_water_transport}).
Nevertheless, individual planets formed in the eJS runs may represent viable pathways to Earth-like planets with suitable water contents.

Our results indicate that the great majority of delivered water arrives by a few -- often only one or two -- decisive collisions, and that those are mostly encounters with other large bodies (embryos and protoplanets).
The picture of (more or less direct) delivery by water-rich planetesimals scattered inwards from the outer disk (beyond the ice line) appears to be unsupported.
Small bodies are rather either removed from the system or accreted by growing, inwards-diffusing protoplanets, which then in turn deliver water to growing planets.
There are particularly two caveats, both with respect to our initial conditions: first, there is initially already more (water) mass in embryos than in planetesimals ($\sim$\,75 vs.\ 25\,\%) -- mainly due to computational constraints -- and second, we assume that embryos have already formed uniformly throughout the whole disk at the start of simulations.
With similar initial conditions like ours and an architecture similar to cJS (but PM), \citet{Raymond2007_High_res_sims_planet_formation_II} found also that the majority (>\,90\%) of water is delivered by larger bodies to planets in the habitable zone.

Based on our results we propose the hypothesis that late-stage water delivery to forming planets, particularly in the habitable zone, is a stochastic process, dominated by a few decisive events instead of a more continuous accretion of water-rich material, at least if embryos have already formed throughout the disk.
The water-delivering impactors are most often at least embryo-sized or larger, and are only rarely small bodies (planetesimals).
Such similar-sized collisions are naturally sometimes in the accretionary and sometimes in the hit-and-run regime, where the typically low efficiency of water transfer in hit-and-run strongly limits the amount of delivered water \citep{Burger2018_Hit-and-run, Burger2019_IAU_proceedings_water_delivery_to_dry_protoplanets_hit_and_run}, and makes accurate modeling thereof particularly important.
The results of \citet{Burger2019_IAU_proceedings_water_delivery_to_dry_protoplanets_hit_and_run} show that in relatively gentle head-on collisions water accretion efficiencies ($\Xi_w$, see Sect.~\ref{sect:collision_characteristics_pot_hab}) can be up to 0.8 (for $v/v_\mathrm{esc}$\,=\,1.5), while for hit-and-run they are less than half of that at most, and often considerably less.
We note that for simple PM, $\Xi_w$ is always 1, and all of the impactor's water inventory is delivered.
After hit-and-run, the second survivor can in principle strike again and deliver more water, as seen for example during accretion of the planet in Fig.~\ref{fig:individual_planet_run100b_675} (see Sect.~\ref{sect:example_pot_hab_planets}), but our results indicate that this does not happen in the majority of cases (Sect.~\ref{sect:collision_characteristics_pot_hab}).
If true, the above outlined hypothesis implies high scatter in water contents when formed under such conditions (cf.\ Table~\ref{tab:pot_hab_planets}), where a few chance events can make the difference between practically dry, moderately water-rich (like Earth), or `ocean planets' \citep[see also][]{Morbidelli2000_Source_regions_for_water}.
The hypothesis can be tested, particularly with a better-resolved planetesimal population (increased numbers and larger initial fraction of mass and water in planetesimals).
In PM simulations with such initial conditions (and $\sim$\,cJS), \citet{OBrien2006_Terrestrial_planet_formation_with_strong_dynamical_friction} found that still only a minority of accreted outer disk material comes from planetesimals, while \citet{Fischer2014_Dynamics_terrestrial_planets_large_number_N-body_sims} find (with cJS and eJS) that more than half of the water-rich material accreted by habitable-zone planets originates in planetesimals (but they do not specify how much is accreted by other embryos or protoplanets before final delivery).
Under the assumption that embryos only form in the inner disk, while the outer disk is initially populated exclusively by planetesimals, \citet{Raymond2007_High_res_sims_planet_formation_II} found (with $\sim$\,cJS) that a considerable amount of water is delivered by planetesimals, a less stochastic mechanism.
\citet{Izidoro2013_Compound_model_origin_water} on the other hand report that only a small fraction of planetesimals from such an outer disk are incorporated into final planets (the rest are presumably ejected or otherwise lost), but they do not specify how many of those are first accreted by other embryos or protoplanets.
It would be interesting to test and compare all these results with realistic collision outcomes instead of simple PM.
%In a dynamically different `Grand Tack' scenario, \citet{OBrien2014_Water_delivery_and_giant_impacts_in_Grand_Tack} found that only few \% of the final planet mass comes from water-rich planetesimals (from beyond Jupiter's orbit initially), and results by \citet{Carter2015_Compositional_evolution_during_accretion} indicate $\sim$\,3\,-\,8\,\% inwards-transported planetesimals from beyond 2.5\,au.

Two of the 18 potentially habitable planets formed in the regular scenarios with our model start accretion as water-rich embryos from beyond 2.5\,au with several tens of $O_\oplus$, then migrate inwards during accretion, and eventually end up at 1.1 and 1.2\,au (in runs SPH03 and SPH08).
Both planets form with cJS, presumably because the chances of such an evolution are simply much lower if the majority of water-rich material is rapidly removed (with eJS) before it can set in.
During the growth of both planets they accrete additional water-rich material, but also experience considerable collisional losses at later times (see Fig.~\ref{fig:pot_hab_planets_over_t}).
However, collisional erosion is difficult once protoplanets have grown to a certain mass and most impactors are significantly smaller \citep{Asphaug2010_Similar_sized_collisions_diversity_of_planets, Burger2018_Hit-and-run}.
Indeed these two planets end up as the most water-rich of the 18 potentially habitable planets -- essentially `ocean planets' with 64 and 50\,$O_\oplus$.
These two examples indicate that if sufficiently large bodies can form in the water-rich outer disk, and the environment is dynamically calm enough, then there is a chance that they will diffuse inwards and retain most of their water inventory despite collisional erosion, or even accrete more on the way -- a viable pathway to very water-rich Earth-sized planets in the inner system.
There is also an interesting connection to PM, where water-rich bodies are always entirely accreted, including their whole water inventory. In some sense this is also the case for these planets, where the majority of the water of their seed embryos is retained.
Their final water contents are indeed relatively close to results of PM runs (Table~\ref{tab:pot_hab_planets}).
Planets that accrete dry until later water delivery on the other hand, receive only a (often small) fraction of the impactors' water inventories with realistic collision treatment, much unlike PM.

Especially with respect to previous work on water delivery during late-stage accretion, questions pertaining to differences between our model and PM and the amount of collisional water loss present themselves.
For eJS scenarios, a direct comparison is not very meaningful due to the very stochastic evolution (but see Tables~\ref{tab:systems} and \ref{tab:pot_hab_planets}).
For cJS, system-wide water contents end up (on average) between three and four times lower than with PM, and for individual potentially habitable planets the difference is still a factor of about two.
Including realistic collision outcomes changes the process of late-stage accretion itself to some degree, which complicates comparison with PM, where frequent hit-and-run encounters lengthen growth times and can lead to complex, sometimes repeated, interaction among protoplanets.
Moreover, high-energy and often erosive collisions can lead to results strongly deviating from PM, especially in the inner disk due to high relative velocities.
To assess the deviations caused by not accounting for realistic collision outcomes we also computed `synthetic PM' accretion tracks from the actual results of our model (Sect.~\ref{sect:example_pot_hab_planets}, and especially Figs.~\ref{fig:individual_planet_run100b_675} and \ref{fig:individual_planet_run63b_656}), which appear to be quite reliable indicators.
In addition, we found that cumulative collisional water losses are mostly somewhat below final inventories (Sects.~\ref{sect:example_pot_hab_planets} and \ref{sect:collision_characteristics_pot_hab}, and especially Fig.~\ref{fig:pot_hab_planets_cum_col_water_losses_over_water}), but considering also the accretion histories of all impactors probably leads to typical losses on the order of the final water content.
Collecting all the evidence together leads to the conclusion that collisional water losses likely reduce final inventories of potentially habitable planets by a factor of two or more in most situations.

\subsection{Initial conditions}
\label{sect:discussion_initial_conditions}
Our initial conditions are intended to represent the onset of late-stage accretion in the terrestrial planet region, once previous growth has led to a system of planetary embryos embedded in a large number of remaining smaller bodies.
We assume that the gas component has already dispersed before larger protoplanets are able to form, which is still an open question also for the early Solar System.
Our giant planet architectures (cJS\,/\,eJS) are in contrast to migrating gas giants, which can influence debris disks to an even greater degree \citep[as in Grand Tack scenarios;][]{Walsh2011_Grand_Tack, OBrien2014_Water_delivery_and_giant_impacts_in_Grand_Tack}, and giant planets that run into dynamical instabilities.
Despite the great variety of known extrasolar systems, similar evolutionary paths might still be frequent.
The initial amount and distribution of water, which strongly depends on the composition and conditions in the protoplanetary disk, is generally not well constrained, where meteorite data provide some information at least for the early Solar System.
Our initial conditions generally follow what is often referred to as dry accretion, that is, no significant amounts of water are incorporated locally into building blocks in the inner disk, and we also assume that no earlier radial mixing has smoothed out the water distribution \citep[see, e.g.,][]{Carter2015_Compositional_evolution_during_accretion}.
Building blocks beyond the ice line on the other hand could be even more water-rich than our 5\% \citep[see][and references therein]{OBrien2018_The_Delivery_of_Water_During_Terrestrial_Planet_Formation}, where for example 10\% is not unreasonably high \citep[for carbonaceous condrites,][]{Morbidelli2000_Source_regions_for_water} and could have a large impact on final water inventories.
In addition, the available budget of radionuclides (e.g., $^{26}$Al) can alter initial water contents over a wide range \citep{Lichtenberg2019_water_budget_dichotomy_from_26Al_heating}.
Even though we chose to limit this study to one particular initial water distribution, our collision model provides the straightforward possibility to apply arbitrary distributions as a simple post-processing step to whole simulations, for example models for local initial water absorption \citep{Izidoro2013_Compound_model_origin_water} or smoothly varying WMF \citep{Quintana2014_Effect_of_planets_beyond_ice_line_on_volatile_accretion}.
The mechanism for that is already applied in the frequently necessary pre-collision upscaling and post-collision correction of water contents in order to reasonably resolve them in SPH simulations (see Sect.~\ref{sect:SPH_numerics}).

\begin{figure}
	\centering{\includegraphics[width=8.8cm]{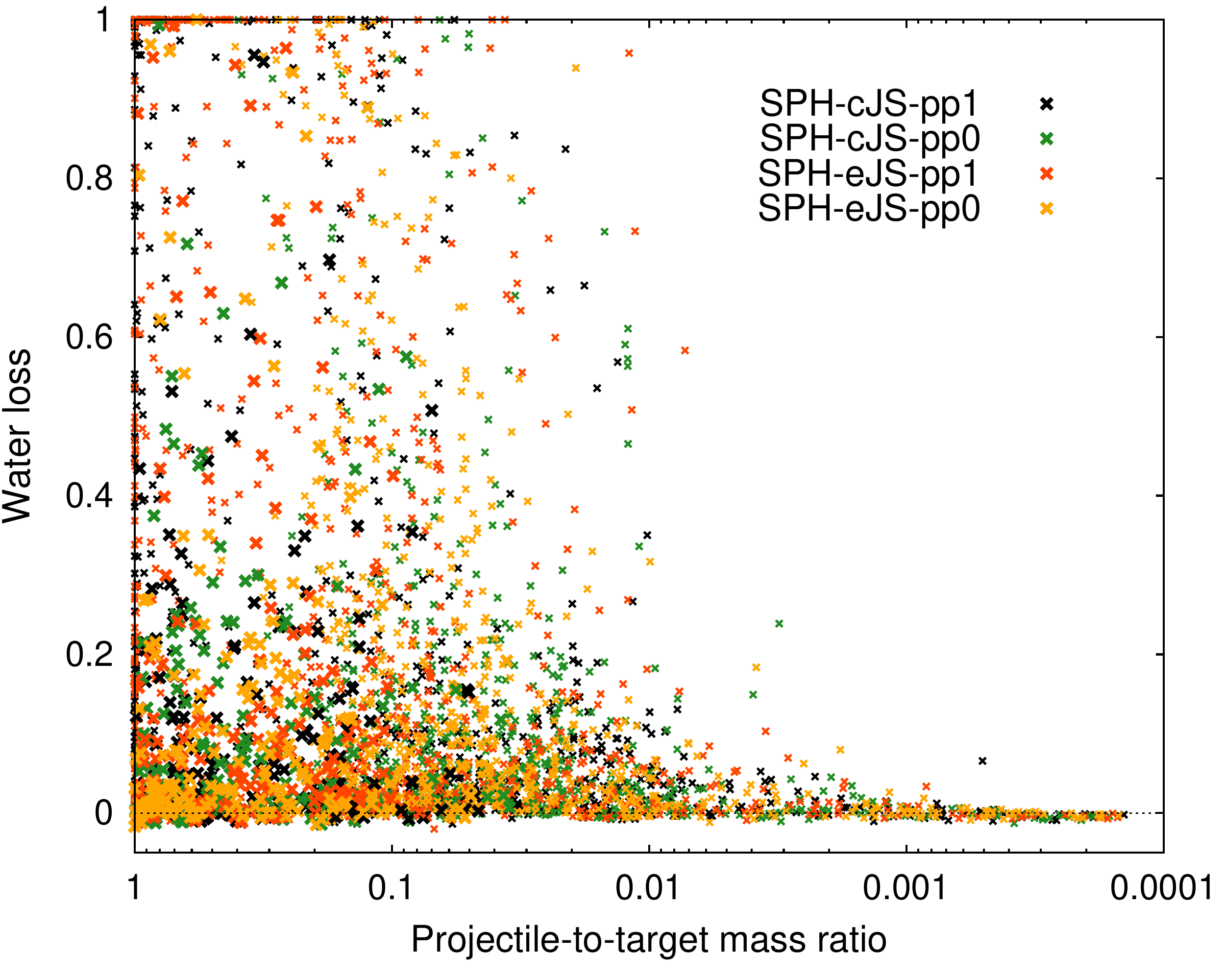}}
	\caption{Water losses in individual collisions as a function of the projectile-to-target mass ratio (where the target is always the larger body), for all regular scenarios with our model. Water loss is defined as the fraction of the (combined) pre-collision water inventory that is not incorporated into either of the post-collision bodies. Larger symbols indicate collisions among embryos and protoplanets alone (i.e., excluding planetesimals). Slightly negative values are merely numerical artifacts and can be assumed to be zero.}
	\label{fig:total_water_loss_over_mass_ratio}
	\end{figure}

Due to naturally limited computational resources our simulations also suffer the common problem of mostly orders-of-magnitude too large and massive planetesimals.
With our model this additionally affects collisions, which typically involve bodies that are considerably too massive (in planetesimal--planetesimal collisions and planetesimal impacts onto larger objects), while their frequency is simultaneously too low.
In some sense this can be considered as sampling only the top most part of the unknown planetesimal mass-distribution and neglecting (collisional) interaction with a large number of smaller bodies, except for the most top-heavy distributions.
It is difficult to assess the consequences of an impactor flux composed additionally of a large number of small-to-medium-sized bodies instead of exclusively high-mass planetesimals.
Particularly for water (and other volatiles), its location and physical state are likely important factors, especially when the bulk of a body's water inventory is vaporized, like after a large collision (further discussed in Sect.~\ref{sect:discussion_physical_state_water_further_losses}).
For impact erosion of (proto)planetary atmospheres, smaller bodies have been found to be much more efficient (per unit mass) than larger ones \citep[see, e.g.,][]{Schlichting2015_Atmospheric_mass_loss_impacts, Schlichting2018_Atmosphere_Impact_Losses, Lammer2019_atmosphere_paper}.
This is mainly because most of their impact energy ejects atmosphere locally, while for larger impactors strong shocks have to traverse the target to eject atmosphere globally.
In this study we performed a large number of more generic (e.g., no atmospheric component) collision simulations, where water is modeled as surface fluid, and evaluated water losses in all these events.
This gives a quite robust sample, covering mass ratios down to $\sim$\,$10^{-3}$, even though the accuracy towards the lowest mass ratios is likely limited by SPH resolution.
The results are plotted in Fig.~\ref{fig:total_water_loss_over_mass_ratio} and appear to support the notion that smaller impactors (per unit mass) lead to higher water losses, even though this is difficult to quantify precisely.
We note that part of this effect could be due to hit-and-run, which becomes more likely in more similar-sized collisions but usually results in lower losses than accretionary or erosive events.
Our results for water delivery to potentially habitable planets suggest that only a small number of decisive collisions with other large bodies are responsible for delivery of the great majority of their final inventories, at least in their immediate collision history (Sect.~\ref{sect:discussion_water_delivery_pot_hab_planets}).
If this hypothesis proves to be correct, then at least for this aspect of water delivery, small bodies and their actual distribution would be of only little consequence.
Ultimately a better resolution and modeling \citep[e.g., via tracer particles;][]{Levison2012_Lagrangian_integrator_planet_formation} of planetesimals is required, which ideally also samples their mass distribution to some degree (instead of uniform masses).
More realistic initial conditions will be a central aspect of future applications of our hybrid framework (see Sect.~\ref{sect:discussion_future_work}).

Besides the regular scenarios, we performed embryo-only simulations where all mass is assumed to have been initially accreted into approximately 30 planetary embryos in order to compare different resolutions and to study the influence of the planetesimal population. Similar setups have been used in the few existing direct hybrid simulations which are comparable to our approach \citep{Genda2011_Direct_coupling_N-body_SPH, Genda2017_Hybrid_code_Ejection_of_iron-bearing_fragments}.
A central aspect of the dynamical evolution in these scenarios is missing dynamical friction by smaller bodies, which leads to generally more excited populations of embryos and protoplanets, and presumably more energetic collisions, eventually leading to planets on often more eccentric orbits.
Additionally, water transport depends exclusively on the evolution of only a few water-rich embryos, which leads to more stochastic water delivery in general, and particularly with eJS (see Sect.~\ref{sect:water_inventories_of_terrestrial_planet_systems}).
Altogether the embryo-only simulations show the same general trends for masses, water contents, collisional losses, and accretion timescales (cf.\ Tabs.~\ref{tab:systems}, \ref{tab:pot_hab_planets} and \ref{tab:pot_hab_planets_collisions}), even though absolute values deviate sometimes considerably and outcomes have higher scatter.
Therefore, results from such scenarios are mostly of limited accuracy, but could nevertheless be useful, for example in parameter studies in a less stochastic environment (like cJS), and with a focus on basic properties and fast computation with run-times of a few weeks instead of several months.

\subsection{The collision model}
\label{sect:discussion_collision_model}
Our physical model aims for realistic treatment of collisional water transport, which means simulation of the dynamical (long-term) evolution with high accuracy and collision of bodies only in actual physical encounters (including tidal collisions to some degree).
Our approach of dedicated SPH simulations has the great advantage of accurate and reliable collision outcomes, and in principle allows the simulation of all desired physics for each event.
Particularly for the application in this study, it is also well-suited (and currently without alternative) for modeling the rather subtle effects of collisional water losses and also transfer in hit-and-run encounters.
The frequently used semi-analytical collision outcome model by \citet{Leinhardt2012_Collisions_between_gravity-dominated_bodies_I, Leinhardt2015_Numerically_predicted_signatures_of_TPF} for example includes scaling laws for both hit-and-run survivors, and also an approximate treatment of changes in (bulk) composition, but appears to be unable to accurately predict collisional water transfer and loss \citep{Burger2018_Hit-and-run}.
It would certainly be a formidable task to derive reasonably precise scaling relations for these subtle processes.
The drawbacks of our approach are increased computational expense and complexity.
Computational limitations make it necessary to allow at most two post-collision bodies (or gravitationally bound aggregates in our model), and to ignore remaining debris.
The error caused by this approach is however limited by the fact that gravity-dominated collisions among similar-sized objects typically result in either \changeslan{0, 1, or 2} large post-collision bodies along with orders-of-magnitude smaller debris, but practically never in three or more.
We are mostly interested in water transport and it may be safe to assume that the water content of these debris fragments will be frequently vaporized after a sufficiently energetic collision, and subsequently lost quickly from small parent bodies \citep[e.g.,][]{Odert2018_Escape_and_fractionation_of_volatiles_from_growing_protoplanets}.
Closer to the star (as around 1\,au), exposed water on small collision fragments is also not long-time stable.
The refractory component of the debris (silicates and iron) on the other hand, can be assumed to further participate in the evolution of the system, and neglecting this material is one of the shortcomings of our model in its current state.
Such `losses' from all $\sim$\,150\,-\,200 collisions in each regular scenario typically amount to $\sim$\,1\,-\,1.5\,$M_\oplus$ -- mostly mantle material with only a relatively small iron fraction of $\sim$\,0.1 (see Table~\ref{tab:systems} and Sect.~\ref{sect:fate_of_building_material}).
Besides increasing the level of dynamical friction and thus eccentricity damping \citep{Genda2017_Hybrid_code_Ejection_of_iron-bearing_fragments}, it might be accreted at some later point and/or lead to further collisional water losses, even though these small bodies can be easily excited and may often be removed from the system.

We applied two distinct models for planetesimal--planetesimal collisions -- either full collisional interaction (pp1), or none at all (pp0).
The main motivation for performing also pp0 simulations was to investigate whether the large masses (and low numbers) of planetesimals influence especially the collisional evolution in a priori unknown ways.
An additional reason was to increase computational speed, especially by omitting close encounters among planetesimals with pp0. However, it turned out that those gains are largely offset by slower decrease in the number of bodies due to lack of planetesimal--planetesimal accretion (cf.\ Fig.~\ref{fig:body_numbers_over_t}) or removal due to mutual scattering \citep{Raymond2006_High_res_sims_planet_formation_I}.
For the most part, our results indicate no obvious significant differences in the main outcome figures between pp1 and pp0 scenarios, even though it is often difficult to distinguish systematic from stochastic variations.
One exception is the longer prevalence of a sizeable planetesimal population with pp0, while with pp1 accretion into embryo-sized or larger bodies proceeds faster (visible, e.g., in Fig.~\ref{fig:pot_hab_planets_time_since_last_col_over_t}, Sect.~\ref{sect:timing_accretion_water_delivery_pot_hab_planets}).
Collision statistics also indicate that events with the highest $v_\mathrm{imp}/v_\mathrm{esc}$ (the most destructive) occur in pp1 scenarios (Fig.~\ref{fig:v_vesc_over_t}), presumably planetesimal--planetesimal collisions which are not possible with pp0.
Those events also dominate the region of highest collisional water losses (Fig.~\ref{fig:total_water_loss_over_t}), but could also be collisions of rather dry material in the inner disk, where orbital dynamics make them more likely. In this case the influence on final water budgets would actually be very low, but perhaps not on mantle stripping (see Sect.~\ref{sect:discussion_system-wide_collisional_evolution_water_transport}).
In principle full planetesimal--planetesimal interaction (pp1) appears preferable, but the gain in realism from including their collisional evolution has to be considered in light of regularly unrealistically high masses.
In addition, treatment with pp1 also includes dynamical planetesimal--planetesimal interaction (at least in close encounters), which was found to be an important part of general dynamical friction \citep{Raymond2006_High_res_sims_planet_formation_I}.
However, to identify the influence of these subtle effects a better resolved planetesimal population and better statistics (more simulations) would be required.

\subsection{Accuracy of collision simulations}
\label{sect:discussion_accuracy_collision_sims}
The question of how accurately our SPH simulations predict collision outcomes has two distinct aspects.
First, there is the uncertainty of the actual physical state of water reservoirs immediately before the collision, which would require detailed modeling of the internal, atmospheric, and environmental evolution (and coupling between them) for all bodies in the intervals between collisions (see Sect.~\ref{sect:discussion_physical_state_water_further_losses}).
Although a formidable task, we believe that including at least the most influential of these processes should be a future direction to extend our model.
Due to the lack of this information, water is modeled as strengthless (surface) fluid, which we believe is the best compromise and is intended to represent roughly mean material properties between the limits of solid ice and hot steam.
\citet{Burger2017_hydrodynamic_scaling} and \citet{Burger2018_Hit-and-run} studied differences in collision outcomes between purely hydrodynamical and full solid-body material models, including icy bodies, from planetesimal- to planet-sized.
For the mass range relevant here, these authors found considerable differences in water transfer and loss up to approximately Mars-sized bodies, beyond which strength effects become negligible.
Final WMFs and losses differed by less than 20\%, but deviations of up to 50\% were found for the amount of transferred water from projectile to target and vice versa.
Including material strength resulted in consistently lower transfer rates, presumably due to the greater amount of energy required to overcome tensile and shear forces.
Differences in transfer and loss for vaporized water in the atmosphere versus liquid or solid reservoirs on or below the surface have not been studied so far to our knowledge.
Without considering any surface volatiles in particular, \citet{Jutzi2015_pressure_dependent_failure_models} and \citet{Emsenhuber2018_Mars-scale_collisions_eos_material_rheologies} also found that rheology can affect collision outcomes up to approximately Mars-sized bodies, where for this size scale however, only subtle effects (for our purposes) like differences in the post-impact temperature distribution were observed.

Second, accuracy of collision outcomes depends on the numerical resolution \citep[see, e.g.,][]{Genda2015_Resolution_dependence_of_collisions}, where
our approach naturally allows only relatively small SPH particle numbers (between 25k and 75k, see Sect.~\ref{sect:SPH_numerics}).
\citet{Burger2018_Hit-and-run} studied resolution dependence in planet-sized collisions of water-rich bodies, including changes in WMF of the largest and second-largest survivor and also water transfer from projectile to target and vice versa, and thus all important aspects relevant to this work. They found that results for all these quantities were within 25\% of approximate resolution convergence, and below 10\% for the post-collision WMF of the target body alone.
One issue directly connected to limited resolution is the modeling of low WMF, which we accomplish by increasing the water layer thickness to a minimum of approximately three SPH particles during the collision simulation, before its outcome is then corrected for this artificial increase (Sect.~\ref{sect:SPH_numerics}).
The error estimated for this procedure is also $\sim$\,25\% or less, including water transfer in hit-and-run \citep{Burger2018_Hit-and-run}.
These combined uncertainties could in principle mean that individual SPH results for water transfer and loss are relatively rough estimates.
However, we believe that in the majority of cases outcomes are fairly robust, and that typically large collision numbers combined with the stochastic nature of accretion tracks means that global results and final planets are too.

Part of our numerical scheme is to run SPH simulations for considerably longer than the immediate collision aftermaths in order for all post-collision fragments to form and clearly separate.
While this works as intended for the vast majority, `graze-and-merge' encounters are particularly demanding, where the colliding bodies loose sufficient energy in a gentle hit-and-run to remain gravitationally bound, and usually immediately re-collide (after one orbit about each other).
The time up to this re-collision depends on the specific geometry and is hard to predict.
Therefore, in a few cases SPH simulations may stop before re-collision, or not continue long enough for the final body to settle.
The latter happens to some degree in the collision illustrated in Fig.~\ref{fig:SPH_snapshots}, where towards the end a tail of material forms (due to fast rotation induced by the re-collision) whose fate is not simulated anymore, but has certainly only minor consequences there.
If the simulation stops while the bodies are still approaching re-collision, they are simply merged (because they are gravitationally bound). However, since the re-collision is necessarily rather gentle (below $v_\mathrm{esc}$), the error can be expected to be small in general.

\subsection{The physical state of water and further losses}
\label{sect:discussion_physical_state_water_further_losses}
In our model we currently combine long-term dynamics and realistic collision outcomes, but these are only two aspects of the evolution of water inventories.
Energy from short-lived radioactive isotopes, gravitational energy released by differentiation, and large collisions can all melt outer silicate layers, leading to local or even global magma oceans. Their evolution and lifetime is strongly coupled to the overlying (steam) atmosphere which can in turn be heavily modified by volatile outgassing.
In addition to direct impact erosion \citep{Genda2003_proto-atmosphere_giant_impacts, Genda2005_atmospheric_loss_giant_impact_phase, DeNiem2012_Atmospheric_erosion_and_replenishment_by_impacts, Schlichting2015_Atmospheric_mass_loss_impacts, Schlichting2018_Atmosphere_Impact_Losses}, vaporized water in an atmosphere is susceptible to hydrodynamic loss (escape of hydrogen from dissociated H$_2$O molecules) driven by high EUV radiation of the young star \citep[strongly depending, e.g., on its rotation state, see][and references therein]{Odert2018_Escape_and_fractionation_of_volatiles_from_growing_protoplanets}.
However, it was found that impact-induced magma oceans might cool and solidify again in $\sim$\,1\,Myr and that subsequent condensation of a dense (outgassed) steam atmosphere and ocean formation can also happen on similar timescales \citep{Lebrun2013_Magma_ocean_with_atmosphere_water_condensation}, mostly shorter than typical intervals between giant impacts.
\citet{Hamano2013_Oceans_after_solidification_of_magma_ocean} also found rapid (a few Myr) solidification and ocean formation beyond a critical distance from the star -- about 0.7\,au for the Sun. This is fast enough for hydrodynamic escape to play only a minor role.
However, their results also indicate that for (proto)planets closer in a very long magma ocean phase ($\sim$\,100\,Myr) can lead to loss of the majority of water via hydrodynamic escape, independent of further collisions.
\citet{Salvador2017_Influence_of_H20_CO2_on_water_evolution_of_rocky_planets} found that degassing and solidification of a post-impact magma ocean and subsequent water ocean formation also depend strongly on the initial H$_2$O (and CO$_2$) contents, and that rapid ocean formation is possible on protoplanets similar to early Earth, and possibly also early Venus.
While the bulk of such work was done for basically already fully grown planets (typically Earth and Venus), \citet{Odert2018_Escape_and_fractionation_of_volatiles_from_growing_protoplanets} studied magma ocean evolution and atmospheric losses for Mars-sized embryos at different orbital distances, atmosphere masses, and (EUV) radiation environments.
The results of these latter authors indicate that at the orbit of Mars the time for steam-atmosphere condensation (e.g., after a giant collision) is very small compared to the timescale of atmospheric escape, and almost all post-collision water content can be retained.
At Earth's orbit, typically a relatively small fraction of water is lost before it can condense into oceans, while at the orbit of Venus virtually all atmosphere is expected to be lost before condensation is able to stop it for these small bodies.
For even lower masses ($\lesssim$ Moon-mass) -- like the collisional debris in our SPH outcomes -- not only EUV-driven hydrodynamic escape but also thermal (Jeans) escape becomes important, which plays only a minor role for Mars-sized bodies and is negligible for Earth-mass objects \citep[e.g.,][]{Odert2018_Escape_and_fractionation_of_volatiles_from_growing_protoplanets}.
Depending on their (post-collision) temperature such small bodies can loose their atmospheres and water very rapidly.
In addition, the evolution after a large collision event is connected to the current (dynamical) environment, where continued energy influx by impacting bodies for example can also delay surface solidification and water condensation \citep{MaindlDvorakLammer2015_Impact_heating_on_early_Mars}.

Our results indicate that, at least for potentially habitable planets, water delivery (by the few decisive events) happens typically late in their accretion histories once they have grown to at least 50\% and often to over 90\% of their final mass (Sect.~\ref{sect:timing_accretion_water_delivery_pot_hab_planets}).
This increases the chances of water retention due to their already high gravity and also because fewer further (giant) collisions lie ahead.
We note however, that this means only the final water-delivering events, where the (mostly smaller) impactors themselves may have complicated collision histories of their own.
The (absolute) times when the bulk of water is delivered can vary strongly, with most between 10 and 100 Myr (Table~\ref{tab:pot_hab_planets}) when intervals between large collisions are typically already several Myr (Sect.~\ref{sect:timing_accretion_water_delivery_pot_hab_planets} and especially Fig.~\ref{fig:pot_hab_planets_time_since_last_col_over_t}), and in some cases also considerably later with typical intervals >\,10\,Myr.
Therefore, repeated ocean formation between (giant) collisions seems to be a plausible scenario, at least for growing planets in the (Sun's) habitable zone and beyond.
The prevalence of hit-and-run adds an additional aspect to these considerations.
Depending mainly on the impact angle, energy deposition can be greatly reduced compared to central (head-on) collisions, with possibly profound consequences for water vaporization \citep{Burger2018_Hit-and-run} and magma ocean formation (or the lack thereof) in ways that remain mostly unstudied.

\subsection{Future work}
\label{sect:discussion_future_work}
Besides the various foreseeable difficulties with other possible methods, for example reliably predicting all the subtleties of hit-and-run with scaling laws or machine learning techniques, we chose this direct computational approach because we believe that it is a solid basis for including further details and physical processes in future work.
The simulations presented here follow only the dynamical and (immediate) collisional evolution of water reservoirs, while at the end of the day it is the coupled dynamical, collisional, geophysical, and atmospherical evolution that shapes a planet's water inventory.
In that sense our current results are only upper limits for final water contents, with the modification roughly decreasing with increasing orbital distance, where collisions are less energetic (due to lower relative velocities) and magma ocean phases are shorter (due to lower stellar flux).
One possibility to incorporate further modeling without having to repeat whole simulations is to use already existing accretion histories of planets of interest as a backbone for modeling of additional physics and/or higher SPH resolution. This could include only a planet's direct accretion track, or those of all impactors (and bodies impacting them, etc.) as well.
To that end, we also offer precise data on the dynamical and collisional evolution of all planets formed in our scenarios to interested researchers (see footnote \ref{footnote:offer_our_data}, page \pageref{footnote:offer_our_data}).

A shortcoming of the simulations in this study is the relatively poorly resolved planetesimal population and we are already running higher-resolution simulations with half the initial mass in (significantly more) planetesimals.
In addition, embryo formation happens on different timescales depending on location, and presumably proceeds considerably faster in the inner part of the disk \citep{Kokubo2002_Formation_of_protoplanet_systems_oligarchic_growth, McNeil2005_Effects_of_typeI_migration_on_terrestrial_planet_formation}.
More realistic initial conditions beyond equal planetesimal masses and uniform embryo formation throughout the whole disk could be obtained from simulating the collisional evolution of a large number of small bodies over a relatively brief period with our hybrid framework \cite[see, e.g.,][]{Raymond2006_High_res_sims_planet_formation_I, Bonsor2015_Collisional_origin_to_Earths_non-chondritic_composition, Carter2015_Compositional_evolution_during_accretion}.
Alternatively, more realistic initial conditions and mass distribution can be based on semi-analytical models for earlier growth phases \citep[e.g.,][]{McNeil2005_Effects_of_typeI_migration_on_terrestrial_planet_formation}.

The first application in this study is in the classical Solar System formation environment, including giant planet formation without migration, but the diversity of detected extrasolar planets shows that this is only one possible path for planetary accretion.
We believe that realistic collision modeling can provide a better understanding of terrestrial planet formation in the great diversity of planetary systems, not least to constrain water contents of detected exoplanets for future observations and assessment of their principal habitability.

\section{Summary}
\label{sect:summary}
We have developed a new hybrid framework to combine the dynamical and collisional evolution of a debris disk, comprising planetary embryos and planetesimals, over several hundred Myr of late-stage terrestrial planet formation.
Each physical collision is evaluated by a dedicated 3D SPH simulation of differentiated, self-gravitating bodies, the results of which are used self-consistently on the fly in the further N-body computation.
The main focus is on collisional water transport, especially to potentially habitable planets, with volatile material like water being particularly susceptible to transfer and loss processes.
Collisions can result in either zero (very destructive), one (accretion or erosion), or two (hit-and-run) post-collision bodies, which allows us to fully track water transfer and loss in hit-and-run encounters as well.

In this first application we chose a Solar System-like architecture, including already formed nonmigrating giant planets and a debris disk between 0.5 and 4\,au.
This well-studied setting allows us to compare and validate our results against existing studies, where at least for water delivery perfect merging (PM) was assumed almost exclusively.
However, our main purpose is not to reproduce the inner Solar System in detail, but to study the importance and implications of realistic modeling of collisional water transport for the early Solar System, and also for planet formation in extrasolar systems.
Results of our PM comparison runs are in general agreement with existing studies, while realistic collision treatment typically leads to lower mass planets with considerably reduced water inventories (Fig.~\ref{fig:final_planets_medN}).
To our knowledge this is the first time that collisional water transport has been modeled self-consistently in combination with the dynamical evolution up to final planets.

Our results have revealed several key implications for water transport, and also for terrestrial planet formation in general:
\begin{itemize}
\item We find that the accretion phase is truly lengthened by up to a factor of two with our collision model compared to PM runs, often to 100\,-\,200\,Myr or more (Sects.~\ref{sect:length_timing_of_accretion_phase} and \ref{sect:timing_accretion_water_delivery_pot_hab_planets}). This is signified by slower mass growth (Fig.~\ref{fig:pot_hab_planets_over_t}), considerably slower decrease in the number of bodies, also for embryo-sized or larger bodies alone (Figs.~\ref{fig:body_numbers_over_t} and \ref{fig:embryo_numbers_over_t}), and significantly later final giant collisions (involving embryos or larger bodies).

\item As also found in previous PM studies, eccentric giant planets (eJS) are detrimental to water delivery due to fast removal of water-rich material from the outer disk by strong resonances, while with circular giant planets (cJS) much more water-rich material initially remains (see Fig.~\ref{fig:N_body_snapshots}). With our model however, the combination of considerable collisional losses and more ejections of water-rich bodies (presumably due to longer accretion times) strongly reduces final water contents of cJS planets as well, and brings them much closer to those of eJS scenarios -- and Earth-like water contents (Sect.~\ref{sect:water_inventories_of_terrestrial_planet_systems}).

\item Due to the quick removal of most water-rich material from systems with eJS architectures, water delivery remains very stochastic even with our realistic collision model, where the (chaotic) fate of individual water-rich embryos often becomes decisive. With cJS water delivery is more robust, since usually a much higher number of water-rich bodies survives and slowly diffuses inwards.

\item We studied water delivery and associated accretion histories of potentially habitable planets (0.75\,-\,1.5\,au) in particular detail and found that only very few -- often one or two -- decisive collisions deliver the vast majority of a planet's final inventory (Sects.~\ref{sect:accretion_water_delivery_pot_hab_planets}, \ref{sect:example_pot_hab_planets}, \ref{sect:discussion_water_delivery_pot_hab_planets} and Fig.~\ref{fig:pot_hab_planets_over_t}), at least if embryos form initially throughout the whole disk and with a relatively small planetesimal fraction of $\sim$\,0.3. These encounters are predominantly with embryo-sized or larger bodies and only rarely with water-rich planetesimals, which either get removed or incorporated into the delivering embryos and protoplanets first. Since only a few collisions (accretionary and hit-and-run alike) seem to clearly dominate final water delivery, it naturally becomes an intrinsically stochastic process.

\item Sixteen of the 18 potentially habitable planets formed with our model accrete from dry material in the inner disk before significant water delivery happens, typically after the planet has reached at least 50\,\%, often already 90\,\%, of its final mass (Sect.~\ref{sect:timing_accretion_water_delivery_pot_hab_planets}). Already long average intervals between giant impacts (several up to tens of Myr) then potentially result in repeated ocean formation between all-vaporizing collisions. The two remaining planets start as water-rich embryos in the outer disk and end up as very water-rich `ocean planets' around 1\,au (see Fig.~\ref{fig:pot_hab_planets_over_t} and Sect.~\ref{sect:discussion_water_delivery_pot_hab_planets}).

\item Water losses in individual collisions are relatively low on average, mostly below 20\,-\,30\,\% (Fig.~\ref{fig:total_water_loss_over_t}, Sect.~\ref{sect:global_collision_characteristics}). Encounters in the direct accretion histories of (potentially habitable) planets generally lead to lower values mostly below 10\,\%, but can nevertheless result in considerable cumulative losses.
Growing planets in our scenarios experience typically 20\,-\,30 collisions overall, and still around 10 encounters involving only embryo-sized or larger bodies (Sect.~\ref{sect:collision_characteristics_pot_hab}).
Collisional erosion clearly preferentially strips water and upper mantle material.

\item Final water inventories are typically reduced by a factor of two or more compared to simple PM. This conclusion is based not only on PM comparison runs, but also on cumulative collision losses (Fig.~\ref{fig:pot_hab_planets_cum_col_water_losses_over_water}) and a `synthetic PM' approach, computed from results with our collision model (Sect.~\ref{sect:example_pot_hab_planets}).

\item Hit-and-run is a frequent outcome during late-stage accretion with occurrence rates around 50\%, consistent with earlier studies (Sects.~\ref{sect:global_collision_characteristics} and \ref{sect:collision_characteristics_pot_hab}). This has strong implications for water transport to growing planets, since often only a small fraction of the impactor's water budget (and mass in general) is transferred to the target body, especially compared to more head-on, accretionary collisions (cf.\ Fig.~\ref{fig:SPH_snapshots}).

\item Our results indicate that (hit-and-run) interactions among those protoplanets that eventually grow into planets are rare, and their final growth phases are mostly rather isolated (from each other). Even though hit-and-run encounters with larger bodies can be very transformative (for the smaller target), this also actually happens only rarely to growing planets in a realistic dynamical environment (see Sect.~\ref{sect:discussion_system-wide_collisional_evolution_water_transport}).

\item The often claimed justification for PM that hit-and-run survivors mostly re-collide (and merge) soon after the initial encounter is not supported by our results and occurs only in the minority of cases (Sect.~\ref{sect:collision_characteristics_pot_hab}). Therefore, it is crucial to include proper treatment of water transfer and loss for hit-and-run, and to track the further independent evolution of both survivors.

\item An interesting additional result is the high formation rate of small, close-in, iron-rich planets resembling Mercury (Fig.~\ref{fig:final_planets_III}), which seem to robustly form in a long sequence of high-velocity mantle stripping collisions instead of one or a few chance events (see Sect.~\ref{sect:discussion_system-wide_collisional_evolution_water_transport}). However, our framework was primarily designed to model collisional water transport, where considerable refractory debris is often neglected.

\end{itemize}

\begin{acknowledgements}
We thank John Chambers for many helpful comments that improved our manuscript.
Simulations in this paper made use of the REBOUND code which can be downloaded freely at \url{http://github.com/hannorein/rebound}.
C.~B. and A.~B. acknowledge support by the FWF Austrian Science Fund projects S11603-N16 and S11608-N16.
The authors acknowledge support by the High Performance and Cloud Computing Group at the Zentrum f{\"u}r Datenverarbeitung of the University of T{\"u}bingen, the state of Baden-W{\"u}rttemberg through bwHPC
and the German Research Foundation (DFG) through grant no.\ INST 37/935-1 FUGG.
\end{acknowledgements}

\bibliographystyle{aa}
%\bibliography{references.bib}
\bibliography{article.bbl}

\begin{thebibliography}{87}
\expandafter\ifx\csname natexlab\endcsname\relax\def\natexlab#1{#1}\fi

\bibitem[{{Asphaug}(2010)}]{Asphaug2010_Similar_sized_collisions_diversity_of_planets}
{Asphaug}, E. 2010, Chemie der Erde / Geochemistry, 70, 199

\bibitem[{{Asphaug} {et~al.}(2006){Asphaug}, {Agnor}, \&
  {Williams}}]{Asphaug2006_hit-and-run}
{Asphaug}, E., {Agnor}, C.~B., \& {Williams}, Q. 2006, \nat, 439, 155

\bibitem[{{Bancelin} {et~al.}(2017){Bancelin}, {Pilat-Lohinger}, {Maindl},
  {Ragossnig}, \&
  {Sch{\"a}fer}}]{Bancelin2017_Orbital_resonances_water_transport_in_binaries}
{Bancelin}, D., {Pilat-Lohinger}, E., {Maindl}, T.~I., {Ragossnig}, F., \&
  {Sch{\"a}fer}, C. 2017, \aj, 153, 269

\bibitem[{{Benz} \&
  {Asphaug}(1999)}]{Benz1999_Catastrophic_disruptions_revisited}
{Benz}, W. \& {Asphaug}, E. 1999, \icarus, 142, 5

\bibitem[{{Bonomo} {et~al.}(2019){Bonomo}, {Zeng}, {Damasso}, {Leinhardt},
  {Justesen}, {Lopez}, {Lund}, {Malavolta}, {Silva Aguirre}, {Buchhave},
  {Corsaro}, {Denman}, {Lopez-Morales}, {Mills}, {Mortier}, {Rice}, {Sozzetti},
  {Vanderburg}, {Affer}, {Arentoft}, {Benbakoura}, {Bouchy},
  {Christensen-Dalsgaard}, {Collier Cameron}, {Cosentino}, {Dressing},
  {Dumusque}, {Figueira}, {Fiorenzano}, {Garc{\'{\i}}a}, {Handberg},
  {Harutyunyan}, {Johnson}, {Kjeldsen}, {Latham}, {Lovis}, {Lundkvist},
  {Mathur}, {Mayor}, {Micela}, {Molinari}, {Motalebi}, {Nascimbeni}, {Nava},
  {Pepe}, {Phillips}, {Piotto}, {Poretti}, {Sasselov}, {S{\'e}gransan}, {Udry},
  \& {Watson}}]{Bonomo2019_GI_likely_origin_Kepler-107_twins}
{Bonomo}, A.~S., {Zeng}, L., {Damasso}, M., {et~al.} 2019, Nature Astronomy
  [\eprint[arXiv]{1902.01316}]

\bibitem[{{Bonsor} {et~al.}(2015){Bonsor}, {Leinhardt}, {Carter}, {Elliott},
  {Walter}, \&
  {Stewart}}]{Bonsor2015_Collisional_origin_to_Earths_non-chondritic_composition}
{Bonsor}, A., {Leinhardt}, Z.~M., {Carter}, P.~J., {et~al.} 2015, \icarus, 247,
  291

\bibitem[{{Burger} {et~al.}(2018){Burger}, {Maindl}, \&
  {Sch{\"a}fer}}]{Burger2018_Hit-and-run}
{Burger}, C., {Maindl}, T.~I., \& {Sch{\"a}fer}, C.~M. 2018, Celestial
  Mechanics and Dynamical Astronomy, 130, 2

\bibitem[{{Burger} {et~al.}(2019){Burger}, {Maindl}, \&
  {Sch{\"a}fer}}]{Burger2019_IAU_proceedings_water_delivery_to_dry_protoplanets_hit_and_run}
{Burger}, C., {Maindl}, T.~I., \& {Sch{\"a}fer}, C.~M. 2019, in IAU Symposium,
  Vol. 345, IAU Symposium, accepted, ed. B.~G. {Elmegreen}, L.~V. {Toth}, \&
  M.~{G{\"u}del}

\bibitem[{{Burger} \& {Sch{\"a}fer}(2017)}]{Burger2017_hydrodynamic_scaling}
{Burger}, C. \& {Sch{\"a}fer}, C.~M. 2017, Proceedings of the First
  Greek-Austrian Workshop on Extrasolar Planetary Systems, 63

\bibitem[{{Cambioni} {et~al.}(2019){Cambioni}, {Asphaug}, {Emsenhuber},
  {Gabriel}, {Furfaro}, \&
  {Schwartz}}]{Cambioni2019_Machine_learning_applied_to_GI}
{Cambioni}, S., {Asphaug}, E., {Emsenhuber}, A., {et~al.} 2019, \apj, 875, 40

\bibitem[{{Canup} {et~al.}(2013){Canup}, {Barr}, \&
  {Crawford}}]{Canup2013_Lunar-forming_impacts_High-resolution_SPH_and_AMR-CTH_simulations}
{Canup}, R.~M., {Barr}, A.~C., \& {Crawford}, D.~A. 2013, \icarus, 222, 200

\bibitem[{{Canup} \&
  {Pierazzo}(2006)}]{Canup2006_Water_planet-scale_collisions}
{Canup}, R.~M. \& {Pierazzo}, E. 2006, in Lunar and Planetary Science
  Conference, Vol.~37, 37th Annual Lunar and Planetary Science Conference, ed.
  S.~{Mackwell} \& E.~{Stansbery}

\bibitem[{{Carter} {et~al.}(2018){Carter}, {Leinhardt}, {Elliott}, {Stewart},
  \& {Walter}}]{Carter2018_Collisional_stripping_of_crusts}
{Carter}, P.~J., {Leinhardt}, Z.~M., {Elliott}, T., {Stewart}, S.~T., \&
  {Walter}, M.~J. 2018, Earth and Planetary Science Letters, 484, 276

\bibitem[{{Carter} {et~al.}(2015){Carter}, {Leinhardt}, {Elliott}, {Walter}, \&
  {Stewart}}]{Carter2015_Compositional_evolution_during_accretion}
{Carter}, P.~J., {Leinhardt}, Z.~M., {Elliott}, T., {Walter}, M.~J., \&
  {Stewart}, S.~T. 2015, \apj, 813, 72

\bibitem[{{Chambers}(2001)}]{Chambers2001_Making_more_terrestrial_planets}
{Chambers}, J.~E. 2001, \icarus, 152, 205

\bibitem[{{Chambers}(2013)}]{Chambers2013_Late-stage_accretion_hit-and-run_fragmentation}
{Chambers}, J.~E. 2013, \icarus, 224, 43

\bibitem[{{Cowan} \&
  {Abbot}(2014)}]{Cowan2014_Water_cycling_ocean_mantle_Super_Earths_need_not_be_waterworlds}
{Cowan}, N.~B. \& {Abbot}, D.~S. 2014, \apj, 781, 27

\bibitem[{{D'Angelo} {et~al.}(2019){D'Angelo}, {Cazaux}, {Kamp}, {Thi}, \&
  {Woitke}}]{DAngelo2019_Monte_Carlo_sims_of_forsterite_hydration}
{D'Angelo}, M., {Cazaux}, S., {Kamp}, I., {Thi}, W.~F., \& {Woitke}, P. 2019,
  \aap, 622, A208

\bibitem[{{de Niem} {et~al.}(2012){de Niem}, {K{\"u}hrt}, {Morbidelli}, \&
  {Motschmann}}]{DeNiem2012_Atmospheric_erosion_and_replenishment_by_impacts}
{de Niem}, D., {K{\"u}hrt}, E., {Morbidelli}, A., \& {Motschmann}, U. 2012,
  \icarus, 221, 495

\bibitem[{{Dvorak} {et~al.}(2015){Dvorak}, {Maindl}, {Burger}, {Sch{\"a}fer},
  \&
  {Speith}}]{DvorakMaindlBurger2015_Nonlinear_phenomena_complex_systems_article}
{Dvorak}, R., {Maindl}, T.~I., {Burger}, C., {Sch{\"a}fer}, C., \& {Speith}, R.
  2015, Nonlinear Phenomena in Complex Systems, Vol.18, No.3, pp.~310-325, 18,
  310

\bibitem[{{Emsenhuber} \&
  {Asphaug}(2019)}]{Emsenhuber2019_Fate_of_Runner_in_har}
{Emsenhuber}, A. \& {Asphaug}, E. 2019, \apj, 875, 95

\bibitem[{{Emsenhuber} {et~al.}(2018){Emsenhuber}, {Jutzi}, \&
  {Benz}}]{Emsenhuber2018_Mars-scale_collisions_eos_material_rheologies}
{Emsenhuber}, A., {Jutzi}, M., \& {Benz}, W. 2018, \icarus, 301, 247

\bibitem[{{Fischer} \&
  {Ciesla}(2014)}]{Fischer2014_Dynamics_terrestrial_planets_large_number_N-body_sims}
{Fischer}, R.~A. \& {Ciesla}, F.~J. 2014, Earth and Planetary Science Letters,
  392, 28

\bibitem[{{Genda} \& {Abe}(2003)}]{Genda2003_proto-atmosphere_giant_impacts}
{Genda}, H. \& {Abe}, Y. 2003, \icarus, 164, 149

\bibitem[{{Genda} \&
  {Abe}(2005)}]{Genda2005_atmospheric_loss_giant_impact_phase}
{Genda}, H. \& {Abe}, Y. 2005, \nat, 433, 842

\bibitem[{{Genda} {et~al.}(2015){Genda}, {Fujita}, {Kobayashi}, {Tanaka}, \&
  {Abe}}]{Genda2015_Resolution_dependence_of_collisions}
{Genda}, H., {Fujita}, T., {Kobayashi}, H., {Tanaka}, H., \& {Abe}, Y. 2015,
  \icarus, 262, 58

\bibitem[{{Genda} {et~al.}(2017){Genda}, {Iizuka}, {Sasaki}, {Ueno}, \&
  {Ikoma}}]{Genda2017_Hybrid_code_Ejection_of_iron-bearing_fragments}
{Genda}, H., {Iizuka}, T., {Sasaki}, T., {Ueno}, Y., \& {Ikoma}, M. 2017, Earth
  and Planetary Science Letters, 470, 87

\bibitem[{{Genda} {et~al.}(2011){Genda}, {Kokubo}, \&
  {Ida}}]{Genda2011_Direct_coupling_N-body_SPH}
{Genda}, H., {Kokubo}, E., \& {Ida}, S. 2011, in Lunar and Planetary Science
  Conference, Vol.~42, Lunar and Planetary Science Conference, 2090

\bibitem[{{Genda} {et~al.}(2012){Genda}, {Kokubo}, \&
  {Ida}}]{Genda2012_Merging_criteria_for_giant_impacts}
{Genda}, H., {Kokubo}, E., \& {Ida}, S. 2012, \apj, 744, 137

\bibitem[{{Haghighipour} {et~al.}(2018){Haghighipour}, {Maindl}, {Sch{\"a}fer},
  \& {Wandel}}]{Haghighipour2018_Main-belt_comets_porosity}
{Haghighipour}, N., {Maindl}, T.~I., {Sch{\"a}fer}, C.~M., \& {Wandel}, O.~J.
  2018, \apj, 855, 60

\bibitem[{{Haghighipour} \&
  {Raymond}(2007)}]{Haghighipour2007_Hab_planet_formation_binaries}
{Haghighipour}, N. \& {Raymond}, S.~N. 2007, \apj, 666, 436

\bibitem[{{Hamano} {et~al.}(2013){Hamano}, {Abe}, \&
  {Genda}}]{Hamano2013_Oceans_after_solidification_of_magma_ocean}
{Hamano}, K., {Abe}, Y., \& {Genda}, H. 2013, \nat, 497, 607

\bibitem[{{Hernandez} \&
  {Dehnen}(2017)}]{Hernandez2017_Symplectic_integrators_for_planetary_systems_error_analysis_and_comparison}
{Hernandez}, D.~M. \& {Dehnen}, W. 2017, \mnras, 468, 2614

\bibitem[{{Inamdar} \&
  {Schlichting}(2016)}]{Inamdar2016_Stealing_the_gas_diversity_exoplanet_densities}
{Inamdar}, N.~K. \& {Schlichting}, H.~E. 2016, \apjl, 817, L13

\bibitem[{{Izidoro} {et~al.}(2013){Izidoro}, {de Souza Torres}, {Winter}, \&
  {Haghighipour}}]{Izidoro2013_Compound_model_origin_water}
{Izidoro}, A., {de Souza Torres}, K., {Winter}, O.~C., \& {Haghighipour}, N.
  2013, \apj, 767, 54

\bibitem[{{Johansen} \&
  {Lambrechts}(2017)}]{Johansen2017_Forming_planets_via_pebble_accretion}
{Johansen}, A. \& {Lambrechts}, M. 2017, Annual Review of Earth and Planetary
  Sciences, 45, 359

\bibitem[{{Jutzi}(2015)}]{Jutzi2015_pressure_dependent_failure_models}
{Jutzi}, M. 2015, \planss, 107, 3

\bibitem[{{Kasting} {et~al.}(1993){Kasting}, {Whitmire}, \&
  {Reynolds}}]{Kasting1993_Habitable_zones}
{Kasting}, J.~F., {Whitmire}, D.~P., \& {Reynolds}, R.~T. 1993, \icarus, 101,
  108

\bibitem[{{Kokubo} \&
  {Genda}(2010)}]{Kokubo2010_Formation_under_realistic_accretion}
{Kokubo}, E. \& {Genda}, H. 2010, \apjl, 714, L21

\bibitem[{{Kokubo} \&
  {Ida}(2002)}]{Kokubo2002_Formation_of_protoplanet_systems_oligarchic_growth}
{Kokubo}, E. \& {Ida}, S. 2002, \apj, 581, 666

\bibitem[{{Kokubo} {et~al.}(2006){Kokubo}, {Kominami}, \&
  {Ida}}]{Kokubo2006_Formation_of_terrestrial_planets_from_protoplanets_I}
{Kokubo}, E., {Kominami}, J., \& {Ida}, S. 2006, \apj, 642, 1131

\bibitem[{{Kopparapu} {et~al.}(2013){Kopparapu}, {Ramirez}, {Kasting}, {Eymet},
  {Robinson}, {Mahadevan}, {Terrien}, {Domagal-Goldman}, {Meadows}, \&
  {Deshpande}}]{Kopparapu2013_Habitable_zones}
{Kopparapu}, R.~K., {Ramirez}, R., {Kasting}, J.~F., {et~al.} 2013, \apj, 765,
  131

\bibitem[{{Lammer} {et~al.}(2019){Lammer}, {Leitzinger}, {Scherf}, {Odert},
  {Burger}, {Kubyshkina}, {Johnstone}, {Maindl}, {Sch{\"a}fer}, {G{\"u}del},
  {Tosi}, {Nikolaou}, {Marcq}, {Erkaev}, {Noack}, {Kislyakova}, {Fossati},
  {Pilat-Lohinger}, {Ragossnig}, \& {Dorfi}}]{Lammer2019_atmosphere_paper}
{Lammer}, H., {Leitzinger}, M., {Scherf}, M., {et~al.} 2019,
  \textnormal{submitted to} \icarus

\bibitem[{{Lebrun} {et~al.}(2013){Lebrun}, {Massol}, {Chassefi{\`e}re},
  {Davaille}, {Marcq}, {Sarda}, {Leblanc}, \&
  {Brandeis}}]{Lebrun2013_Magma_ocean_with_atmosphere_water_condensation}
{Lebrun}, T., {Massol}, H., {Chassefi{\`e}re}, E., {et~al.} 2013, Journal of
  Geophysical Research (Planets), 118, 1155

\bibitem[{{L{\'e}cuyer} {et~al.}(1998){L{\'e}cuyer}, {Gillet}, \&
  {Robert}}]{Lecuyer1998_hydrogen_isotope_composition_seawater_and_global_water_cycle}
{L{\'e}cuyer}, C., {Gillet}, P., \& {Robert}, F. 1998, Chemical Geology, 145,
  249

\bibitem[{{Leinhardt} {et~al.}(2015){Leinhardt}, {Dobinson}, {Carter}, \&
  {Lines}}]{Leinhardt2015_Numerically_predicted_signatures_of_TPF}
{Leinhardt}, Z.~M., {Dobinson}, J., {Carter}, P.~J., \& {Lines}, S. 2015, \apj,
  806, 23

\bibitem[{{Leinhardt} \&
  {Stewart}(2012)}]{Leinhardt2012_Collisions_between_gravity-dominated_bodies_I}
{Leinhardt}, Z.~M. \& {Stewart}, S.~T. 2012, \apj, 745, 79

\bibitem[{{Levison} {et~al.}(2012){Levison}, {Duncan}, \&
  {Thommes}}]{Levison2012_Lagrangian_integrator_planet_formation}
{Levison}, H.~F., {Duncan}, M.~J., \& {Thommes}, E. 2012, \aj, 144, 119

\bibitem[{{Lichtenberg} {et~al.}(2019){Lichtenberg}, {Golabek}, {Burn},
  {Meyer}, {Alibert}, {Gerya}, \&
  {Mordasini}}]{Lichtenberg2019_water_budget_dichotomy_from_26Al_heating}
{Lichtenberg}, T., {Golabek}, G.~J., {Burn}, R., {et~al.} 2019, Nature
  Astronomy [\eprint[arXiv]{1902.04026}]

\bibitem[{{Maindl} {et~al.}(2015){Maindl}, {Dvorak}, {Lammer}, {G{\"u}del},
  {Sch{\"a}fer}, {Speith}, {Odert}, {Erkaev}, {Kislyakova}, \&
  {Pilat-Lohinger}}]{MaindlDvorakLammer2015_Impact_heating_on_early_Mars}
{Maindl}, T.~I., {Dvorak}, R., {Lammer}, H., {et~al.} 2015, \aap, 574, A22

\bibitem[{{Maindl} {et~al.}(2014){Maindl}, {Dvorak}, {Sch{\"a}fer}, \&
  {Speith}}]{Maindl2014_Fragmentation_of_planetesimals_water}
{Maindl}, T.~I., {Dvorak}, R., {Sch{\"a}fer}, C., \& {Speith}, R. 2014, in IAU
  Symposium, Vol. 310, IAU Symposium, 138--141

\bibitem[{{Maindl} {et~al.}(2017){Maindl}, {Sch{\"a}fer}, {Haghighipour},
  {Burger}, \& {Dvorak}}]{MaindlSchaeferHaghighipourBurger2017_water_transport}
{Maindl}, T.~I., {Sch{\"a}fer}, C.~M., {Haghighipour}, N., {Burger}, C., \&
  {Dvorak}, R. 2017, Proceedings of the First Greek-Austrian Workshop on
  Extrasolar Planetary Systems, 137

\bibitem[{{Malamud} {et~al.}(2018){Malamud}, {Perets}, {Sch{\"a}fer}, \&
  {Burger}}]{Malamud2018_Moonfalls}
{Malamud}, U., {Perets}, H.~B., {Sch{\"a}fer}, C., \& {Burger}, C. 2018,
  \mnras, 479, 1711

\bibitem[{{Marcus} {et~al.}(2010){Marcus}, {Sasselov}, {Stewart}, \&
  {Hernquist}}]{Marcus2010_Icy_Super_Earths_max_water_content}
{Marcus}, R.~A., {Sasselov}, D., {Stewart}, S.~T., \& {Hernquist}, L. 2010,
  \apjl, 719, L45

\bibitem[{{Marcus} {et~al.}(2009){Marcus}, {Stewart}, {Sasselov}, \&
  {Hernquist}}]{Marcus2009_Collisional_stripping_of_Super-Earths}
{Marcus}, R.~A., {Stewart}, S.~T., {Sasselov}, D., \& {Hernquist}, L. 2009,
  \apjl, 700, L118

\bibitem[{{Marty}(2012)}]{Marty2012_Origins_of_volatiles_on_Earth}
{Marty}, B. 2012, Earth and Planetary Science Letters, 313, 56

\bibitem[{{McNeil} {et~al.}(2005){McNeil}, {Duncan}, \&
  {Levison}}]{McNeil2005_Effects_of_typeI_migration_on_terrestrial_planet_formation}
{McNeil}, D., {Duncan}, M., \& {Levison}, H.~F. 2005, \aj, 130, 2884

\bibitem[{{Melosh}(1989)}]{Melosh1989_Impact_cratering}
{Melosh}, H.~J. 1989, {Impact cratering: A geologic process}

\bibitem[{{Meng} {et~al.}(2014){Meng}, {Su}, {Rieke}, {Stevenson}, {Plavchan},
  {Rujopakarn}, {Lisse}, {Poshyachinda}, \&
  {Reichart}}]{Meng2014_Large_impacts_around_solar-analog_star}
{Meng}, H.~Y.~A., {Su}, K.~Y.~L., {Rieke}, G.~H., {et~al.} 2014, Science, 345,
  1032

\bibitem[{{Morbidelli} {et~al.}(2000){Morbidelli}, {Chambers}, {Lunine},
  {Petit}, {Robert}, {Valsecchi}, \&
  {Cyr}}]{Morbidelli2000_Source_regions_for_water}
{Morbidelli}, A., {Chambers}, J., {Lunine}, J.~I., {et~al.} 2000, Meteoritics
  and Planetary Science, 35, 1309

\bibitem[{{Morishima} {et~al.}(2010){Morishima}, {Stadel}, \&
  {Moore}}]{Morishima2010_N-body_sims_including_gas_and_giant_planets}
{Morishima}, R., {Stadel}, J., \& {Moore}, B. 2010, \icarus, 207, 517

\bibitem[{{Morlok} {et~al.}(2014){Morlok}, {Mason}, {Anand}, {Lisse},
  {Bullock}, \&
  {Grady}}]{Morlok2014_Dust_from_collisions_probe_composition_exoplanets}
{Morlok}, A., {Mason}, A.~B., {Anand}, M., {et~al.} 2014, \icarus, 239, 1

\bibitem[{{Noack} {et~al.}(2016){Noack}, {H{\"o}ning}, {Rivoldini},
  {Heistracher}, {Zimov}, {Journaux}, {Lammer}, {Van Hoolst}, \&
  {Bredeh{\"o}ft}}]{Noack2016_Water-rich_planets_how_habitable}
{Noack}, L., {H{\"o}ning}, D., {Rivoldini}, A., {et~al.} 2016, \icarus, 277,
  215

\bibitem[{{O'Brien} {et~al.}(2018){O'Brien}, {Izidoro}, {Jacobson}, {Raymond},
  \&
  {Rubie}}]{OBrien2018_The_Delivery_of_Water_During_Terrestrial_Planet_Formation}
{O'Brien}, D.~P., {Izidoro}, A., {Jacobson}, S.~A., {Raymond}, S.~N., \&
  {Rubie}, D.~C. 2018, \ssr, 214, 47

\bibitem[{{O'Brien} {et~al.}(2006){O'Brien}, {Morbidelli}, \&
  {Levison}}]{OBrien2006_Terrestrial_planet_formation_with_strong_dynamical_friction}
{O'Brien}, D.~P., {Morbidelli}, A., \& {Levison}, H.~F. 2006, \icarus, 184, 39

\bibitem[{{O'Brien} {et~al.}(2014){O'Brien}, {Walsh}, {Morbidelli}, {Raymond},
  \& {Mandell}}]{OBrien2014_Water_delivery_and_giant_impacts_in_Grand_Tack}
{O'Brien}, D.~P., {Walsh}, K.~J., {Morbidelli}, A., {Raymond}, S.~N., \&
  {Mandell}, A.~M. 2014, \icarus, 239, 74

\bibitem[{{Odert} {et~al.}(2018){Odert}, {Lammer}, {Erkaev}, {Nikolaou},
  {Lichtenegger}, {Johnstone}, {Kislyakova}, {Leitzinger}, \&
  {Tosi}}]{Odert2018_Escape_and_fractionation_of_volatiles_from_growing_protoplanets}
{Odert}, P., {Lammer}, H., {Erkaev}, N.~V., {et~al.} 2018, \icarus, 307, 327

\bibitem[{{Quintana} {et~al.}(2016){Quintana}, {Barclay}, {Borucki}, {Rowe}, \&
  {Chambers}}]{Quintana2016_Frequency_giant_impacts_Earth-like_worlds}
{Quintana}, E.~V., {Barclay}, T., {Borucki}, W.~J., {Rowe}, J.~F., \&
  {Chambers}, J.~E. 2016, \apj, 821, 126

\bibitem[{{Quintana} \&
  {Lissauer}(2014)}]{Quintana2014_Effect_of_planets_beyond_ice_line_on_volatile_accretion}
{Quintana}, E.~V. \& {Lissauer}, J.~J. 2014, \apj, 786, 33

\bibitem[{{Raymond} {et~al.}(2004){Raymond}, {Quinn}, \&
  {Lunine}}]{Raymond2004_Dynamical_sims_water_delivery}
{Raymond}, S.~N., {Quinn}, T., \& {Lunine}, J.~I. 2004, \icarus, 168, 1

\bibitem[{{Raymond} {et~al.}(2006){Raymond}, {Quinn}, \&
  {Lunine}}]{Raymond2006_High_res_sims_planet_formation_I}
{Raymond}, S.~N., {Quinn}, T., \& {Lunine}, J.~I. 2006, \icarus, 183, 265

\bibitem[{{Raymond} {et~al.}(2007){Raymond}, {Quinn}, \&
  {Lunine}}]{Raymond2007_High_res_sims_planet_formation_II}
{Raymond}, S.~N., {Quinn}, T., \& {Lunine}, J.~I. 2007, Astrobiology, 7, 66

\bibitem[{{Rein} \& {Liu}(2012)}]{Rein2012_REBOUND_introduction}
{Rein}, H. \& {Liu}, S.-F. 2012, \aap, 537, A128

\bibitem[{{Rein} \& {Spiegel}(2015)}]{Rein2015_IAS15_integrator}
{Rein}, H. \& {Spiegel}, D.~S. 2015, \mnras, 446, 1424

\bibitem[{{Rein} \& {Tamayo}(2015)}]{Rein2015_WHFAST_integrator}
{Rein}, H. \& {Tamayo}, D. 2015, \mnras, 452, 376

\bibitem[{{Reinhardt} \&
  {Stadel}(2017)}]{Reinhardt2017_Numerical_aspects_of_giant_impact}
{Reinhardt}, C. \& {Stadel}, J. 2017, \mnras, 467, 4252

\bibitem[{{Reufer} {et~al.}(2012){Reufer}, {Meier}, {Benz}, \&
  {Wieler}}]{Reufer2012_Hit-and-run_Moon_formation}
{Reufer}, A., {Meier}, M.~M.~M., {Benz}, W., \& {Wieler}, R. 2012, \icarus,
  221, 296

\bibitem[{{Salvador} {et~al.}(2017){Salvador}, {Massol}, {Davaille}, {Marcq},
  {Sarda}, \&
  {Chassefi{\`e}re}}]{Salvador2017_Influence_of_H20_CO2_on_water_evolution_of_rocky_planets}
{Salvador}, A., {Massol}, H., {Davaille}, A., {et~al.} 2017, Journal of
  Geophysical Research (Planets), 122, 1458

\bibitem[{{Sch{\"a}fer} {et~al.}(2016){Sch{\"a}fer}, {Riecker}, {Maindl},
  {Speith}, {Scherrer}, \& {Kley}}]{Schaefer2016_miluphcuda}
{Sch{\"a}fer}, C., {Riecker}, S., {Maindl}, T.~I., {et~al.} 2016, \aap, 590,
  A19

\bibitem[{{Sch{\"a}fer} {et~al.}(2017){Sch{\"a}fer}, {Scherrer}, {Buchwald},
  {Maindl}, {Speith}, \& {Kley}}]{Schaefer2017_Regolith_sampling}
{Sch{\"a}fer}, C.~M., {Scherrer}, S., {Buchwald}, R., {et~al.} 2017, \planss,
  141, 35

\bibitem[{{Schlichting} \&
  {Mukhopadhyay}(2018)}]{Schlichting2018_Atmosphere_Impact_Losses}
{Schlichting}, H.~E. \& {Mukhopadhyay}, S. 2018, \ssr, 214, 34

\bibitem[{{Schlichting} {et~al.}(2015){Schlichting}, {Sari}, \&
  {Yalinewich}}]{Schlichting2015_Atmospheric_mass_loss_impacts}
{Schlichting}, H.~E., {Sari}, R., \& {Yalinewich}, A. 2015, \icarus, 247, 81

\bibitem[{{Stewart} \&
  {Leinhardt}(2012)}]{Stewart2012_Collisions_between_gravity-dominated_bodies_II}
{Stewart}, S.~T. \& {Leinhardt}, Z.~M. 2012, \apj, 751, 32

\bibitem[{{Tillotson}(1962)}]{Tillotson1962_Metallic_EOS}
{Tillotson}, J.~H. 1962, Metallic Equations of State for Hypervelocity Impact,
  Tech. Rep. General Atomic Report GA-3216, General Dynamics, San Diego, CA

\bibitem[{{Tsiganis} {et~al.}(2005){Tsiganis}, {Gomes}, {Morbidelli}, \&
  {Levison}}]{Tsiganis2005_Nice_model}
{Tsiganis}, K., {Gomes}, R., {Morbidelli}, A., \& {Levison}, H.~F. 2005, \nat,
  435, 459

\bibitem[{{Walsh} {et~al.}(2011){Walsh}, {Morbidelli}, {Raymond}, {O'Brien}, \&
  {Mandell}}]{Walsh2011_Grand_Tack}
{Walsh}, K.~J., {Morbidelli}, A., {Raymond}, S.~N., {O'Brien}, D.~P., \&
  {Mandell}, A.~M. 2011, \nat, 475, 206

\bibitem[{{Wyatt} \&
  {Jackson}(2016)}]{Wyatt2016_Insights_into_Planet_Formation_from_debris_disks_II_GI_exosystems}
{Wyatt}, M.~C. \& {Jackson}, A.~P. 2016, \ssr, 205, 231

\end{thebibliography}

\end{document}